\documentclass[acmsmall,screen, nonacm]{acmart}

\usepackage{pifont}
\usepackage{xcolor}
\usepackage{amsmath}
\usepackage{tabularx, booktabs}
\usepackage{multirow}
\usepackage{float}
\usepackage{algorithm}
\usepackage{algorithmic}
\usepackage{graphicx}
\usepackage{tikz}
\usetikzlibrary{positioning}
\usetikzlibrary{arrows.meta,positioning,shapes.geometric}

\newcolumntype{R}[1]{>{\hsize=#1\hsize\raggedright\arraybackslash}X}

\usepackage{tcolorbox}
\tcbuselibrary{skins, breakable}

\newcommand{\cmark}{\ding{51}}%
\newcommand{\xmark}{\ding{55}}%

\definecolor{step1}{HTML}{185FA5}
\definecolor{step2}{HTML}{3B6D11}
\definecolor{step3}{HTML}{854F0B}
\definecolor{step4}{HTML}{534AB7}
\definecolor{step5}{HTML}{0F6E56}
\definecolor{step6}{HTML}{993556}
\definecolor{step7}{HTML}{5F5E5A}

\acmJournal{CSUR}

\begin{document}

\title[Classical, Machine Learning, and Hybrid PDE Solvers]{Classical, Machine Learning, and Hybrid Solvers for Partial Differential Equations: An Evidence-Weighted Review and Decision Framework}
\thanks{This research was supported by funding from Saskatchewan Polytechnic, Faculty of Digital Innovation, Arts and Sciences, and by the 2026 Applied Research Release Time (ARRT) Fund of Saskatchewan Polytechnic.}

\author{Yingchao Huang}
\authornote{Corresponding author.}
\affiliation{%
  \institution{University of Alberta}
  \department{Faculty of Engineering, Electrical \& Computer Engineering Department}
  \city{Edmonton}
  \state{AB}
  \postcode{T6G 2H5}
  \country{Canada}
}
\email{yingcha1@ualberta.ca}

\author{Xin Wang}
\affiliation{%
  \institution{Saskatchewan Polytechnic}
  \department{Faculty of Digital Innovation, Arts \& Sciences}
  \city{Regina}
  \state{SK}
  \postcode{S4S 5X1}
  \country{Canada}
}

\author{Shanshan Yao}
\affiliation{%
  \institution{University of Alberta}
  \department{Civil and Environmental Engineering}
  \city{Edmonton}
  \state{AB}
  \postcode{T6G 2H5}
  \country{Canada}
}

\author{Fanhua Zeng}
\affiliation{%
  \institution{University of Regina}
  \department{Energy Systems Engineering, Faculty of Engineering and Applied Science}
  \city{Regina}
  \state{SK}
  \postcode{S4S 0A2}
  \country{Canada}
}

\author{Wei Peng}
\affiliation{%
  \institution{University of Regina}
  \department{Industrial Systems Engineering, Faculty of Engineering and Applied Science}
  \city{Regina}
  \state{SK}
  \postcode{S4S 0A2}
  \country{Canada}
}



\begin{abstract}
Machine learning for partial differential equations raises a practical question: when are classical, learned, or hybrid solvers justified? This survey compares the three families in a decision framework organized by data availability, query volume, and certification needs. An amortization law, tested on parametric Darcy flow, shows that surrogates trained on solver-generated data break even only after many queries, beyond thirty thousand for errors below one percent. An evidence ladder, a benchmark reanalysis, and three case studies show that claims of speedup and generalization carry the weakest evidence. Solver choice should follow regime, query volume, and evidence quality, not method novelty.
\end{abstract}

\maketitle

\section{Introduction}
\label{sec:introduction}

Partial differential equations (PDEs) remain a central modeling language for continuum mechanics, transport, waves, diffusion, electromagnetism, and materials, and their numerical solution has long been dominated by finite difference, finite volume, finite element, spectral, multigrid, domain decomposition, and reduced order methods \citep{leveque2002finitevolume,ern2004theory,quarteroni2015reducedbasis}. Machine learning (ML) has added residual based neural approximations, variational neural formulations, supervised neural operators, and physics informed operator variants to this landscape \citep{sirignano2018dgm,raissi2019pinn,lu2021deeponet}, and a practitioner can now find successful examples of classical, ML only, and hybrid solvers on apparently similar PDE tasks \citep{karniadakis2021piml,kovachki2023neuraloperator,takamoto2022pdebench}. Broad surveys document both the breadth of this activity and the recurring difficulty of turning method demonstrations into reliable engineering tools \citep{brunton2020machine,willard2022integrating}, and a recent perspective stresses that traditional numerical algorithms are so mature and scalable that learned solvers are expected to compete with decades of progress \citep{brunton2024promising}.

\subsection{The practitioner's dilemma}
\label{subsec:practitioner_dilemma}

The literature reached an identifiable inflection point between 2022 and 2025. Standardized benchmark suites appeared because systematic cross family comparison was missing, including PDEBench \citep{takamoto2022pdebench}, CFDBench \citep{luo2023cfdbench}, PINNacle \citep{hao2024pinnacle}, and APEBench \citep{koehler2024apebench}, and they revealed that neural solver performance is strongly regime dependent. The most direct evidence is the audit of McGreivy and Hakim \citep{mcgreivy2024weakbaselines}, who found that 60 of 76 ML papers claiming to outperform a standard numerical solver on fluid related PDEs, about 79 percent, compared against a weak baseline, and that reporting biases further inflated the apparent advantage. A controlled study likewise reported that physics informed neural networks (PINNs) did not beat the finite element method on combined solution time and accuracy, although trained models could be cheaper at evaluation \citep{grossmann2023canpinn}. These results point to a credibility problem, not merely to rapid growth.

The consequence is a decision problem rather than only a methodological one. The best solver family depends on the PDE regime, the geometry, the required reliability, the query volume, and the availability of trusted solution data \citep{quarteroni2015reducedbasis,karniadakis2021piml,kovachki2023neuraloperator}. A one-time certified elliptic solve on a complex geometry differs from a many-query parametric surrogate inside optimization or uncertainty quantification \citep{ern2004theory,hesthaven2016certified}; a residual trained function approximator needs no solution corpus, whereas a supervised operator learner usually does \citep{raissi2019pinn,lu2021deeponet}; and one hybrid may retain a conservative discretization with a learned correction while another is a neural surrogate regularized by a PDE residual \citep{li2021pino,barsinai2019learning,kochkov2021mlaccelerated}. The open question is therefore not whether ML can approximate PDE solutions \citep{raissi2019pinn,li2021fno}, but when a learning based solver should replace, augment, or remain subordinate to a classical method under fair cost accounting, counting offline data generation, training, inference, amortized query volume, and out-of-distribution behaviour \citep{takamoto2022pdebench,hao2024pinnacle}.

\subsection{What existing reviews cover and what they leave open}
\label{subsec:existing_reviews_gap}

The review literature is strong but fragmented. The classical corpus is mature and distributed across venues such as \emph{Acta Numerica} and \emph{SIAM Review} \citep{thomee2001fdtofe,tadmor2012review,xu2017amg}. The dedicated ML review corpus is essentially a post 2022 phenomenon and is already crowded: Cuomo et al.\ \citep{cuomo2022scientific} gave the most influential broad treatment of PINNs, Farea et al.\ \citep{farea2024understanding} and Luo et al.\ \citep{luo2025pinnreview} mapped the PINN landscape, Liu et al.\ \citep{liu2025neuraloperators} surveyed neural operator architectures, and Kovachki et al.\ \citep{kovachki2023neuraloperator} supplied a function space framework for operator learning. Only a handful of reviews bridge the paradigms, chiefly at the ML and domain decomposition interface \citep{heinlein2021combining,klawonn2024mldd}. The limitation these reviews share is orientation rather than quality: they organize the field by architecture, loss formulation, and application area rather than by the decision a practitioner faces. They rarely force comparison against classical solvers and reduced order models (ROMs) under matched cost accounting \citep{karniadakis2021piml,cuomo2022scientific}, the paired corpus behind supervised operator learning can hide a substantial offline solver cost \citep{lu2021deeponet,li2021fno}, and benchmark suites support solver selection only to the extent that they include strong classical and ROM baselines, statistical reporting, out of distribution tests, and transparent data generation cost \citep{takamoto2022pdebench,hao2024pinnacle}. In many-query parametric regimes ROMs are a genuine competitor to operator learning, since both require an offline stage before cheap online evaluation \citep{hesthaven2016certified,quarteroni2015reducedbasis}.

Table~\ref{tab:review_landscape} positions representative reviews against the two properties a solver selection reader most needs: whether they compare across the classical, ML, and hybrid families, and whether they grade evidence against fair baselines and transparent cost accounting. Classical reviews are authoritative within one family but do not engage learning based solvers; the ML reviews are architecture centric; the bridging reviews are confined to one slice of the hybrid space; and the two most comparative works are an evidence audit and a controlled study rather than a selection framework \citep{mcgreivy2024weakbaselines,grossmann2023canpinn}. The closest work is the recent preprint of Nooraiepour et al.\ \citep{nooraiepour2026pdeage}, which spans all three families within a broader epistemological framework that contrasts deductive and inductive paradigms, treats data availability as a design dimension, decomposes hybrid error into discretization, neural, and coupling terms, and states a break-even formula for training cost. It discusses weak baselines and amortization but does not grade individual claims or reproduce published results, and these two activities, per-claim evidence grading and replication-based auditing, are where the present review differs. A further architecture centric survey would add little \citep{cuomo2022scientific,luo2025pinnreview,liu2025neuraloperators}. The under occupied position is the last row of the table, where cross family scope meets per-claim evidence grading and replication, and this is the position the present review is designed to fill.

\subsection{Scope, research questions, and contribution}
\label{subsec:scope_questions_contribution}

This review focuses on forward deterministic and parametric PDE solving, which allows a direct comparison of classical discretization, reduced order modeling, residual trained neural solvers, supervised operator learning, autoregressive emulation, and hybrid designs. Inverse problems are treated only where they clarify data scarcity, supervision type, or solver choice \citep{karniadakis2021piml,raissi2019pinn}; PDE discovery, pure control, and weather or climate modeling are outside the scope. The review is a structured narrative review curated by domain expertise rather than a protocol-driven systematic review. Coverage was assembled by forward and backward citation search from anchor papers in classical numerical analysis, physics informed learning, operator learning, and scientific ML benchmarking, supplemented by targeted searches of SIAM journals, the Journal of Computational Physics, NeurIPS, ICLR, ICML, AAAI, and arXiv (cs.NA, cs.LG, math.NA) for 2017--2025, with emphasis on 2022--2025. Preprints are included when widely cited or when they are the only source for a specific finding, and are flagged in the text. Because the classical, PINN, and operator literatures are already well surveyed (Table~\ref{tab:review_landscape}), the solver families are summarized only to the depth the decision framework requires, and readers are referred to those reviews for method detail.

Six research questions organize the review. \textbf{RQ1} asks when a practitioner should prefer a classical, ML only, or hybrid solver for a given PDE regime, including where hybrid designs are most promising. \textbf{RQ2} asks which PDE characteristics most strongly determine success or failure, such as dimensionality, geometry, stiffness, shocks, chaos, and boundary complexity, and which failure modes recur across PINNs, operator learners, and hybrids. \textbf{RQ3} asks how much training data or simulator generated supervision each family requires, and when amortized inference justifies the training and data generation cost. \textbf{RQ4} asks how credible published claims of accuracy, speedup, generalization, and resolution transfer are once baseline strength, cost accounting, and out of distribution tests are considered. \textbf{RQ5} asks which reporting and reproducibility standards are minimally necessary for trustworthy comparison. \textbf{RQ6} asks which open problems are most defensible for the next three to five years, linked to unresolved decision barriers rather than to new architectures. Table~\ref{tab:rq_map} maps each question to the sections that answer it.

\begin{table}[H]
\centering
\caption{Landscape of representative PDE solver reviews positioned against this review. A check mark denotes substantive coverage, a partial mark denotes coverage limited to one slice or paradigm, and a cross denotes no coverage. ``Cross family'': classical, ML only, and hybrid solvers compared within one framework (\cmark), only two families or one slice (\textsl{partial}), or one family only (\xmark). ``Evidence graded'': claims weighed against fair baselines and cost accounting (\cmark), for a subset only (\textsl{partial}), or not at all (\xmark). Supplement S7 gives the section reference or quotation supporting each mark.}
\label{tab:review_landscape}
\scriptsize
\begin{tabular}{p{0.26\linewidth}p{0.08\linewidth}p{0.28\linewidth}cc}
\hline
\textbf{Representative review (year)} & \textbf{Paradigm} & \textbf{Primary focus} & \textbf{Cross family} & \textbf{Evidence graded} \\
\hline
Thom{\'e}e (2001) \citep{thomee2001fdtofe} & Classical & Historical synthesis from finite differences to finite elements & \xmark & \xmark \\
Xu \& Zikatanov (2017) \citep{xu2017amg} & Classical & Algebraic multigrid theory and scalable linear solvers & \xmark & \xmark \\
Cuomo et al.\ (2022) \citep{cuomo2022scientific} & ML only & Broad physics informed neural network review and open issues & \xmark & \xmark \\
Hafiz et al.\ (2024) \citep{hafiz2024largedata} & ML only & Data driven and large data PDE solvers & \xmark & \xmark \\
Farea et al.\ (2024) \citep{farea2024understanding} & ML only & PINN techniques, applications, and challenges & \xmark & \xmark \\
Luo et al.\ (2025) \citep{luo2025pinnreview} & ML only & Comprehensive PINN review for PDE problems & \xmark & \xmark \\
Liu et al.\ (2025) \citep{liu2025neuraloperators} & ML only & Neural operator architectures and variants & \xmark & \xmark \\
Huang et al.\ (2025) \citep{huang2025pdemeetdnn} & ML only & Deep learning for PDE taxonomy with classical baselines noted & \textsl{partial} & \xmark \\
Heinlein et al.\ (2021) \citep{heinlein2021combining} & Hybrid & ML and domain decomposition methods & \textsl{partial} & \xmark \\
Klawonn et al.\ (2024) \citep{klawonn2024mldd} & Hybrid & Survey of ML and domain decomposition & \textsl{partial} & \xmark \\
McGreivy \& Hakim (2024) \citep{mcgreivy2024weakbaselines} & Evidence & Weak baseline audit of ML for fluid PDEs & \textsl{partial} & \cmark \\
Grossmann et al.\ (2024) \citep{grossmann2023canpinn} & Evidence & Controlled comparison of PINNs and the finite element method & \textsl{partial} & \cmark \\
Nooraiepour et al.\ (2026)$^{*}$ \citep{nooraiepour2026pdeage} & All three & Critical synthesis of classical, learning, and hybrid solvers; deductive versus inductive paradigms & \cmark & \textsl{partial} \\
\textbf{This review} & \textbf{All three} & \textbf{Regime based decision framework with data axis, evidence grading, and replication audit} & \cmark & \cmark \\
\hline
\multicolumn{5}{l}{\footnotesize $^{*}$Preprint (arXiv:2603.07655), not yet peer reviewed.} \\
\end{tabular}
\end{table}

The review makes four contributions. First, it provides a cross family decision framework that compares classical, ML only, and hybrid solvers under shared decision regimes rather than separate methodological narratives, and treats data availability and offline solver budget as first-class selection dimensions, building on reviews that discuss data needs per method \citep{karniadakis2021piml,kovachki2023neuraloperator} and a recent preprint that separates equation-driven, data-driven, and multi-fidelity regimes \citep{nooraiepour2026pdeage}. We distinguish four data regimes and use them as the primary selection axis of the decision matrix: no training data (classical methods and ROMs), residual or variational signals only (function learners such as PINNs), a paired input--output solution corpus (supervised operator learners such as Fourier neural operators (FNO) and DeepONet), and trajectory data (autoregressive emulators); this chain also organizes the taxonomy of Section~\ref{sec:taxonomy}. Second, it states an amortization law, a dimensionless form of the break-even condition of McGreivy and Hakim \citep{mcgreivy2024weakbaselines}, from which break-even exceeds the training-set size whenever each training solution costs at least one classical solve and the inference speedup acts mainly through the training cost it implies. It tests the law in a measured break-even study of FNO against multigrid on parametric Darcy flow, and tests the reduced basis comparator on the same data by a proper orthogonal decomposition (POD) analysis (Sections~\ref{subsec:cost_comparison}--\ref{subsec:measured_nstar}). Third, it orders evidence on a ladder from training loss to workflow-level evidence, identifies the proxy gap between each family's commonly reported metric and its decision quantity (Section~\ref{subsec:evidence_ladder}), and grades each claim on two separate axes, evidence quality (Strong, Moderate, Weak, or Red Flag, according to baseline fairness, statistics, cost accounting, out of distribution testing, data generation transparency, and reproducibility) and verdict (Supported, Not supported, Mixed, or Not established). Fourth, it applies this methodology to published benchmark tables, measuring how consistently the suites rank the same methods (Section~\ref{subsec:benchmark_suites}), and to three case studies (Section~\ref{sec:case_studies}): a full computational replication of a Green's function guided operator network and evidence extraction audits of the foundational Fourier neural operator and PINN papers. One rubric applied to these three methods yields differentiated, claim-specific grades rather than a uniform verdict. Full protocols and extraction records are provided in the electronic supplement.

\begin{figure}[H]
\centering
\includegraphics[width=0.8\linewidth]{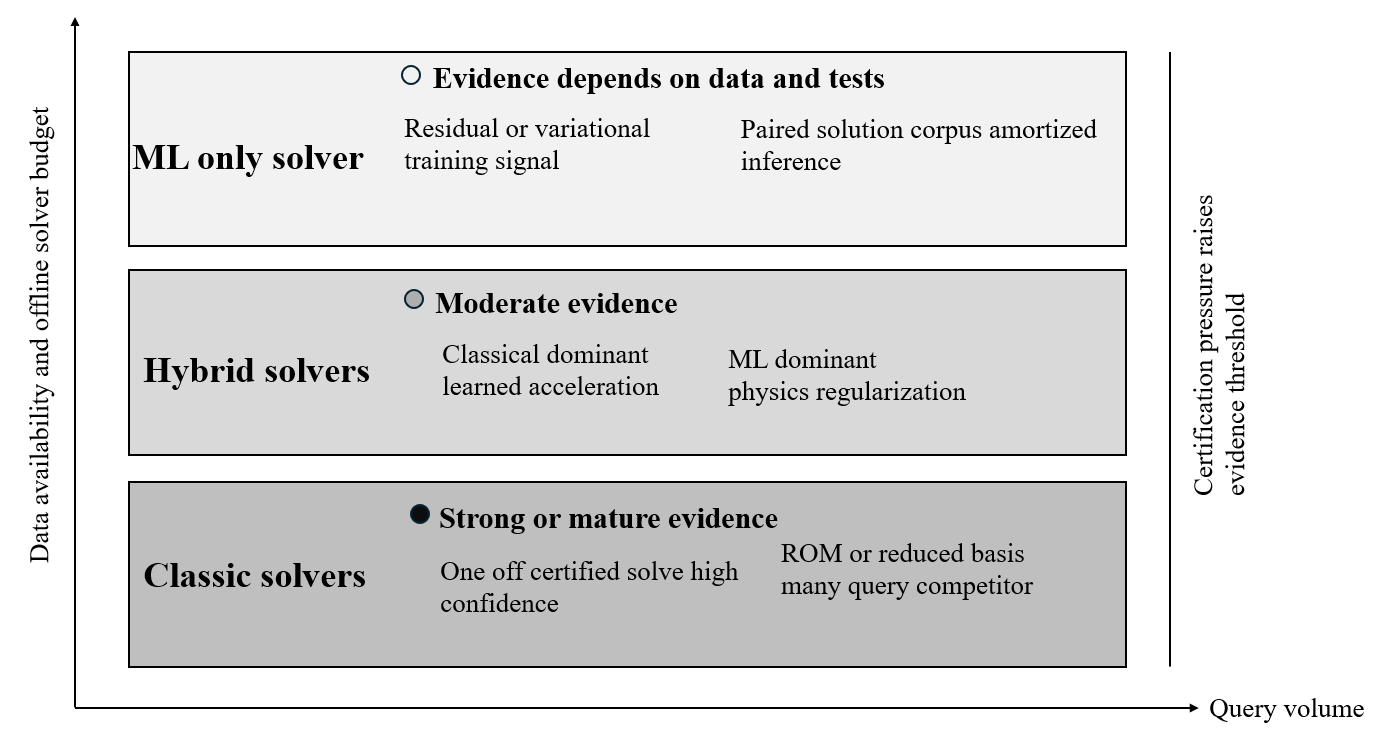}
\caption{Decision landscape for PDE solver selection. The horizontal direction represents query volume, the vertical direction represents data availability and offline solver budget, and the side constraint represents certification pressure. The figure foreshadows the taxonomy in Section~\ref{sec:taxonomy}, the benchmark evidence framework in Section~\ref{sec:benchmark_landscape}, and the decision matrix in Section~\ref{sec:decision_matrix}.}
\label{fig:decision_landscape}
\end{figure}

The practical payoff is easy to illustrate. An engineer choosing a surrogate for a many-query parametric elliptic problem will often see an FNO achieve low relative error on Darcy flow \citep{li2021fno,takamoto2022pdebench}. Those tables rarely surface the offline cost of generating the training corpus, so at the engineer's actual query volume a direct classical solve, or a certified reduced basis method where the parameter dependence allows one, may deliver lower total cost and stronger guarantees. Figure~\ref{fig:decision_landscape} summarizes this premise: classical solvers occupy the low data and high reliability region, supervised operator learning becomes plausible when paired data and many queries justify amortization, and hybrids must be placed according to where learning enters the solver \citep{quarteroni2015reducedbasis,lu2021deeponet,li2021pino}. Certification pressure and out of distribution deployment raise the evidence required before a neural or hybrid method is recommended \citep{takamoto2022pdebench,hao2024pinnacle,hesthaven2016certified}.

\begin{table}[H]
\centering
\scriptsize
\setlength{\tabcolsep}{4pt}
\renewcommand{\arraystretch}{1.1}
\caption{Map of the six research questions to the sections that address them.}
\label{tab:rq_map}
\begin{tabularx}{\linewidth}{R{0.45}R{1.9}R{0.65}}
\hline
\textbf{RQ} & \textbf{Question (abbreviated)} & \textbf{Section(s)} \\
\hline
RQ1 & When to prefer a classical, ML only, or hybrid solver, and where are hybrids most promising? & \S\ref{sec:taxonomy}, \S\ref{sec:solver_families}, \S\ref{sec:decision_matrix} \\
RQ2 & Which PDE characteristics determine success, and which failure modes recur? & \S\ref{sec:solver_families}, \S\ref{sec:decision_matrix} \\
RQ3 & How much data or simulation is required; when does amortization justify cost? & \S\ref{sec:taxonomy}, \S\S\ref{subsec:cost_comparison}--\ref{subsec:measured_nstar}, \S\ref{sec:case_studies} \\
RQ4 & How credible are accuracy, speedup, generalization, and resolution transfer claims, given baseline strength? & \S\ref{sec:benchmark_landscape}, \S\ref{sec:case_studies} \\
RQ5 & What reporting and reproducibility standards are minimally necessary? & \S\ref{sec:benchmark_landscape}, Supplement S1 \\
RQ6 & What are the most defensible open problems for the next 3--5 years? & \S\ref{sec:open_problems} \\
\hline
\end{tabularx}
\end{table}

\begin{tcolorbox}[enhanced, breakable, sharp corners, colback=gray!4, colframe=gray!45, boxrule=0.5pt,
  title={Key findings}, fonttitle=\bfseries\small, coltitle=black, colbacktitle=gray!15, left=4pt, right=4pt]
\small
Findings marked counter-intuitive run against the common reading of the literature. Brackets give the evidence basis, as in the decision trace (Supplement S9): [M] a quantity measured or derived in this review, [G] a claim graded in our case studies, and [X] an external study.
\begin{description}\setlength{\itemsep}{2pt}
\item[F1] \emph{Break-even exceeds the training-set size.} [M] A surrogate trained on $N_{tr}$ solutions, each costing at least one classical solve, must be queried more than $N_{tr}$ times before it saves any cost (Section~\ref{subsec:amortization_law}). Implication: a many-query claim must state its query volume.
\item[F2] \emph{The headline speedup does not measure its own value} [M, X] (counter-intuitive). Break-even is governed by data and training cost, the quantities most often unreported; in the original FNO paper's own showcase application, a $440\times$ per-query speedup becomes about $1.5\times$ end to end. Speed matters through training cost: in our measured Darcy study training costs a fixed $E_{\mathrm{eff}}\approx360$--$1300$ network evaluations per training sample, so a faster network lowers break-even mainly by training faster: at $64^2$ a network ten times faster would cut break-even about eightfold (Sections~\ref{subsec:amortization_law} and~\ref{subsec:measured_nstar}). Implication: report data and training cost before speedup.
\item[F3] \emph{Whether a reduced basis beats an operator depends on parametric structure, and the standard Darcy benchmark is the unfavorable case.} [M, X] On the thresholded Darcy problem of our measured study, a POD basis needs 125--335 modes to match FNO's error, not the 30 of the order-of-magnitude estimate, and the coefficient itself does not compress, so no certified reduced basis with a cheap online stage exists; reassembly alone caps its per-query speedup at about $20\times$ (Section~\ref{subsec:cost_comparison}). Implication: reduced basis is the comparator for many-query claims when the parameter dependence is affine and low dimensional; otherwise its absence should be justified by a decay test.
\item[F4] \emph{Cheap classical solves are hostile to learning.} [M, G] In our replication of a Green's function guided operator network (GON) the training cost equals about $5.7\times10^{6}$ classical solves, so in the regime tested the method never breaks even end to end (Section~\ref{subsec:case_gon}).
\item[F5] \emph{Accurate prediction does not establish structure} [G] (counter-intuitive). A GON model with test error $1.1\times10^{-7}$ embeds a kernel that is asymmetric and violates the boundary condition, even on a problem built so that a correct operator must encode the Green's function (Sections~\ref{subsec:case_gon} and~\ref{subsec:evidence_ladder}).
\item[F6] \emph{Each family fails where its reported metric departs from its decision quantity} [G, X]: residual versus solution error, in-distribution versus deployment error, one-step versus rollout error, and per-query versus end-to-end cost (Table~\ref{tab:proxy_gap}).
\item[F7] \emph{The claims most often cited for solver replacement carry the weakest evidence.} [G] In all three case studies the lowest quality grades or negative verdicts fall on speedup, generalization, efficiency, or structure, while in-distribution accuracy is supported for at least some cases in each (Section~\ref{sec:case_studies}).
\item[F8] \emph{Reported accuracy may not survive faithful reproduction.} [G, M, X] With the original code, architecture, and metric, GON's mean squared errors are 1.8 to 40 times larger than reported on every non-trivial case (Section~\ref{subsec:case_gon}). The ratio is not one-sided: our re-implementation of the FNO Darcy study reached 0.6--0.7 times the reported error (Supplement S4).
\item[F9] \emph{The benchmark can decide the winner.} [M] In published result tables, FNO beats U-Net on every PDEBench task but on only $43\%$ of CFDBench and $20\%$ of APEBench tasks, and eleven PINN variants have eight different winners across eighteen PINNacle tasks, where the task-dependent part of their log error is five to seven times their average difference (Section~\ref{subsec:benchmark_suites}).
\item[F10] \emph{At matched accuracy, finer resolution works against the surrogate} [M] (counter-intuitive). In our measured Darcy study, FNO's error is set by its training data, not by the grid, so the matched classical solve stays coarse while FNO's cost grows: with $10^3$ training samples, break-even rises from about $3\times10^4$ queries at $64^2$ to $9\times10^5$ at $512^2$, and with fewer samples at $512^2$ FNO never breaks even (Section~\ref{subsec:measured_nstar}).
\end{description}
\end{tcolorbox}

\section{Decision regimes and taxonomy}
\label{sec:taxonomy}

The taxonomy is organized by decision regime rather than by algorithm name, because the same family can be appropriate or inappropriate depending on query volume, data availability, regularity, geometry, reliability requirements, and offline budget \citep{quarteroni2015reducedbasis,hesthaven2016certified,karniadakis2021piml}. Five primary regimes are distinguished (Table~\ref{tab:decision_regimes}): a one off high fidelity solve for a fully specified problem \citep{ern2004theory,leveque2002finitevolume}; many-query amortized inference over a parametric family, where an offline stage pays off only if online evaluation is cheap and reliable \citep{quarteroni2015reducedbasis,lu2021deeponet}; a high dimensional or mesh free solve \citep{sirignano2018dgm,raissi2019pinn}; complex or irregular geometry, where meshing, geometry encoding, and cross geometry transfer can dominate cost \citep{li2023geofno,li2023gino}; and long time rollout of nonlinear or chaotic dynamics, where accumulated error and attractor fidelity matter more than one step accuracy \citep{brandstetter2022message,kochkov2021mlaccelerated}. Data informed inverse modeling is a bounded secondary regime, included because it raises the value of data scarce methods that combine sparse observations with PDE constraints \citep{raissi2019pinn,cuomo2022scientific}. Its natural classical competitors are adjoint-based PDE-constrained optimization and variational data assimilation, which are mature tools with established convergence and regularization theory, so a PINN inverse claim is weakened when it omits such a baseline.

Two further axes cut across the regimes. Solution regularity matters because smooth elliptic or parabolic problems, convection dominated flows, shocks, multiscale coefficients, and chaos stress different failure modes: conservative finite volume methods build shock handling into the discretization \citep{leveque2002finitevolume}, whereas smooth network bias, residual imbalance, and boundary treatment make sharp layers difficult for neural function approximators \citep{wang2022ntk,krishnapriyan2021characterizing}. The data regime is treated as a solver characteristic rather than an implementation detail (Table~\ref{tab:data_requirements}): classical solvers consume the mathematical model directly, function learners use residual or variational signals, operator learners usually require a paired solution corpus, autoregressive emulators require trajectories, and hybrids must be classified by the learned component and its supervision source \citep{ern2004theory,lu2021deeponet,kovachki2023neuraloperator}. Whether paired data already exist, can be generated by a trusted offline solver at a counted cost, or cannot be obtained at all then determines which of these families is feasible before any accuracy comparison is made \citep{quarteroni2015reducedbasis,li2021fno}.

A taxonomy figure in Supplement S11 places the families by the role they play in a decision problem, including Green's function guided operator networks (GON) under interpretability guided operator learning, whose decision relevance includes whether the learned structure supports a credible Green's function claim \citep{gu2025explainable}. Table~\ref{tab:decision_regimes} also lists regime level red flags, conditions that make a positive recommendation risky even when a method performed well in nearby examples: an operator learning claim without a feasible corpus budget, a certified use case without error control or a mature classical baseline, a shock dominated problem evaluated only by pointwise error, and a long horizon surrogate evaluated only by short rollout error. The resulting decision rule is conservative and is applied throughout: a family is not recommended outside the regime where its data source, cost structure, and reliability evidence are credible, residual trained solvers are not treated as cost free because optimization and validation can dominate \citep{krishnapriyan2021characterizing}, and hybrids are not treated as automatically robust (Section~\ref{sec:hybrid_solvers}).

\begin{table}[tp]
\centering
\caption{Decision regimes, constraints, and feasible solver families.}
\label{tab:decision_regimes}
\scriptsize
\setlength{\tabcolsep}{3pt}
\renewcommand{\arraystretch}{1.18}
\begin{tabularx}{\linewidth}{R{1.0}R{1.05}R{0.9}R{0.9}R{0.95}R{1.1}R{1.1}}
\hline
\textbf{Decision regime} & \textbf{Typical PDE setting} & \textbf{Data regime} & \textbf{Cost structure} & \textbf{Reliability requirement} & \textbf{Candidate solvers} & \textbf{Red flags} \\
\hline
One off high fidelity solve & Elliptic, parabolic, hyperbolic, or multiphysics with specified domain and data & No training data required & Direct solve cost dominates & Error control and reproducibility often central & FEM, FVM, FDM, adaptive FEM, multigrid, domain decomposition & Neural speedup claim without classical cost anchor or error control \\
Many query amortized inference & Parametric PDE family used in design, inference, optimization, or UQ & Trusted offline solver or reusable paired corpus often needed & Offline--online tradeoff dominates & Stable accuracy across parameter range & ROMs, reduced basis, DeepONet, FNO, PINO, GON variants & Offline data generation cost omitted or query volume too low \\
High dimensional or mesh free solve & High dimensional PDE, scattered collocation, or mesh-difficult representation & Often limited paired data or residual signal & Training and sampling cost dominate & Generalization and optimization reliability central & Deep Galerkin Method (DGM), PINN, Deep Ritz, weak neural formulations, selected hybrid variants & Residual loss reported without independent error or failure cases \\
Complex or irregular geometry & Curved, moving, heterogeneous, or cross-geometry domains & Geometry data or solver-generated corpus & Mesh generation, geometry encoding, and transfer cost dominate & Boundary accuracy and geometry transfer central & FEM, adaptive FEM, GeoFNO, GINO, graph operators, geometry-aware hybrids & Boundary error unreported or train geometry too close to test geometry \\
Long time nonlinear or chaotic rollout & Burgers, Navier--Stokes, Kuramoto--Sivashinsky, reaction-diffusion, or turbulent surrogate & Trajectory data or solver-in-the-loop data often needed & Rollout cost and error accumulation dominate & Stability, conservation, and attractor fidelity central & FVM, spectral solvers, autoregressive emulators, ML-accelerated CFD, hybrid correction & Only one step error reported or no rollout diagnostics \\
Data informed inverse regime & Unknown coefficient, source, BC, material field, or parameter & Sparse observations plus residual or variational signal & Optimization and identifiability cost dominate & Uncertainty and regularization central & PINN-style inverse solvers, adjoint-based methods, hybrid data assimilation & Forward accuracy assumed to imply inverse reliability \\
\hline
\end{tabularx}
\end{table}

\begin{table}[tp]
\centering
\caption{Data requirements by solver family.}
\label{tab:data_requirements}
\scriptsize
\setlength{\tabcolsep}{3pt}
\renewcommand{\arraystretch}{1.18}
\begin{tabularx}{\linewidth}{R{0.85}R{1.2}R{1.05}R{0.85}R{0.9}R{1.15}}
\hline
\textbf{Solver family} & \textbf{Practical data requirement} & \textbf{Training or solve signal} & \textbf{Offline cost} & \textbf{Feasible without paired data} & \textbf{Decision implication} \\
\hline
Classical solvers & No training data required; PDE, geometry, coefficients, BCs, ICs, and discretization needed & Discrete governing equations, variational form, flux balance, or algebraic systems & Mesh construction, assembly, solver setup, preconditioning & High for well-posed forward problems & Default for one-off solves, certification needs, unavailable paired data \\
ROMs and reduced basis & Snapshot solutions or intrusive reduced operators over a parametric family usually needed & Projection, empirical interpolation, greedy sampling, a posteriori error estimators & Offline basis construction and error estimator construction & Low when no solver or snapshots are available & Strong classical many-query competitor when intrusive access and parametric structure exist \\
Function learning solvers & Collocation, boundary, initial data, and PDE residual access needed; labeled pairs may not be required & Residual or variational training signal with BC and IC penalties or constraints & Neural optimization, sampling, loss balancing, and validation & Moderate when PDE is known and optimization is tractable & Useful in data-scarce or inverse settings; not automatically cheap or reliable \\
Operator learning solvers & Paired input--output solution corpus across a parametric PDE family usually required & Supervised map learning from input functions, coefficients, sources, or histories to solution fields & Reference solution generation, corpus storage, training, and validation & Low without simulation or measurement budget & Attractive only when many queries amortize corpus generation and training \\
Autoregressive emulators & Trajectory data from simulations or measurements usually required & One-step or multistep prediction with rollout validation & Long trajectory generation, training, and stability testing & Low without trajectories or simulator access & Candidate for repeated dynamic prediction only when rollout evidence is strong \\
Hybrid solvers & Highly variable: residual signals, paired solutions, solver states, correction targets, matrices, or trajectories & Physics-informed loss, learned updates, corrections, discretizations, or solver-coupled training & Case dependent; often includes both numerical and neural costs & Case dependent & Must be classified by learned component, retained structure, data source, and reliability risk \\
\hline
\end{tabularx}
\end{table}

\section{Solver families through the decision lens}
\label{sec:solver_families}

This section summarizes each solver family only to the depth the decision framework requires: its data profile, the classical comparator it must beat, its recurring failure modes, and the evidence needed before it can be recommended. Method detail is available in the dedicated reviews of classical discretization and multilevel solvers \citep{thomee2001fdtofe,xu2017amg,quarteroni2015reducedbasis}, PINNs \citep{cuomo2022scientific,farea2024understanding,luo2025pinnreview}, neural operators \citep{kovachki2023neuraloperator,liu2025neuraloperators}, and ML and domain decomposition hybrids \citep{heinlein2021combining,klawonn2024mldd}.

\subsection{Classical solvers as regime-specific baselines}
\label{subsec:classical_solvers}

Classical solvers are treated here as rigorous baselines and many-query competitors rather than as background. They need no training data, which is decisive when paired data do not exist or when certification and traceable error matter more than inference speed \citep{ern2004theory,ainsworth2000aposteriori,hesthaven2016certified}. The appropriate comparator is set by the regime (Table~\ref{tab:classical_capability_matrix}): conservative finite volume or discontinuous Galerkin (DG) schemes for shocks and convection \citep{leveque2002finitevolume,cockburn2001rkdg,toro2009riemann}, mixed finite element methods (FEM) for Stokes, Darcy, and Maxwell problems with conservation claims \citep{boffi2013mixed}, adaptive FEM for singularities and heterogeneous coefficients \citep{ainsworth2000aposteriori,bonito2024adaptive}, variational and heterogeneous multiscale methods for learned closures \citep{hughes1998vms,e2003hmm}, and symplectic integrators for long-horizon conservative dynamics \citep{hairer2006geometric}. For elliptic and parabolic solves, multigrid gives near linear complexity \citep{trottenberg2001multigrid,xu2017amg} and the choice of Krylov preconditioner can change solve time by one to two orders of magnitude \citep{saad2003iterative,benzi2002preconditioning}, so a speedup that does not name the preconditioner, hierarchy, tolerance, and parallel hardware is graded as a Red Flag (Section~\ref{sec:benchmark_landscape}).

Reduced order models and reduced basis methods are the main classical competitors to supervised operator learning, because both use an offline stage to make online evaluation cheap \citep{quarteroni2015reducedbasis,hesthaven2016certified,benner2015survey}. Projection based reductions inherit stability and, in favorable cases, certified a posteriori error estimators with an analyzable offline cost \citep{rozza2008reduced}; their limitation is slow Kolmogorov $n$-width decay for transport dominated or sharply parameter dependent problems, including the discontinuous coefficients of the standard Darcy benchmark (Section~\ref{subsec:cost_comparison}), together with the need for intrusive solver access, which are the conditions under which operator learning becomes most compelling \citep{ohlberger2015reduced,kovachki2023neuraloperator}. Classical solvers are not universal winners: mesh generation can dominate complex-geometry workflows \citep{brennerscott2008mathematical}, grid based discretization fails in genuinely high dimension \citep{han2018solving,sirignano2018dgm}, and spectral methods lose exponential convergence for low regularity solutions \citep{canuto2007spectral}. They remain the default for one off high fidelity solves, unavailable paired data, certification needs, or regimes with a mature solver stack, and a learned or hybrid solver becomes credible when it addresses a real limitation of that stack. For the same reason, a speedup or accuracy number is uninterpretable unless the paper names which of the baselines in Table~\ref{tab:classical_capability_matrix} it used.

\subsection{ML only solvers}
\label{sec:ml_only_solvers}

ML only solvers are evaluated here by training signal, data requirement, optimization reliability, generalization, and amortized cost rather than by architectural novelty. Table~\ref{tab:ml_solver_comparison} summarizes the four families, their evidence requirements, and their named failure modes.

\subsubsection{Function learning solvers}
\label{subsec:function_learning}

Function learning solvers approximate the solution of a specified PDE instance from strong form residuals, weak or variational formulations, boundary and initial data, or sparse observations, and need no paired corpus \citep{raissi2019pinn,weinan2018deepritz,zang2020wan}. Their clearest motivation is high dimension, where deep BSDE and Deep Galerkin methods solved nonlinear parabolic PDEs in hundreds of dimensions, an advantage that erodes when sharp gradients or complex geometry reintroduce dense sampling \citep{han2018solving,sirignano2018dgm}. PINNs, the most widely used member, fail on benchmark PDEs of increasing difficulty under naive residual optimization \citep{krishnapriyan2021characterizing}, suffer from loss imbalance and spectral bias \citep{wang2022ntk,rahaman2019spectral}, and can leave boundary errors that dominate the physical error \citep{cuomo2022scientific}. Numerical analysis explains these failures through the spectral and stability properties of the differential operator \citep{deryck2024numerical,mishra2022estimates}, and convergence to the true solution has been proven for linear second-order elliptic and parabolic problems \citep{shin2020convergence}. A small training residual therefore guarantees nothing about solution accuracy (rung P0 of Section~\ref{subsec:evidence_ladder}). Function learning is most credible when the PDE is known, paired data are unavailable, dimension or geometry makes meshing difficult, and the study reports independent reference error, boundary error, multiple seeds, and a classical baseline cost \citep{hao2024pinnacle}.

\begin{table}[tp]
\centering
\caption{Classical solvers as baselines and many query competitors.}
\label{tab:classical_capability_matrix}
\tiny
\setlength{\tabcolsep}{2pt}
\renewcommand{\arraystretch}{1.0}
\begin{tabularx}{\linewidth}{R{0.9}R{0.95}R{1.1}R{0.95}R{1.15}R{1.0}R{1.0}R{0.55}}
\hline
\textbf{Method} & \textbf{Data requirement} & \textbf{Accuracy / certification} & \textbf{Cost profile} & \textbf{Benchmark anchor} & \textbf{Best-suited regimes} & \textbf{Limitations} & \textbf{Refs.} \\
\hline
Finite difference methods & Structured domain, BCs & Strong consistency/stability on structured grids & Efficient on regular grids; fast solvers compatible & Structured grid reference for diffusion, wave, transport & One-off and repeated solves on simple domains & Complex geometry and local adaptivity difficult & \citep{leveque2007finite,quarteroni2007numerical} \\
\hline
Finite volume methods & Mesh/control volumes, fluxes, BCs & Conservation built in; shock capturing by design & Depends on mesh, Riemann solvers, time stepping, limiters & Reference for hyperbolic conservation laws and compressible flow & Convection-dominated, shocks, conservation-critical dynamics & High-order accuracy and multidimensional limiting complex & \citep{leveque2002finitevolume,toro2009riemann} \\
\hline
Discontinuous Galerkin & Element patches, numerical flux, stabilization & High-order accuracy with conservation and upwinding; natural for discontinuities & Element-local solves; block-diagonal mass matrix & High-order reference for convection-dominated and wave problems & High-order convection, wave propagation, shocks & Increased DOF per cell; interface flux design requires care & \citep{cockburn2001rkdg,leveque2002finitevolume} \\
\hline
Mixed FEM & Compatible function space pairs, weak form, saddle-point solver & Local conservation via inf-sup stable element pairs & Mixed/Krylov solver with saddle-point preconditioner & Reference for Stokes, Darcy, Maxwell with structure preservation & Incompressible flow, porous media, electromagnetics & Stable pair selection and saddle-point preconditioning complex & \citep{boffi2013mixed,ern2004theory} \\
\hline
Finite element methods & Weak form, mesh, coefficients, BCs & Flexible geometry; mature variational error analysis & Assembly, mesh generation, linear/nonlinear solve, preconditioning & Reference for elliptic, parabolic, structural, multiphysics & One-off high-fidelity, complex geometry, certified workflows & Mesh generation and solver setup can dominate & \citep{ern2004theory,brennerscott2008mathematical} \\
\hline
Adaptive FEM & Error indicators/estimators, refinement strategy & A posteriori estimation; proven convergence and optimality & Estimator and refinement cost; reduced when error is localized & Framework for singularities, boundary layers, heterogeneous fields & High-accuracy local features; certification-oriented solves & Estimator reliability problem-dependent; implementation-intensive & \citep{ainsworth2000aposteriori,verfurth2013aposteriori,bonito2024adaptive} \\
\hline
Spectral / high-order & Smooth solution, regular domain, suitable basis & Exponential convergence for smooth problems & Dense operators; very high accuracy per DOF & High-accuracy reference for smooth/periodic problems & Smooth high-accuracy solves; spectrally accurate references & Accuracy degrades for low regularity, shocks, complex geometry & \citep{canuto2007spectral,quarteroni2007numerical} \\
\hline
Multigrid / iterative solvers & Discretized algebraic system, hierarchy or preconditioner & Near-optimal linear complexity; mesh-independent convergence & Setup plus iterative cost; decisive at high resolution & Acceleration layer for elliptic and parabolic discretizations & Large-scale one-off and repeated linearized solves & Degrades under strong heterogeneity, indefiniteness, poor hierarchy & \citep{hackbusch1985multigrid,trottenberg2001multigrid,benzi2002preconditioning} \\
\hline
Domain decomposition & Subdomain partition, interface treatment, parallel implementation & Scalable parallel solution; condition number bounds with coarse spaces & Communication, interface iteration, local solver costs & High-performance route for large PDE systems on distributed memory & Large-scale simulation, multiphysics coupling, parallel solves & Interface conditions and coarse spaces require problem-specific design & \citep{toselli2005domain,smith1996domain,dolean2015domain} \\
\hline
ROMs / reduced basis & Snapshots, intrusive structure, or offline solver for parametric family & Certified online error estimates; projection-based stability & Offline basis construction expensive; online solves cheap & Classical many-query alternative for parametric PDEs and optimization & Many-query regimes with reusable parametric structure & Slow Kolmogorov width decay for transport-dominated problems & \citep{quarteroni2015reducedbasis,hesthaven2016certified,benner2015survey,antoulas2005approximation} \\
\hline
\end{tabularx}
\end{table}

\begin{table}[H]
\centering
\caption{ML only solver families, evidence strengths, and failure modes.}
\label{tab:ml_solver_comparison}
\tiny
\setlength{\tabcolsep}{3pt}
\renewcommand{\arraystretch}{1.2}
\begin{tabularx}{\linewidth}{p{0.09\linewidth}p{0.09\linewidth}p{0.11\linewidth}p{0.09\linewidth}p{0.11\linewidth}p{0.11\linewidth}p{0.13\linewidth}X}
\hline
ML family & Representative methods & Training signal & Paired data need & Best use cases & Quantitative failure examples or named stress tests & Required evaluation tests & Representative references \\
\hline
Function learning & PINN, DGM, Deep Ritz, WAN, DFVM & Strong residual, weak form, variational objective, boundary data, initial data, or sparse observations & Usually low for labeled solution pairs, but PDE access is required & Data scarce forward solves, high dimensional settings, and inverse settings with sparse observations & PINN failure modes on benchmark PDEs, neural tangent kernel imbalance, spectral bias, and multiscale training difficulty & Reference error, residual diagnostics, boundary error, sensitivity to collocation, random seeds, and classical baseline cost & \citep{raissi2019pinn,sirignano2018dgm,krishnapriyan2021characterizing} \\
\hline
Operator learning & DeepONet, FNO, GeoFNO, GNOT, U-NO & Supervised input output solution pairs over a parametric PDE family & Usually high because a reusable solution corpus is needed & Many query parametric inference with high amortized query volume & Train test proximity, geometry shift, boundary shift, discontinuities, and uncounted corpus generation cost & Data accounting, OOD tests, resolution tests, baseline comparison with ROMs, and amortized cost & \citep{lu2021deeponet,li2021fno,kovachki2023neuraloperator} \\
\hline
Interpretability guided operator variants & Green's function guided operator networks and related Green's function approaches & Usually paired solution data plus structural guidance or architectural constraints & Usually high unless analytic or measured Green's function information is available & Linear or structured PDE regimes where Green's function interpretation is meaningful & Structural interpretability may not match predictive accuracy unless kernel properties are audited & Predictive error, cost accounting, data efficiency, OOD tests, and Green's function structure checks & \citep{gu2025explainable,lu2021deeponet,kovachki2023neuraloperator} \\
\hline
Autoregressive emulators & Message passing neural PDE solvers and learned rollout models & Trajectory data and one step or multistep prediction losses & High when long simulated trajectories are required & Repeated dynamic prediction and learned acceleration over fixed regimes & Error accumulation, rollout instability, attractor drift, and conservation violation & Long horizon rollout, stability, conservation, physical diagnostics, and statistical behavior & \citep{brandstetter2022message,kochkov2021mlaccelerated,takamoto2022pdebench} \\
\hline
\end{tabularx}
\end{table}

\subsubsection{Operator learning and interpretability guided variants}
\label{subsec:operator_learning}

Operator learners map input functions such as coefficients, sources, boundary data, or previous states to solution functions across a parametric family, and most supervised versions require a paired corpus generated by a classical solver or by measurement \citep{lu2021deeponet,li2021fno}. Generating the corpus can cost more than a small number of direct solves, so a credible claim must report that cost and the query volume that amortizes it (Section~\ref{subsec:amortization_law}). Approximation theory bounds the error in terms of operator regularity and training set size \citep{lanthaler2022deeponet,kovachki2021universal,deryck2023generic}, and geometry aware, transformer, wavelet, and convolutional variants extend the design space without removing the corpus dependence \citep{li2023geofno,li2023gino,hao2023gnot,raonic2023cno}. A cross-family law follows from this theory. Operators that reconstruct the output linearly in a learned basis of dimension $p$, such as DeepONet and PCA-Net, are bounded below by the tail of the eigenvalues of the output covariance \citep{lanthaler2022deeponet}, which is the same Kolmogorov-width barrier that limits linear reduced order models; for linear advection and inviscid Burgers the bound gives an error of order $1/p$ \citep{lanthaler2023nonlinear}. Operators with nonlinear reconstruction, such as FNO and shift-DeepONet, are not subject to it, and on these problems DeepONet errors of $7.95\%$ and $28.5\%$ compare with $0.71\%$ and $1.57\%$ for FNO \citep{lanthaler2023nonlinear}. For transport-dominated or shock-forming families, the decision therefore turns on linear versus nonlinear reconstruction rather than on classical versus learned: a linear-reconstruction operator inherits the weakness of a linear reduced basis, while nonlinear reductions and nonlinear-reconstruction operators need not \citep{ohlberger2015reduced}. Interpretability guided variants such as the Green's function guided operator network (GON) add a structural claim that must be audited separately from predictive accuracy \citep{gu2025explainable}; Section~\ref{sec:case_studies} reports such an audit.

The key risk is that interpolation within a well covered corpus looks like broad generalization, so out of distribution tests should shift parameters, coefficients, boundary conditions, geometry, and noise whenever these are part of the deployment claim \citep{takamoto2022pdebench,hao2024pinnacle}. Resolution transfer has two senses that are often conflated. In the \emph{mathematical} sense, a well-posed neural operator is discretization invariant in the limit and can be evaluated on any discretization without architectural change \citep{kovachki2023neuraloperator}; in the \emph{empirical} sense, a model trained at one resolution must predict accurately at another on held-out data. Benchmarks show that zero-shot transfer degrades once the input distribution departs from training \citep{takamoto2022pdebench,luo2023cfdbench}, so a resolution transfer claim is graded Strong only when demonstrated under an input distribution shift.

\subsubsection{Autoregressive emulators}
\label{subsec:autoregressive_emulators}

Autoregressive emulators learn temporal update maps from trajectory data \citep{brandstetter2022message,pfaff2021meshgraphnets}. Because their own predictions become future inputs, small one step errors accumulate into long-time drift, especially in chaotic regimes \citep{kochkov2021mlaccelerated,takamoto2022pdebench}; refinement based rollout \citep{lippe2023pderefiner} and Hamiltonian networks that conserve energy by construction \citep{greydanus2019hnn} are partial remedies. Credible studies report rollout horizons, stability failures, conserved quantities, and attractor statistics rather than one step error \citep{koehler2024apebench}. Across the ML only families, absent paired data does not mean low compute cost for PINNs, and fast inference does not mean low total cost for operators.

\subsection{Hybrid solvers}
\label{sec:hybrid_solvers}

Hybrid solvers are a heterogeneous design space rather than a single safer category, and they are classified here by where learning enters the pipeline (Table~\ref{tab:hybrid_classification}). Classical dominant hybrids retain a discretization, residual equation, or iterative method and learn a local component, correction, closure, prolongation, or preconditioner; ML dominant hybrids start from a neural surrogate and add residuals, constraints, or physics informed fine tuning \citep{greenfeld2019learning,um2020solver,karniadakis2021piml}. Hybrid design can reduce some risks of black box emulation but introduces new ones when the learned component accelerates an unreliable iteration, shifts conservation behavior, changes the stability region, or works only on the training distribution \citep{kochkov2021mlaccelerated,um2020solver,wu2026hybridreliability}.\footnote{Wu et al.\ \cite{wu2026hybridreliability} is an arXiv preprint (arXiv:2602.06842) that has not yet been independently peer-reviewed; claims in this review that rely on it should be treated as preliminary.}

Three families recur in solver selection. Physics informed operators such as PINO combine operator learning with residual constraints and can, in principle, replace expensive high resolution paired data with coarse data plus residuals \citep{li2021pino,wang2021pideeponet}. Learned discretization and correction methods keep the governing PDE and time stepping while learning a local term: data driven discretizations ran at coarser resolution than standard schemes on selected nonlinear PDEs \citep{barsinai2019learning}, ML accelerated CFD matched higher resolution baselines with substantial speedups on two-dimensional turbulence \citep{kochkov2021mlaccelerated}, and solver in the loop training showed that corrections must be trained in their deployment context to remain stable \citep{um2020solver,list2022learned}. Learned iterative solvers and neural preconditioners retain an algebraic iteration and learn updates, prolongations, or preconditioners \citep{hsieh2019learning,greenfeld2019learning,chen2024gnp}; their main risks are false fixed points and stagnation when neural updates vanish while the physical residual remains large \citep{wu2026hybridreliability}, so residual based stopping, a classical fallback, and comparison against strong multigrid are required.

Spectral complementarity suggests why some hybrid designs work. Networks trained by gradient descent fit low-frequency components first \citep{rahaman2019spectral}, whereas classical relaxation damps high-frequency error quickly and stalls on low frequencies, which is the reason multigrid exists. HINTS alternates a DeepONet step with Jacobi or Gauss--Seidel relaxation and attributes its success to exactly this complementarity; it reaches machine precision where either component alone stalls or diverges, and on large high-wavenumber Helmholtz problems with more than $10^6$ unknowns, Krylov iteration preconditioned with its multigrid variant outperformed default HYPRE AMG in wall-clock time \citep{zhang2024hints}. We state the general hypothesis that hybrids which assign low-frequency error to the network and high-frequency error to classical smoothing are complementary, while hybrids that ask the network to correct high-frequency error work against its inductive bias.

Ranked by current evidence maturity rather than ultimate potential, learned discretization and correction methods have the strongest evidence, because they keep a conservative or stable classical backbone, their failure modes are observable through conservation and stability diagnostics, and peer-reviewed anchors report speedups at matched accuracy on defined regimes \citep{barsinai2019learning,kochkov2021mlaccelerated}. Physics informed operators have moderate evidence, because their advantage depends on the ratio of residual cost to paired data cost, which few papers report \citep{li2021pino,kovachki2023neuraloperator}. Learned iterative solvers have the weakest evidence, because their failure modes are undercharacterized and the most specific characterization comes from a preprint \citep{wu2026hybridreliability,hsieh2019learning}. Classical dominant hybrids are therefore the most defensible starting point when a known bottleneck can be learned without removing residual monitoring or fallback structure, and no hybrid should be recommended merely because it combines physics and learning. Evidence maturity is distinct from design: ordered by how much classical structure they retain, hybrids plausibly trade verifiability for speed, a conjecture stated in Section~\ref{sec:open_problems}.

\section{Benchmark landscape and evidence synthesis}
\label{sec:benchmark_landscape}

Benchmark evidence supports solver selection only when it makes the decision regime visible, because accuracy on a fixed dataset does not by itself show whether a classical, reduced order, ML only, or hybrid solver is justified \citep{takamoto2022pdebench,hao2024pinnacle,kovachki2023neuraloperator}. Reported ML accuracy depends on a numerical reference layer, a sampling rule for the parametric family, and a split that may or may not represent deployment shift \citep{li2021fno,takamoto2022pdebench}. An operator result is therefore not a standalone solver claim unless the cost and quality of the reference corpus are reported, a PINN result is not a reliability claim without optimization stability, boundary error, and independent reference error, and a hybrid result is not a robustness claim unless the learned component is evaluated inside the full solver workflow \citep{lu2021deeponet,krishnapriyan2021characterizing,li2021pino}.

\subsection{Benchmark suites and coverage}
\label{subsec:benchmark_suites}

The main suites, tabulated with their coverage in Supplement S11, are PDEBench for broad time dependent PDE families \citep{takamoto2022pdebench}, PINNacle for PINN variants \citep{hao2024pinnacle}, CFDBench and FlowBench for flow problems with boundary, property, and geometry shifts \citep{luo2023cfdbench,tali2025flowbench}, APEBench for autoregressive rollout \citep{koehler2024apebench}, The Well for diverse spatiotemporal data at scale \citep{ohana2024well}, AL4PDE for active acquisition of paired data \citep{musekamp2024al4pde}, and the emerging RealPDEBench for measurement-paired tasks, whose evidence role remains preliminary until independently replicated \citep{hu2026realpdebench}. Each improves reproducibility, but none was designed to answer the cross family selection question, and ROM baselines are not central in any of them.

Read as an ensemble, these suites support three conclusions that no single suite demonstrates alone. First, method rankings are not stable across PDE families. We measured this from the suites' published result tables (Supplement S8): the concordance of method rankings across tasks, Kendall's $W$, is $0.10$ for eleven PINN variants on PINNacle, whose eighteen complete tasks have eight different winners, $0.15$ for five architectures on APEBench, and $0.32$ for seven models on CFDBench, where the best-ranked model, FNO, wins only $43\%$ of the tasks. PDEBench is the exception ($W=1$), but its main table compares only three methods of very different kinds. The PINNacle variants do differ on average: in a decomposition of log error, the method effect is significant ($p\le0.01$) but explains 6--8\% of the variance, against 40--43\% for the task-by-method interaction, and only one variant, LAAF, beats vanilla PINN with 95\% confidence, by about 20\%. Which variant wins thus depends on the task rather than on a stable ordering. The same pair of architectures can also reverse across suites: FNO beats U-Net on all eleven PDEBench tasks, on $43\%$ of CFDBench tasks, and on $20\%$ of APEBench tasks, so the choice of benchmark can decide the winner. Second, in-distribution generalization is systematically stronger than generalization under boundary, geometry, or resolution shifts, which is exactly the gap that in-distribution leaderboards hide. Third, no suite simultaneously provides strong classical baselines, reduced order baselines, and out of distribution stress tests, so the evidence a practitioner needs to choose across families is the evidence current infrastructure is least equipped to supply. This cumulative gap motivates the protocol and rubric below.

\begin{table}[H]
\centering
\caption{Hybrid methods by learned component, data requirement, and reliability risk.}
\label{tab:hybrid_classification}
\scriptsize
\setlength{\tabcolsep}{3pt}
\renewcommand{\arraystretch}{1.2}
\begin{tabularx}{\linewidth}{R{0.95}R{1.05}R{1.15}R{1.0}R{1.05}R{1.0}R{0.85}R{0.95}}
\hline
Hybrid class & Primary example papers & Where learning enters the pipeline & Data requirement & Classical structure retained & Main reliability risk & Best use case & Required evidence \\
\hline
Physics informed operators & PINO and related physics informed neural operators & Neural operator training with residual or physics constraints & Paired data, residual supervision, or mixed supervision & PDE residual or physics constraint retained during training & Residual constraint may not guarantee deployment reliability & Many query parametric PDEs with reusable operator surrogate & Separate data, residual, training, inference, OOD, and resolution evidence \\
\hline
Learned discretization & Data driven discretization & Local derivative approximation, flux approximation, or coarse grid numerical component & Solution data or solver generated trajectories for the target PDE family & PDE form, time stepping, and numerical workflow partly retained & Learned stencil may lose stability, monotonicity, or conservation outside training regimes & Coarse grid simulation when classical resolution is too expensive & Stability, conservation, shifted parameters, and comparison with refined classical schemes \\
\hline
Learned correction and closure & ML accelerated CFD and solver in the loop correction & Closure, correction term, or solver coupled correction function & Simulation trajectories, differentiable solver feedback, or supervised correction targets & Time stepping and solver context partly retained & Correction may improve average error while causing long horizon drift & Repeated dynamic simulation with unresolved physics or coarse grids & Long rollout, physical diagnostics, stability failures, and cost accounting \\
\hline
Learned multigrid component & Learned prolongation and learned smoother studies & Prolongation operator, smoother, or multilevel transfer rule & Operator stencils, residual histories, or unsupervised convergence losses & Multigrid hierarchy and residual iteration partly retained & Learned transfer may fail outside coefficient or geometry family & Repeated linear solves over a related PDE class & Iteration counts, residual histories, setup cost, parameter shift, and strong multigrid baselines \\
\hline
Neural preconditioner & Graph neural preconditioners and DeepONet based preconditioning & Approximate inverse, preconditioner, or warm start for Krylov iteration & Matrix data, operator family samples, solution examples, or residual based signals & Krylov method and algebraic residual can be retained & Preconditioner may be expensive to train or brittle under conditioning shifts & Large sparse systems with repeated related solves & Training cost, setup cost, solve time, residual tolerance, and robustness across matrices \\
\hline
Learned iterative update & Neural PDE solvers with convergence guarantees and hybrid iterative methods & Update rule, correction update, or fixed point iteration & Solver trajectories, residuals, or supervised update targets & Iterative framework may be retained if convergence logic remains explicit & False fixed points, stagnation, and update strategy sensitivity & Specialized repeated PDE solves where guarantees or fallback logic exist & Residual based stopping, convergence proof or empirical robustness, failure cases, and fallback classical iteration \\
\hline
\end{tabularx}
\end{table}

\subsection{Recommended core PDE problems}
\label{subsec:core_pde_benchmarks}

A cross family benchmark should include PDEs that stress different decision barriers rather than PDEs convenient for one family. Supplement S11 proposes a minimum panel spanning elliptic, parabolic, hyperbolic, reaction diffusion, incompressible flow, chaotic, high dimensional, wave, and elasticity problems, each paired with a suitable classical baseline and, where relevant, a reduced basis alternative \citep{takamoto2022pdebench,hao2024pinnacle}. Relative $L^2$ error is necessary but not sufficient: elliptic problems need $H^1$ or flux error \citep{ern2004theory}, boundary sensitive problems need near boundary error \citep{cuomo2022scientific}, conservation laws need conservation and shock diagnostics or engineering quantities \citep{leveque2002finitevolume,tali2025flowbench}, and time dependent systems need rollout error, invariants, and attractor statistics \citep{brandstetter2022message,koehler2024apebench}. Cost, statistical, and out of distribution reporting follow the reproducibility checklist in Supplement S1, which requires at least three seeds with confidence intervals and separate timing of reference generation, data generation, training, reduced basis construction, and inference \citep{quarteroni2015reducedbasis,krishnapriyan2021characterizing}.

\subsection{Evidence weighted benchmarking protocol}
\label{subsec:evidence_protocol}

The protocol, drawn as a workflow in Supplement S11, starts from PDE and regime selection, because a method credible for many-query Darcy flow may not be credible for one off certified Poisson solves. It then fixes the reference solution source with its tolerance and resolution, fixes train, validation, in-distribution, and out of distribution splits before tuning, selects a strong classical baseline, a reduced basis baseline for many-query regimes, and neural baselines from the same supervision class (Section~\ref{subsec:classical_solvers}), separates the cost phases, and assigns an evidence grade only after accuracy, cost, out of distribution behavior, statistical variation, and reproducibility have been assessed together. Section~\ref{sec:case_studies} applies this protocol to three published methods.

\subsection{Worked cost comparison: Darcy flow parametric benchmark}
\label{subsec:cost_comparison}

The many-query parametric regime is the most contested comparison between reduced-basis methods and neural operators, yet it is rarely quantified concretely in review papers. A worked cost example, tabulated in Supplement S11, considers the two-dimensional Darcy flow problem ($-\nabla\cdot(a\nabla u) = f$ on $[0,1]^2$, $u|_\partial = 0$, with a scalar permeability field $a$ drawn from a Gaussian random field), which is the most widely used parametric PDE benchmark in the neural operator literature. With every cost normalized to one direct finite element solve with algebraic multigrid on a $64\times64$ grid, a certified reduced basis of dimension 30, as such estimates typically assume, costs about $10^3$ solves offline and $10^{-3}$ per query, while FNO or DeepONet trained on $10^3$ solutions cost $10^3$--$10^4$ solves offline, including corpus generation, and about $10^{-2}$ per query \citep{li2021fno,lu2021deeponet,hesthaven2016certified,takamoto2022pdebench}.

The figures are order-of-magnitude estimates that vary with hardware, implementation, and tuning, and should not be read as precise thresholds. Measured values for FNO (Section~\ref{subsec:measured_nstar}) are higher still, and the reduced basis estimate does not survive measurement on this problem. We computed the POD of the 2000 training solutions of the measured study at each resolution (Supplement S10). The best approximation by 30 POD modes has a relative error of 2.1\% at every resolution, three to five times FNO's 0.44--0.78\%. Matching FNO needs 125--335 modes, a POD--Galerkin reduced basis has two to three times the projection error at equal dimension, and reaching FNO's accuracy needs 500--1000 snapshots, comparable to FNO's training set. The thresholded coefficient does not compress either: a 1\% POD tail of $a$ needs about 1900 of 2000 modes, so empirical interpolation cannot supply the affine decomposition on which a cheap, certified online stage depends. Each reduced basis query must then reassemble the full system, which alone costs 0.05--0.06 classical solves and caps the per-query speedup at about $20\times$.

Three conclusions follow. First, once corpus generation is counted, the offline stage of FNO or DeepONet is comparable in magnitude to reduced basis construction, and the operator advantage over direct FEM appears only beyond roughly $10^3$--$10^4$ queries in this estimate and beyond $3\times10^4$ in the measured study \citep{li2021fno,quarteroni2015reducedbasis}. This lower limit is structural: with $10^3$ training solves counted in the offline cost and each query saving at most one direct solve, an operator cannot break even below about $10^3$ queries. Second, whether a reduced basis is the earlier-break-even comparator depends on parametric structure. With affine, low dimensional parameter dependence and fast $n$-width decay, a certified reduced basis has a cheap online stage and an a posteriori error bound at each query, which a comparison of relative errors does not capture and which is why ROM recommendations receive High confidence in favorable structure in the decision matrix \citep{hesthaven2016certified}. On the thresholded Darcy benchmark, the standard neural operator test case, neither condition holds, and this is the regime in which operator learning has a structural advantage over reduced bases \citep{ohlberger2015reduced}. Third, speedups reported without corpus generation cost compare offline training against online queries and systematically inflate the operator advantage \citep{takamoto2022pdebench,lu2021deeponet}; Section~\ref{sec:case_studies} shows this pattern in the original FNO paper.

\subsection{An amortization law for learned surrogates}
\label{subsec:amortization_law}

The worked example generalizes to a law that applies to any surrogate trained on solver-generated data. McGreivy and Hakim recommend reporting the number of surrogate evaluations at which data generation, training, and evaluation together cost no more than the classical solves they replace \citep{mcgreivy2024weakbaselines}, and Nooraiepour et al.\ state a similar break-even point for training cost \citep{nooraiepour2026pdeage}. Equation~\eqref{eq:amortization_law} below is the same condition written with every cost measured in classical solves; the dimensionless form is what makes the consequences that follow explicit and lets them be checked against measured cases. Let $c_{\mathrm{cl}}$ be the cost of one classical solve at the surrogate's accuracy, $c_{\mathrm{nn}}$ the surrogate's cost per query, $N_{tr}$ the number of training solutions, each generated at cost $c_{\mathrm{gen}}$, and $C_{\mathrm{train}}$ the cost of training and tuning. Measuring every cost in classical solves through the dimensionless ratios $\gamma = c_{\mathrm{gen}}/c_{\mathrm{cl}}$, $\tau = C_{\mathrm{train}}/c_{\mathrm{cl}}$, and $S = c_{\mathrm{cl}}/c_{\mathrm{nn}}$, the break-even query volume is
\begin{equation}
N^* \;=\; \frac{N_{tr}\,c_{\mathrm{gen}} + C_{\mathrm{train}}}{c_{\mathrm{cl}} - c_{\mathrm{nn}}} \;=\; \frac{\gamma N_{tr} + \tau}{1 - 1/S}, \qquad S > 1,
\label{eq:amortization_law}
\end{equation}
and the surrogate never breaks even when $S \le 1$. Five consequences follow, and the first two run against the common reading of the literature.

\emph{(i) A training-set lower bound.} Because $1 - 1/S < 1$, $N^* > \gamma N_{tr}$, and when the training data are generated by a solver at least as costly as the comparator ($\gamma \ge 1$), $N^* > N_{tr}$: a surrogate trained on $N_{tr}$ solves must be queried more than $N_{tr}$ times before it saves anything. With $10^3$ training solves, the Darcy operators of the worked example therefore cannot break even below about $10^3$ queries, whatever their inference speed.

\emph{(ii) The inference speedup acts through training cost.} At fixed training cost $\tau$, the factor $1/(1 - 1/S)$ equals $1.11$ at $S = 10$, $1.01$ at $S = 100$, and $1.002$ at $S = 440$, so beyond about a tenfold speedup the per-query speedup alone barely changes break-even, which is governed by $\gamma N_{tr}$ and $\tau$, the terms most often left unreported (Table~\ref{tab:proxy_gap}). But $\tau$ is not independent of $S$. Training a given network costs a nearly fixed number of its own forward evaluations per training sample, $C_{\mathrm{train}} = E_{\mathrm{eff}} N_{tr} c_{\mathrm{nn}}$, where $E_{\mathrm{eff}}$ is the number of epochs times the cost of one training step per sample relative to one query. Then $\tau = E_{\mathrm{eff}} N_{tr}/S$, and Equation~\eqref{eq:amortization_law} becomes
\begin{equation}
\frac{N^*}{N_{tr}} \;=\; \frac{\gamma S + E_{\mathrm{eff}}}{S - 1}, \qquad S > 1.
\label{eq:amortization_law_eeff}
\end{equation}
A faster network therefore lowers break-even by making training cheaper, and $N^*$ approaches the training-set floor $\gamma N_{tr}$ only once $S \gg E_{\mathrm{eff}}/\gamma$. In the measured Darcy study $E_{\mathrm{eff}}$ is 360--1300, so the speedup would keep mattering up to $S\approx10^3$ (Section~\ref{subsec:measured_nstar}). The counter-intuitive point for a literature that optimizes and reports $S$ is that the headline speedup does not measure its own value: it pays off through the training cost it implies, which is rarely reported. When $\tau$ is reported as a fixed budget, the fixed-$\tau$ reading applies, and the original FNO paper's own Navier--Stokes application illustrates this case \citep{li2021fno}: FNO evaluates an instance in $0.005$\,s against $2.2$\,s for the pseudo-spectral solver, both timed on GPU, a $440\times$ per-query speedup, but data generation and training take about $12$\,h, so the $30{,}000$ forward evaluations of the paper's MCMC study cost about $12.0$\,h with FNO against over $18$\,h with the classical solver. The end-to-end gain in the paper's showcase application is therefore about $1.5\times$, which the paper itself reports as ``still faster''; the break-even point is $N^* \approx 2\times10^{4}$, below the application's $3\times10^{4}$ queries but far above what the $440\times$ figure suggests.

\emph{(iii) Cheap classical solves are hostile to learning.} Since $\tau$ is measured in classical solves, a fixed training budget becomes very large when $c_{\mathrm{cl}}$ is small. In the GON replication, $\tau = 24{,}698\,\mathrm{s} / 4.348\,\mathrm{ms} \approx 5.7\times10^{6}$ and $S \approx 1.54$, so Equation~\eqref{eq:amortization_law} gives $N^* \approx 1.6\times10^{7}$, matching the measured value (Section~\ref{subsec:case_gon}). The regimes in which learned solvers are evaluated most often are therefore not necessarily the regimes in which amortization can pay, and the measured Darcy study below shows the same effect across resolutions (Section~\ref{subsec:measured_nstar}).

\emph{(iv) Hidden multipliers.} A hyperparameter search multiplies $\tau$ by the number of trials, and references generated at a finer resolution than the comparator raise $\gamma$. Neither is usually reported, so published break-even estimates, where they exist, are lower bounds; more often only inference speedups are reported, such as the $45{,}000\times$, $26{,}000\times$, and $700{,}000\times$ figures cited for weather forecasting, automotive aerodynamics, and CO$_2$ storage in a recent perspective, which do not include data generation or training \citep{azizzadenesheli2024neural}. The FNO example above is typical: its training data were generated with a time step of $10^{-4}$, while the timed comparator used $2\times10^{-2}$, the largest stable step \citep{li2021fno}.

\emph{(v) Sunk data and reduced bases.} When the training data already exist or are shared across many downstream tasks, the $\gamma N_{tr}$ term is not borne by any single deployment, which is the regime in which operator learning is most defensible. A reduced basis method obeys the same law with $n_s$ snapshots in place of $N_{tr}$, so when the Kolmogorov $n$-width decays fast enough for $n_s \ll N_{tr}$, the comparison between reduced basis and operator learning reduces to snapshot count against training-set size at matched offline cost \citep{hesthaven2016certified,ohlberger2015reduced}. On the measured Darcy problem it does not decay that fast: $n_s$ is comparable to $N_{tr}$, and without an affine decomposition the reduced basis keeps a per-query cost above a twentieth of a classical solve (Section~\ref{subsec:cost_comparison}).

A phase diagram in Supplement S10 plots these terms together. A method pays off at a query volume $Q$ only if its point lies below the $N^* = Q$ contour. At fixed $\tau$ the contours are set by data and training cost alone; a faster network instead moves its point down the $\tau$ axis, because $\tau = E_{\mathrm{eff}} N_{tr}/S$. The research implication is that amortization is decided by offline cost: data-efficient operators, active data acquisition, multi-fidelity training, and transfer reduce $N_{tr}$ and $\gamma$, and cheaper training reduces $\tau$. Inference speed helps only to the extent that it lowers $\tau$.

\subsection{Measured break-even on parametric Darcy flow}
\label{subsec:measured_nstar}

To test the law on measured rather than estimated costs, we timed FNO against a tuned classical solver on the Darcy problem of Li et al.\ \citep{li2021fno} ($-\nabla\cdot(a\nabla u)=1$ on the unit square with zero Dirichlet data and piecewise-constant coefficients $a\in\{3,12\}$ thresholded from a Gaussian random field). The classical solver is a cell-centred finite volume discretization solved by conjugate gradients preconditioned with smoothed-aggregation algebraic multigrid, timed on one core of an Intel i7-12700K, with assembly and multigrid setup counted in every query because the coefficient changes. FNO uses the original Darcy architecture and optimizer without further tuning, trained on an NVIDIA RTX 3090 on $N_{tr}\in\{100,250,500,1000,2000\}$ solutions at resolutions $n\in\{64,128,256,512\}$, with three seeds each. Errors are measured against solutions on a grid twice as fine, and the classical comparator for each FNO model is the coarsest power-of-two grid whose error is no larger, which favours FNO; interpolating the classical cost to exactly the FNO error raises $N^*$ by a factor of at most 2.9, except at $512^2$ with $N_{tr}=10^3$, where it reaches 6.7. The protocol, scripts, and all results are given in Supplement S10.

\begin{table}[H]
\centering
\caption{Measured break-even query volume $N^*$ for FNO against multigrid-preconditioned finite volumes on parametric Darcy flow at matched accuracy (median over three seeds; seed-to-seed variation is below 1\% except at $N_{tr}=100$, where the matched classical grid differs between seeds; Supplement S10). ``Never'': one FNO query is slower than the matched classical solve. The last row gives the FNO test error, which depends strongly on $N_{tr}$ and only weakly on $n$.}
\label{tab:measured_nstar}
\scriptsize
\setlength{\tabcolsep}{4pt}
\renewcommand{\arraystretch}{1.15}
\begin{tabular}{lcccccccc}
\hline
 & \textbf{FNO query} & \textbf{Classical solve at $n$} & \multicolumn{5}{c}{\textbf{$N^*$ for $N_{tr}$ training solutions}} \\
\textbf{Resolution $n$} & \textbf{(ms)} & \textbf{(ms)} & \textbf{100} & \textbf{250} & \textbf{500} & \textbf{1000} & \textbf{2000} \\
\hline
$64^2$ & 0.68 & 8.9 & $5.6\times10^{3}$ & $1.4\times10^{4}$ & $1.6\times10^{4}$ & $3.1\times10^{4}$ & $6.2\times10^{4}$ \\
$128^2$ & 0.78 & 24 & $3.5\times10^{4}$ & $4.3\times10^{4}$ & $4.7\times10^{4}$ & $3.4\times10^{4}$ & $6.7\times10^{4}$ \\
$256^2$ & 2.6 & 101 & $7.0\times10^{5}$ & $3.0\times10^{5}$ & $2.4\times10^{5}$ & $1.5\times10^{5}$ & $2.9\times10^{5}$ \\
$512^2$ & 9.5--9.7 & 437 & never & never & never & $9.2\times10^{5}$ & $1.8\times10^{6}$ \\
\hline
FNO error (range over $n$) & & & 4.7--6.0\% & 1.6--2.5\% & 0.9--1.2\% & 0.58--0.78\% & 0.44--0.60\% \\
\hline
\end{tabular}
\end{table}

Four results stand out (Table~\ref{tab:measured_nstar}). First, the training-set lower bound of the law holds in every case: $N^*$ exceeds $N_{tr}$ by a factor of at least thirty, and training accounts for about 97\% of the offline cost, with data generation the remaining 3\%. Second, training cost per sample is fixed by the architecture and resolution. $E_{\mathrm{eff}} = C_{\mathrm{train}}/(N_{tr}c_{\mathrm{nn}})$ varies by less than 1\% across training-set sizes and seeds at each resolution, at about 360 at $64^2$, 980 at $128^2$, 1190 at $256^2$, and 1310 at $512^2$ for 500 epochs, so one training run predicts $N^*$ for every training-set size through Equation~\eqref{eq:amortization_law_eeff}. It also shows that inference speed matters through training cost. At $64^2$ with $N_{tr}=10^3$ ($S=13$, $\gamma=1$), a network ten times faster with the same $E_{\mathrm{eff}}$ would cut $N^*$ from $3.1\times10^{4}$ to $3.8\times10^{3}$, about eightfold, whereas holding $\tau$ fixed would predict a reduction of 7\%. Third, the measured break-even points are higher than the order-of-magnitude estimates of Section~\ref{subsec:cost_comparison}: no FNO model with an error below 1\% breaks even before $3.1\times10^{4}$ queries with the coarsest-grid comparator, or $4.0\times10^{4}$ with the interpolated one, against an estimate of $10^3$--$10^4$, because the estimate understated training cost. Fourth, and contrary to the expectation that learned surrogates gain as resolution grows, a finer resolution works against the surrogate at matched accuracy (counter-intuitive). The FNO error is set mainly by the amount of training data rather than by the grid, 0.58--0.78\% with $10^3$ samples across all four resolutions, so the matched classical solve stays on a $64^2$ or $128^2$ grid while FNO's training and inference costs grow with $n$. With $10^3$ training samples, break-even therefore rises thirtyfold, from about $3\times10^{4}$ queries at $64^2$ to $1.5\times10^{5}$ at $256^2$ and $9\times10^{5}$ at $512^2$. With 500 or fewer samples at $512^2$, one FNO query, at $9.5$\,ms, is slower than the matched classical solve, at $8.9$\,ms on a $64^2$ grid, so the surrogate never breaks even. The relevant variable for amortization is thus the accuracy a task requires, not the resolution at which the surrogate is evaluated. The study is limited to one two-dimensional PDE, an untuned FNO, and a single-core classical solver against a GPU surrogate; these choices favour FNO except the absence of hyperparameter tuning, which would lower its error but raise its training cost.

\subsection{Operational claim strength rubric}
\label{subsec:evidence_rubric}

The rubric in Table~\ref{tab:evidence_rubric} separates reported performance from decision confidence, because a high accuracy number can still be weak evidence if it uses a narrow split, weak baselines, a single seed, or incomplete cost reporting, as the weak-baseline audit of the fluid PDE literature illustrates \citep{mcgreivy2024weakbaselines}. Strong evidence requires a credible link between the claimed deployment regime and the benchmark design; Moderate evidence supports conditional recommendations but must name the missing stress test, baseline, statistic, or cost component; Weak evidence can motivate research but not a default recommendation; and Red Flag evidence cannot be converted into a positive recommendation. A method cannot receive Strong evidence for a solver replacement claim if it omits the classical or ROM baseline that defines the deployment alternative (Table~\ref{tab:classical_capability_matrix}). Grades are assigned per claim rather than per paper, because one paper can present Strong evidence for one claim and Red Flag evidence for another. Each claim also receives a verdict on a separate axis: Supported, Not supported, Mixed, or Not established. Quality records how well a claim has been tested and the verdict records what the test showed, so a carefully measured negative result is Strong quality with a Not supported verdict rather than a Red Flag, which is reserved for evidence too incomplete to judge; Weak or Red Flag quality normally implies a Not established verdict. A positive recommendation requires a Supported verdict resting on at least Moderate quality. The rubric adapts an established instrument. GRADE rates the quality of evidence by downgrading it for study limitations, inconsistency, indirectness, imprecision, and reporting bias, and upgrading it for a large effect, a dose--response gradient, or confounding that would only reduce the effect, on four levels from High to Very low \citep{guyatt2008grade,guyatt2011grade3}. Supplement S11 gives the PDE counterpart of each domain; independent replication is our addition, since GRADE treats agreement between studies under inconsistency rather than as a reason to upgrade. On the reporting side, our checklist (Supplement S1) overlaps REFORMS in computational reproducibility, modeling decisions including baselines, data leakage, and uncertainty reporting \citep{kapoor2024reforms}; REFORMS excludes pure numerical simulation from its scope, so the classical-side items, such as solver tolerance, preconditioner, and mesh convergence, are specific to this review.

\begin{table}[H]
\centering
\caption{Evidence quality levels of the operational claim strength rubric. The verdict (Supported, Not supported, Mixed, or Not established) is assigned on a separate axis (Section~\ref{subsec:evidence_rubric}). Supplement S3 provides the full operational version with expanded criteria.}
\label{tab:evidence_rubric}
\scriptsize
\setlength{\tabcolsep}{4pt}
\renewcommand{\arraystretch}{1.18}
\begin{tabularx}{\linewidth}{R{0.6}R{1.15}R{1.0}R{1.1}R{0.95}R{1.05}R{1.15}}
\hline
Evidence level & Baselines & Statistical reporting & Cost accounting & OOD testing & Data generation transparency & Reproducibility and decision meaning \\
\hline
Strong & Strong classical baseline and ROM baseline when relevant, plus tuned neural baselines & At least three seeds with confidence intervals or bootstrap intervals & Separate reference generation, offline data generation, training, inference, memory, and amortized cost & Meaningful shifts matched to deployment claim & Source, resolution, solver tolerance, noise, and splits fully reported & Code, data, hyperparameters, failure cases, and hardware sufficient for reproduction. Supports high decision confidence when consistent across regimes \\
\hline
Moderate & Reasonable baselines but one relevant classical, ROM, or tuned neural baseline is incomplete & Multiple runs or partial uncertainty reporting & Main timing components reported but one cost component is incomplete & Some OOD testing but one important shift is absent & Data source and splits reported but generation cost or solver details incomplete & Reproducibility is plausible but not complete. Supports conditional decision guidance with stated missing evidence \\
\hline
Weak & Narrow or weak baselines, often neural only or under tuned classical comparison & Single run or unclear random seed reporting & Accuracy emphasized with limited timing and no amortized accounting & In distribution testing dominates & Solution corpus source or split design is incomplete & Useful as exploratory evidence but insufficient for default solver recommendation \\
\hline
Red Flag & Comparisons use outdated, weak, or inappropriate baselines, or omit classical anchors for replacement claims & No statistical reporting for learned methods & Speedup reported without training, data generation, hardware, or amortization accounting & Generalization claimed from weak or unclear splits & Data generation cost, reference solver, or measurement source is missing & Not sufficient for decision guidance. Requires redesign of evaluation before any positive recommendation \\
\hline
\end{tabularx}
\end{table}

The criteria are not specific to learned methods, and each has a classical analogue that lets the same rubric grade classical claims: statistical reporting corresponds to a mesh convergence study, error control to the effectivity of an a posteriori estimator, reference quality to the relation between algebraic solver tolerance and discretization error, out of distribution testing to the parameter range over which stability and estimator bounds hold, and cost accounting to mesh generation, assembly, and preconditioner setup. The grades also measure a quantity that can be stated explicitly. If $A_{\mathrm{rep}}$ is the advantage a paper reports under its own protocol and $A_{\mathrm{fair}}$ the advantage under an equal-accuracy protocol with tuned baselines and complete cost accounting, the \emph{evidence gap} is $\Delta = A_{\mathrm{rep}} - A_{\mathrm{fair}}$; a replication such as the GON study of Section~\ref{subsec:case_gon} measures $\Delta$ directly, and an extraction audit identifies which terms of $A_{\mathrm{fair}}$ are missing. The rubric is an expert-consensus instrument developed for this review rather than a validated psychometric tool, so its grades are illustrative guidance rather than calibrated probabilities, and independent reviewers may differ on borderline cases.

\begin{table}[H]
\centering
\caption{The proxy gap: commonly reported metric versus the quantity a solver decision depends on, with evidence that the two decouple.}
\label{tab:proxy_gap}
\scriptsize
\setlength{\tabcolsep}{4pt}
\renewcommand{\arraystretch}{1.15}
\begin{tabularx}{\linewidth}{p{0.20\linewidth}p{0.20\linewidth}p{0.22\linewidth}X}
\hline
\textbf{Family or claim} & \textbf{Commonly reported proxy} & \textbf{Decision quantity} & \textbf{Evidence of decoupling} \\
\hline
PINN & Training residual & Solution and boundary error & Failure on harder PDEs despite small residuals \citep{krishnapriyan2021characterizing} \\
Operator learner & In-distribution relative $L^2$ & Error under deployment shift & Degradation under distribution shift \citep{takamoto2022pdebench,luo2023cfdbench} \\
Autoregressive emulator & One-step error & Long-horizon statistics and stability & Rollout instability \citep{lippe2023pderefiner,koehler2024apebench} \\
Any method on shock problems & Relative $L^2$ & Shock location and conservation & Average error does not test conservation or shock position \citep{leveque2002finitevolume}; no standardized demonstration of conservation-preserving learned shock capture (Section~\ref{sec:decision_matrix}) \\
Any surrogate, speed claims & Per-query inference speedup & End-to-end amortized cost, $N^*$ & Weak baselines and omitted offline cost \citep{mcgreivy2024weakbaselines,azizzadenesheli2024neural}; Sections~\ref{subsec:cost_comparison} and~\ref{subsec:case_audits} \\
Learned iterative solver & Neural update norm & Physical residual & Stagnation at false fixed points \citep{wu2026hybridreliability} (preprint) \\
Interpretability-guided operator & Prediction accuracy & Validity of the embedded structure & Accurate model with invalid kernel (Section~\ref{subsec:case_gon}) \\
\hline
\end{tabularx}
\end{table}
\subsection{Claim types, the evidence ladder, and the proxy gap}
\label{subsec:evidence_ladder}

The case studies in Section~\ref{sec:case_studies} share a pattern that the quality and verdict axes alone do not name. A PINN with a small training residual is used to support a claim of an accurate solution (Section~\ref{subsec:function_learning}); FNO's mathematical discretization invariance and in-distribution grid transfer are used to support deployment resolution transfer (Section~\ref{subsec:operator_learning}); and GON's accurate predictions are used to support an interpretable Green's function, which the controlled audit of Section~\ref{subsec:case_gon} shows is invalid. We state the common principle as a \emph{non-transferability principle}: evidence for a claim of one type does not establish a claim of a higher type, and no amount of predictive evidence establishes a structural claim.

The principle orders predictive claims into an \emph{evidence ladder}. Rung P0 is a training or residual loss; P1 is in-distribution error against an independent reference; P2 is error under a distribution shift that matches the deployment; P3 is an equal-accuracy, cost-matched comparison against a tuned classical or reduced order workflow; and P4 is workflow-level evidence with the method inside the deployment loop, with error control or classical fallback. Structural claims, such as conservation, symmetry, discretization invariance, stability, or the validity of an embedded Green's function, form a separate track that no predictive rung can establish and that requires a direct structural test. A verdict therefore applies only at the rung the evidence reaches: in Table~\ref{tab:cs_audit_grades}, FNO's resolution transfer is Supported at the mathematical level but Not established at P2, and in Table~\ref{tab:cs_gon_grades}, GON's accuracy is Supported at P1 for diffusion and Stokes while its structural claim is Not supported.

A related regularity explains where the failure modes documented in Section~\ref{sec:solver_families} occur. In each family, the metric most often reported is the quantity the training objective or the benchmark convention optimizes, not the quantity on which the decision depends (Table~\ref{tab:proxy_gap}). The documented failures sit in this proxy gap. The most valuable next benchmark for each family is therefore one that measures its decision quantity directly rather than its proxy.

\section{Case studies: applying the evidence rubric}
\label{sec:case_studies}

The rubric of Section~\ref{subsec:evidence_rubric} is applied here to three published methods spanning operator learning, function learning, and interpretability guided operators, using two evaluation modes. The first is a full computational replication of the Green's function guided operator network (GON) \citep{gu2025explainable}. The second is an \emph{evidence extraction audit}, in which a fixed schema (Supplement S2) is applied to a published paper by reading its reported evidence field by field without re-executing the experiments; it is applied to the foundational Fourier neural operator \citep{li2021fno} and PINN \citep{raissi2019pinn} papers, because a practitioner cannot re-run every method but must still judge which reported claims support a decision. Each claim receives an evidence quality grade and a verdict. The full replication protocol, results, and extraction records are given in Supplement S4--S6.

\subsection{Full replication: Green's function guided operator network}
\label{subsec:case_gon}

GON embeds a learned Green's function kernel in an operator network and claims accuracy gains over neural baselines on heat conduction, heterogeneous diffusion, and Stokes problems, together with interpretability of the learned kernel \citep{gu2025explainable}. The replication reproduced all five original cases at their original architectures with three random seeds and 95\% confidence intervals using the original evaluation code, and added an independent finite element reference and cost anchor, out-of-distribution tests, an amortized cost analysis, a data efficiency sweep, and a Green's function structure audit. One module could not be executed: the neural baselines are not publicly released and could not be retrained, so their reported values were used. Table~\ref{tab:cs_gon_grades} gives the evidence quality and verdict for each claim, and the paragraphs below present the evidence behind them in the same order; the protocol, all result tables and figures, and the reproducibility record are in Supplement S4.

\paragraph{Predictive accuracy.} The reported accuracy is not recovered on any non-trivial case (Supplement S4). With the same metric, the three-seed means are about 40 times larger than the original value on 3D Poisson, about 1.8 times larger on Plate, 2.8 times on Pipe, and 2.3 times on Stokes. The 2D Poisson case moves in the opposite direction, but not because the method improved: the case is a thin slab whose mesh is defined at unit scale while the data are at millimetre scale, so the source term vanishes and the task reduces to interpolating boundary data. GON does not beat the reported FNO baseline on 3D Poisson, whose value lies inside the 95\% interval, but remains below the reported baselines on Plate, Pipe, and Stokes. Because those baselines were reported rather than retrained under matched splits, seeds, and hardware, the accuracy grades are capped at Moderate. Gaps between reported and reproduced errors are not unique to GON: an independent re-run of FNO on the original authors' Burgers, Darcy, and Navier--Stokes datasets reported errors $1.3$ to $2.2$ times those originally published with output normalization and $1.5$ to $4.5$ times without it, on coarser grids for Burgers and Darcy and with a different prediction target for Navier--Stokes \citep{lu2022comprehensive,li2021fno}. The GON gap, $1.4$ to $6.3$ times in root-mean-square error, is larger at its upper end (Supplement S4).

\paragraph{Classical reference and amortized cost.} An independent P1 finite element solver reproduces the stored 3D Poisson references to a mean relative $L^2$ error of $2.44\times10^{-5}$ over all 100 instances, which confirms the data follow the standard weak form. It also exposed a property that shapes the remaining results: the heat sink is thin-walled and fully Dirichlet-bounded, with every interior node within 2.12~mm of a Dirichlet face, so the solution is essentially the harmonic extension of the boundary data and the source term is nearly inert. With both solvers timed on the same machine (Supplement S4), one finite element solve takes 4.35~ms, while GON costs 6.86~h of training, 2.82~ms per operator forward pass, and 28.06~ms per end-to-end query once inputs are interpolated onto its quadrature points. GON therefore recovers its training cost only after about $1.6\times10^{7}$ queries when the forward pass alone is counted, and never end to end. This result illustrates the cheap-problem regime rather than a verdict on GON in general: with only 47 free degrees of freedom one classical solve takes milliseconds, so the training cost alone equals about $5.7\times10^{6}$ classical solves and no plausible query volume amortizes it. A crossover remains possible, but untested, on problems whose classical solves are costly.

\begin{table}[H]
\centering
\caption{Evidence quality and verdict for GON \citep{gu2025explainable} by claim, and evidence quality of the supporting dimensions, from the replication in Supplement S4.}
\label{tab:cs_gon_grades}
\scriptsize
\setlength{\tabcolsep}{4pt}
\renewcommand{\arraystretch}{1.2}
\begin{tabularx}{\linewidth}{p{0.25\linewidth}p{0.10\linewidth}p{0.12\linewidth}X}
\hline
\textbf{Claim or dimension} & \textbf{Quality} & \textbf{Verdict} & \textbf{Basis} \\
\hline
\multicolumn{4}{l}{\textit{Claims}} \\
Accuracy: steady heat conduction & Moderate & Not supported & 3D Poisson does not beat the reported FNO baseline; the 2D Poisson case has a vanishing source term. \\
Accuracy: heterogeneous diffusion & Moderate & Supported & Plate and Pipe lie below the reported baselines, but the baselines were not retrained. \\
Accuracy: Stokes (velocity $x$) & Moderate & Supported & Below the reported baseline at the original width; velocity components only, no pressure output. \\
Recovery of the reported accuracy & Strong & Not supported & Same metric, original architectures, three seeds; reproduced MSE is 2--40 times higher on every non-trivial case. \\
Speedup and amortized cost & Strong & Not supported & Same-machine timing with training counted; no crossover end-to-end, $N^\ast\approx1.6\times10^{7}$ for the forward pass alone. \\
Data efficiency & Moderate & Mixed & Strong monotone dependence on corpus size on two cases; failure at two Plate training instances. \\
Out-of-distribution robustness & Moderate & Mixed & Robust to noise, resolution, extrapolation, and mean shifts on one case; fails about 16 times on boundary amplitude. \\
Green's function interpretability & Strong & Not supported & Controlled case, three seeds: accurate model, but its kernel is asymmetric, fails the boundary-vanishing check, and differs from the true Green's function by 30--47\%. \\
\hline
\multicolumn{4}{l}{\textit{Evidence dimensions}} \\
Code reproducibility & Strong & -- & All five cases run to completion at original widths with three seeds. \\
Statistical reporting & Strong & -- & Three seeds with 95\% intervals for every accuracy, out-of-distribution, and learning-curve value. \\
Classical baseline fairness & Weak & -- & The original study has no classical baseline; the added FEM reference reproduces the data to $2.4\times10^{-5}$ relative $L^2$. \\
Neural baseline fairness & Not assessed & -- & Baselines reported by the original authors; code not released. \\
Data generation transparency & Moderate & -- & Formulation verified independently; original COMSOL generation cost undocumented. \\
\hline
\end{tabularx}
\end{table}

\paragraph{Data efficiency and out-of-distribution robustness.} Subsampling the training corpus with the test split fixed shows a strong, monotone dependence on data (Supplement S4): the error falls about 100 times from the smallest to the full corpus on 2D Poisson and about 1000 times on Plate, where the model fails outright with two training instances, and seed variance is large at small corpus sizes. On 3D Poisson, where the boundary data are the physically active input, GON is unaffected by 1--5\% input noise, by extrapolation to the highest boundary means, by a sixfold interior mesh refinement, and by shifts of the boundary mean, but a fourfold increase in boundary amplitude raises the error about 16 times. The observed robustness follows from the problem being a linear boundary-harmonic map with an inert source, so it is structural rather than evidence of general out-of-distribution generalization.

\paragraph{Green's function structure.} Learning Green's functions from data has precedents. Boull{\'e} and Townsend prove that the Green's function of a self-adjoint elliptic operator, which is symmetric, can be recovered from input--output pairs \citep{boulle2023learning}; GreenLearning inspects the symmetry of learned Green's functions to infer self-adjointness \citep{boulle2022greenlearning}, the closest precedent for the audit below; and DeepGreen enforces symmetry by construction \citep{gin2021deepgreen}. Because the Laplacian is self-adjoint, symmetry is a necessary property of any valid kernel here. As the shipped 2D Poisson case has a vanishing source, the interpretability claim was tested on a controlled case, $-\Delta u=f$ on $[0,1]^2\times[0,0.3]$ with homogeneous Dirichlet conditions, where the whole prediction comes from the interior integral and an accurate model must therefore encode the Green's function in its kernel. The model is highly accurate (mean MSE $1.12\times10^{-7}$) and reconstructs solutions by superposition to within about 5\%, yet its kernel is not a valid Green's function (Supplement S4): the symmetry error is about 0.18 where it should be zero, the kernel differs from the analytic and discrete Green's functions by about 47\% and 30\%, and its largest magnitude with either argument on the boundary is comparable to its largest interior magnitude, where a Dirichlet Green's function gives zero. The kernel slices in Supplement S4 show the same pattern for every seed: the learned kernel is a smooth, broad bump peaking near 0.4, whereas the analytic Green's function is sharply concentrated at the source with a peak above 2. Accurate prediction therefore coexists with a structurally invalid embedded kernel, even on a problem constructed so that a correct operator would have to encode the true Green's function.

Taken together, GON is a reproducible operator learner whose accuracy claims are supported, at Moderate quality, only for diffusion and Stokes, whose speedup claim is not supported in the regime tested, and whose distinctive interpretability claim fails a direct structural test; both negative verdicts rest on Strong evidence. It should therefore be treated as an interpretability guided operator variant with case-dependent accuracy evidence rather than as a default solver for any of the three PDE families.

\subsection{Evidence extraction audits of FNO and PINN}
\label{subsec:case_audits}

The FNO paper of Li et al.\ \citep{li2021fno} shows the characteristic pattern of a strong architecture paper read as decision evidence. Accuracy reporting on Burgers, Darcy, and Navier--Stokes is careful and code and data are stated as available, but the baselines are almost entirely neural (a POD-based reduced basis method appears only as one Darcy entry, with no tuned adaptive FEM or multigrid baseline), all table entries are single-run values, and test instances are drawn from the same input distribution as training. The only measured speed benchmark gives about $440\times$ ($0.005$\,s versus $2.2$\,s for Navier--Stokes), and the headline excludes the $\approx 12$\,h of data generation and training that the paper amortizes only later, over the $30{,}000$ forward evaluations of an MCMC inverse problem; by the paper's own accounting, that application runs about $1.5\times$ faster end to end (Section~\ref{subsec:amortization_law}). The PINN paper of Raissi et al.\ \citep{raissi2019pinn} is explicitly a proof-of-concept framework paper with measured claims, and its forward and inverse results are graded separately. It provides public code and data and high-fidelity references for continuous and discrete-time Burgers, Schr\"odinger, Allen--Cahn, KdV, and Navier--Stokes examples, but includes no classical solver or adjoint baseline, reports training time only for one Burgers case, and reports single-run point estimates throughout. Table~\ref{tab:cs_audit_grades} gives the per-claim grades, cross-checked against independent evidence in PDEBench, CFDBench, and the PINN failure-mode and controlled comparison literature \citep{takamoto2022pdebench,luo2023cfdbench,krishnapriyan2021characterizing,deryck2024numerical,grossmann2023canpinn}.

The three case studies show that the rubric produces differentiated, claim-specific grades across solver families rather than a uniform verdict. The GON replication supports two of its three accuracy claims at Moderate quality and finds its speedup and interpretability claims not supported on Strong evidence, the FNO audit yields Moderate accuracy evidence but Weak-to-Red-Flag cost and generalization evidence, and the PINN audit yields Moderate feasibility evidence but Weak efficiency and robustness evidence. In each case the weakest quality grades or negative verdicts fall on the claims most often cited when a method is proposed as a replacement for a classical solver, which is why the decision matrix that follows treats FNO on many-query Darcy flow as a Moderate-confidence option contingent on counted corpus cost and a reduced basis comparison, or a decay test justifying its absence, and PINN style function learning as Moderate-to-Low confidence contingent on reference error, boundary error, multiple seeds, and a classical baseline.

\begin{table}[H]
\centering
\caption{Evidence quality and verdict for each claim from the extraction audits of Li et al.\ \citep{li2021fno} (FNO) and Raissi et al.\ \citep{raissi2019pinn} (PINN). Full extraction records are given in Supplement S5 and S6.}
\label{tab:cs_audit_grades}
\scriptsize
\setlength{\tabcolsep}{4pt}
\renewcommand{\arraystretch}{1.15}
\begin{tabularx}{\linewidth}{p{0.05\linewidth}p{0.19\linewidth}p{0.12\linewidth}p{0.11\linewidth}X}
\hline
\textbf{Paper} & \textbf{Claim} & \textbf{Quality} & \textbf{Verdict} & \textbf{Reasoning} \\
\hline
FNO & In-distribution accuracy on Darcy, Burgers, Navier--Stokes & Moderate & Supported & Low relative $L^2$ error against several neural baselines, but only one RBM entry, which is the strongest baseline on Darcy at 2.3--2.6 times FNO's error rather than the stated order of magnitude; no tuned classical baseline and no multi-seed intervals. \\
FNO & Speedup versus traditional solvers & Weak $\to$ Red Flag & Mixed & About $440\times$ per query is measured once, below the stated three orders of magnitude; an end-to-end advantage is not established because the headline excludes the $\approx 12$\,h offline cost, which is amortized only over $30{,}000$ MCMC evaluations. \\
FNO & Resolution transfer & Moderate (math.) / Weak (deployment) & Supported (math.) / Not established (deployment) & Grid transfer is shown on a fixed input distribution, which supports the mathematical sense of Section~\ref{subsec:operator_learning} but not deployment shift. \\
FNO & Generalization across the parametric family & Weak & Not established & Training and test inputs come from the same Gaussian random field; no coefficient, forcing, or geometry shift is evaluated. \\
\hline
PINN & Forward solution feasibility & Moderate & Supported & Low error against high-fidelity references for the instances shown, but no tuned classical baseline, cost comparison, or multi-seed reporting. \\
PINN & Inverse parameter identification & Moderate & Supported & Striking Navier--Stokes and KdV discovery examples, but no adjoint-based or variational assimilation baseline. \\
PINN & Efficiency / solver replacement & Weak & Not established & No accuracy-matched wall-clock comparison; a later controlled study finds PINNs do not beat FEM on combined time and accuracy \citep{grossmann2023canpinn}. \\
PINN & Robustness across PDE difficulty & Weak & Not established & Results are shown on configurations that train successfully; failure modes on harder PDEs are documented later \citep{krishnapriyan2021characterizing,deryck2024numerical}. \\
\hline
\end{tabularx}
\end{table}

\section{Decision matrix}
\label{sec:decision_matrix}

The decision matrix translates the evidence synthesis of Sections~\ref{sec:taxonomy}--\ref{sec:case_studies} into solver selection guidance. A recommendation receives High confidence when it is supported by Strong evidence from at least two PDE families or by mature classical theory plus benchmark evidence with clear limitations; Moderate confidence when it rests on Moderate evidence, on Strong evidence from one PDE family, or on evidence lacking one important stress test; and Low confidence when it rests mainly on narrow benchmarks, incomplete cost accounting, weak out of distribution testing, or unresolved deployment behavior. Red Flag evidence is never converted into a positive recommendation, and neither is a Not supported verdict, however well measured. For example, adaptive FEM for one-off elliptic problems receives High confidence from mature a posteriori theory and diverse benchmark evidence \citep{ainsworth2000aposteriori,bonito2024adaptive}; FNO on many-query Darcy flow receives Moderate confidence because in-distribution accuracy is credibly demonstrated but distribution shift evidence and ROM comparison are incomplete (Section~\ref{subsec:case_audits}); and learned iterative solvers on general linear systems receive Low confidence because their failure mode evidence relies substantially on one unreviewed preprint \citep{wu2026hybridreliability} (Section~\ref{sec:hybrid_solvers}).

Six decision rules follow from Sections~\ref{sec:taxonomy}--\ref{sec:benchmark_landscape}. (i) Classical solvers remain the High confidence default for one off high fidelity solves of fully specified problems requiring error control, and a learned alternative must beat a tuned classical workflow rather than a uniform or unpreconditioned baseline (Section~\ref{subsec:classical_solvers}). (ii) ROMs and operator learners become attractive only when many queries amortize an offline stage; ROMs should be considered first when intrusive access, low dimensional parametric structure, and error estimation are available, and a many-query recommendation should state the query volume that justifies the offline stage, at least $3\times10^4$ for FNO errors below 1\% in our measured Darcy study (Table~\ref{tab:measured_nstar}), and recomputed for each regime. (iii) Supervised operator learning, including GON style variants, is infeasible when paired data cannot be generated or measured, leaving classical, residual trained, inverse, or residual-supervised hybrid options (Table~\ref{tab:data_requirements}). (iv) PINN style function learning is a Moderate or Low confidence option for data scarce, inverse, high dimensional, or mesh difficult settings, conditional on reference error, boundary error, multiple seeds, and a classical comparison (Section~\ref{subsec:function_learning}). (v) Autoregressive emulators are recommended only for repeated dynamic prediction with long horizon rollout diagnostics (Section~\ref{subsec:autoregressive_emulators}). (vi) Hybrids are classified by learned component and retained structure, with classical dominant hybrids most credible when residual monitoring or a fallback solver remains, and hybrid evidence downgraded when only average error is reported (Table~\ref{tab:hybrid_classification}).

\begin{table}[tp]
\centering
\caption{Solver family recommendation by decision regime. The first seven rows are indexed by PDE structure and use case. The final row (``Limited paired data but known PDE'') is a data-axis complement: it cross-cuts the PDE-regime rows above it for the specific case where the governing equation is known but paired solution data are scarce. Readers should consult both the relevant PDE-regime row and this row when data availability is the primary constraint. Supplement S9 traces every confidence level to the theory, external studies, graded claims, and measurements that support it.}
\label{tab:decision_matrix}
\tiny
\setlength{\tabcolsep}{2pt}
\renewcommand{\arraystretch}{1.0}
\begin{tabularx}{\linewidth}{R{0.9}R{0.9}R{0.8}R{0.9}R{0.9}R{0.9}R{0.9}R{0.9}R{0.9}R{0.8}R{1.2}}
\hline
Decision regime & Classical solvers & ROMs & Function learning & Operator learning & Autoregressive emulators & Classical dominant hybrids & ML dominant hybrids & Data requirement & Decision confidence & Failure conditions \\
\hline
One off high fidelity solve & Default choice with FEM, FVM, FDM, adaptive solvers, and strong linear algebra & Usually not needed unless related queries follow & Conditional when mesh free or high dimensional structure dominates & Usually not justified because paired corpus and training cost are hard to amortize & Not appropriate unless the task is dynamic prediction & Useful when learned acceleration retains residual checks & Usually weak unless residual training solves a real bottleneck & PDE, geometry, boundary data, and no paired corpus required for classical solve & High for classical, Low for most ML replacement claims & Certification need, weak baseline, missing error control, or omitted solver tolerance \\
\hline
Many query parametric family & Strong reference layer and data generator & Strong candidate when reduced basis structure and error estimation exist & Limited role unless each instance is solved separately or residual fine tuning is needed & Strong candidate when paired corpus and query volume justify amortization & Useful only for time dependent repeated prediction & Useful when learned components accelerate repeated solves & Useful when physics informed operator improves data efficiency or transfer & Offline solver, snapshots, or paired solution corpus usually needed & High for ROMs in favorable structure, Moderate for neural operators with full evidence & Query volume too low, corpus cost omitted, weak ROM comparison, or OOD shift untested \\
\hline
High dimensional or mesh free solve & Problem dependent and often limited by dimensionality & Usually difficult unless low dimensional structure exists & Candidate through DGM, PINN, Deep Ritz, or weak formulations & Candidate only if paired high dimensional data exist & Usually not central unless high dimensional dynamics are sampled & Possible if solver in the loop sampling is feasible & Possible when residual and data signals are combined & Residual signal may be more feasible than paired corpus & Moderate to Low depending on reference validation & No trusted reference, unstable optimization, weak sampling study, or no dimension sensitivity \\
\hline
Complex or irregular geometry & Default when meshing and geometry treatment are mature & Candidate when geometry parameterization supports reduction & Candidate when boundary and geometry encoding are reliable & Candidate with geometry aware operators and adequate geometry corpus & Limited unless trajectory geometry data exist & Candidate through learned correction within FEM or FVM workflow & Candidate when physics informed geometry transfer is tested & Geometry representation and boundary data are decisive & High for mature FEM, Moderate to Low for geometry learning & Boundary error omitted, test geometries too close to training, or mesh cost ignored \\
\hline
Long time nonlinear or chaotic rollout & Default reference with FVM, spectral, FEM, or CFD solvers & Limited unless dynamics have reducible structure & Usually difficult unless residual training is stable over time & Candidate for solution operators over fixed horizons & Candidate when rollout evidence is strong & Candidate through learned correction or ML accelerated CFD & Candidate when residual regularization improves rollout stability & Trajectory data and rollout diagnostics usually needed for learned methods & High for classical reference, Moderate to Low for emulators and hybrids & One step error only, no conservation test, no attractor statistics, or unstable long horizon behavior \\
\hline
Data informed inverse regime & Strong through adjoint, variational, and PDE constrained optimization when available & Useful for repeated inverse solves with reduced models & Candidate because residual and sparse observations can be combined & Usually secondary unless paired inverse data or surrogate training data exist & Not central unless inverse dynamics are learned from trajectories & Candidate with solver in the loop or differentiable correction & Candidate when physics informed surrogate supports inference & Sparse observations, PDE residual, and regularization dominate & Moderate for PINN style inverse use with validation, High for mature adjoint workflows & Identifiability ignored, uncertainty omitted, sparse data overinterpreted, or forward accuracy assumed to imply inverse reliability \\
\hline
Shock dominated or conservation critical regime & Default with conservative FVM and suitable limiters & Possible for smooth parametric summaries but difficult near shocks & Risky unless conservation and discontinuity treatment are explicit & Candidate only with shock aware training and stress tests & Risky for long rollout without conservation diagnostics & Candidate with learned discretization inside conservative workflow & Usually Low unless physics constraints preserve conservation & Conservation diagnostics and near shock metrics are required & High for classical FVM, Low to Moderate for learned methods & Average error only, shock location omitted, conservation ignored, or limiter comparison absent \\
\hline
Limited paired data but known PDE & Classical default if direct solve is affordable & Possible if snapshots can be chosen efficiently & Candidate through residual, weak, or variational training & Usually not feasible unless active data generation expands corpus & Not feasible without trajectories & Candidate through learned correction with solver in the loop & Candidate through physics informed fine tuning & Residual signal may be available while paired corpus is limited & High for classical, Moderate for residual methods with evidence & Treating no labels as no cost, no collocation sensitivity, or no independent reference \\
\hline
\end{tabularx}
\end{table}

Table~\ref{tab:decision_matrix} is the central practical output of this review, mapping decision regimes to feasible families, data requirements, confidence levels, and failure conditions. Each confidence level is traceable: Supplement S9 lists, for every regime, the evidence behind the entry and the missing evidence that keeps it from being higher, so a reader can dispute any cell on its evidence rather than on its label. Its value lies in evidence specific qualifications rather than generic cautions. For example, the appropriate classical baselines for learned closures are variational and heterogeneous multiscale methods (Section~\ref{subsec:classical_solvers}). In the shock dominated row, conservative finite volume methods retain High confidence because no neural operator has yet demonstrated conservation-preserving shock capture at finite volume accuracy on a standardized cross family benchmark; specialized conservation-aware neural methods exist in constrained settings \citep{discacciati2020controlling} but do not yet meet that threshold, so a learned method there should report shock location, conservation, and near-discontinuity error against a limiter-equipped baseline. When studies conflict on a similar PDE family, the matrix follows the more conservative result unless the stronger result has better baselines, transparent data generation, multiple seeds, cost accounting, and out of distribution testing, which keeps confidence tied to evidence quality rather than to the most optimistic report.

Figure~\ref{fig:solver_selection_flowchart} gives the procedural reading of the matrix. Data availability is the first branch, because operator learning cannot be recommended without a feasible paired corpus; the second separates one off from many-query use, because direct solution, reduced basis construction, operator training, and emulation amortize differently; and the third raises the evidence threshold when certification, conservation, or safety critical reliability is required.

\section{Open problems and research priorities}
\label{sec:open_problems}

Each open problem is treated as an unresolved decision barrier: can the method be trusted, afforded, reproduced, and deployed in the regime where it is recommended? A table in Supplement S11 links each barrier to the affected families, the current evidence status, and the evidence needed to strengthen future recommendations. Several barriers deserve emphasis because the decision matrix depends on them most.

Out of distribution trustworthiness is the central barrier for operator learning, since current evidence supports interpolation within benchmark distributions more strongly than deployment shifts, and future benchmarks should distinguish interpolation, moderate extrapolation, and deployment shift with fixed split definitions (Sections~\ref{subsec:operator_learning} and~\ref{subsec:case_audits}). Uncertainty quantification matters whenever learned solvers enter design, risk assessment, or inverse inference. Neural methods such as Bayesian PINNs and ensemble or dropout estimators exist \citep{yang2021bpinns,psaros2023uncertainty}, so the open problem is their calibration against the classical stochastic baselines, generalized polynomial chaos with spectral convergence in the stochastic dimension and multilevel Monte Carlo with near-optimal cost for elliptic problems, and their link to numerical reference error, where reduced basis methods already provide computable estimators. Geometry generalization, shock preservation, and long horizon reliability each ask whether learned solvers can match mature finite element, finite volume, and spectral workflows once boundary error, mesh and data generation cost, conservation and shock location, and rollout statistics are counted. Hybrid convergence guarantees require learned components whose effect on fixed points, stability, and tolerance can be analyzed or reliably tested. Data efficiency calls for learning curves against training instances, reference resolution, and generation cost for operators and against collocation samples and seeds for PINNs, together with solver in the loop acquisition that uses residuals, uncertainty, or validation error to decide where new reference solves are generated, as AL4PDE begins to benchmark \citep{musekamp2024al4pde}. Reproducibility infrastructure remains a cross family problem, and the case studies of Section~\ref{sec:case_studies} show how much a claim's grade can move once reproduction, baseline fairness, cost anchoring, and shift tests are added. High dimensional PDEs still lack accepted protocols with trusted references, dimension scaling studies, and comparisons against sparse grid or stochastic numerical baselines.

Three emerging directions bear directly on the data cost axis of this review. Multi-fidelity operator learning combines a few expensive high-fidelity solves with many cheap low-fidelity ones: multi-fidelity DeepONet reached accuracy competitive with high-fidelity-only training at substantially reduced high-fidelity budgets \citep{lu2022multifidelity}, and a physics-informed multi-fidelity operator network analyzed the bias--variance trade-off on parametric Darcy and Navier--Stokes problems with the kind of cost accounting the rubric requires \citep{howard2023multifidelity}, although the evidence is still concentrated on a narrow set of problems. Foundation models pretrain across physical systems; Poseidon matched or exceeded specialized architectures on held-out PDEs at lower data requirements while degrading when the target PDE was structurally distant from pretraining \citep{herde2024poseidon}, and related studies show both promise and open questions about transfer across genuinely different operators \citep{mccabe2023multiple,subramanian2024foundation}. Generative and diffusion based solvers are increasingly used for field reconstruction and probabilistic prediction, but their evidence for fair cost accounting and out of distribution reliability is still thin. Each direction should be evaluated with the same data, cost, and shift accounting as the established families, and whether its advantages persist at matched budget against a tuned specialized model and, where the parameter dependence allows one, a certified reduced basis is the defining question.

The central research priority is to make solver evidence conditional and auditable. Future studies should state the decision regime, data regime, query volume, baseline class, cost model, and failure conditions before presenting superiority claims, which would make the field less dependent on architecture centered comparison and more useful to practitioners choosing among classical, ML only, and hybrid solvers under real constraints.

Several of the regularities identified in this review can be stated as conjectures that future work can confirm or refute; a fifth candidate, that method rankings are unstable across PDE families, has already been measured (Section~\ref{subsec:benchmark_suites}). \emph{Conjecture 1 (accuracy, not resolution).} Because a surrogate's error is set by its training data, a finer evaluation grid raises its cost without lowering its error, while the classical solve that matches its accuracy stays coarse; operator learning can then pay only where reaching the required accuracy classically is itself expensive. The measured Darcy study supports the first part (Section~\ref{subsec:measured_nstar}), and its data make the second quantitative. If the surrogate's error falls as $N_{tr}^{-\alpha}$ and the classical cost at matched accuracy grows as $\varepsilon^{-q}$, with $q\to d/p$ for an optimal solver of order $p$ in $d$ dimensions, then at fixed surrogate resolution and $S\gg1$, Equation~\eqref{eq:amortization_law_eeff} gives $N^*\propto\varepsilon^{\,q-1/\alpha}$, so a tighter tolerance favors learning only if $\alpha > p/d$. On the Darcy data, $p\approx1.4$, and $\alpha$ falls from 0.7--0.9 over $N_{tr}=100$--$2000$ to 0.4--0.6 above $N_{tr}=1000$; the resulting exponents, $-1.0$ at $64^2$ and $-0.6$ at $128^2$, agree with the measured $-0.9$ and $-0.5$ (interpolated comparator; Supplement S10). In two dimensions the criterion $\alpha>0.7$ fails at the accurate end, so tightening the tolerance raises break-even. In three dimensions it becomes $\alpha>0.47$, which the same $\alpha$ straddles. The conjecture is therefore that operator learning gains from tighter tolerances exactly where $\alpha$ exceeds $p/d$, and the test is the matched-accuracy protocol on a three-dimensional problem, measuring $\alpha$ and $p$ directly. \emph{Conjecture 2 (proxy gap).} For each family, method rankings by the commonly reported proxy and by the decision quantity of Table~\ref{tab:proxy_gap} disagree; a test is to report both in benchmark suites and compute their rank correlation. \emph{Conjecture 3 (non-transferability).} Interpretability-guided operators that do not enforce structure architecturally will generally fail direct structural tests even when accurate; a test is to apply the symmetry and boundary checks of Section~\ref{subsec:case_gon} to other Green's-function operators on controlled problems. \emph{Conjecture 4 (retained structure).} Among hybrids, claimed speedup rises and verifiability falls as less classical structure is retained (Section~\ref{sec:hybrid_solvers}); a test is a coded set of hybrid studies plotted on these two axes.

\section{Conclusion}
\label{sec:conclusion}

This review has argued that no solver family wins universally, because practical PDE solver choice depends on the PDE regime, the available data, the offline budget, the query volume, the geometry, the solution regularity, the time horizon, and the need for certification. Classical methods remain the strongest default for one off high fidelity solves, conservation critical problems, and certified workflows, where mature discretizations, adaptive refinement, multigrid, preconditioning, and domain decomposition provide reliable performance without a training corpus. This does not dismiss ML solvers: residual trained methods, neural operators, autoregressive emulators, and hybrids address genuine bottlenecks in data scarce inverse settings, high dimensional approximation, many-query parametric inference, and learned acceleration of repeated simulation.

\begin{figure}[H]
\centering
\includegraphics[width=0.75\linewidth]{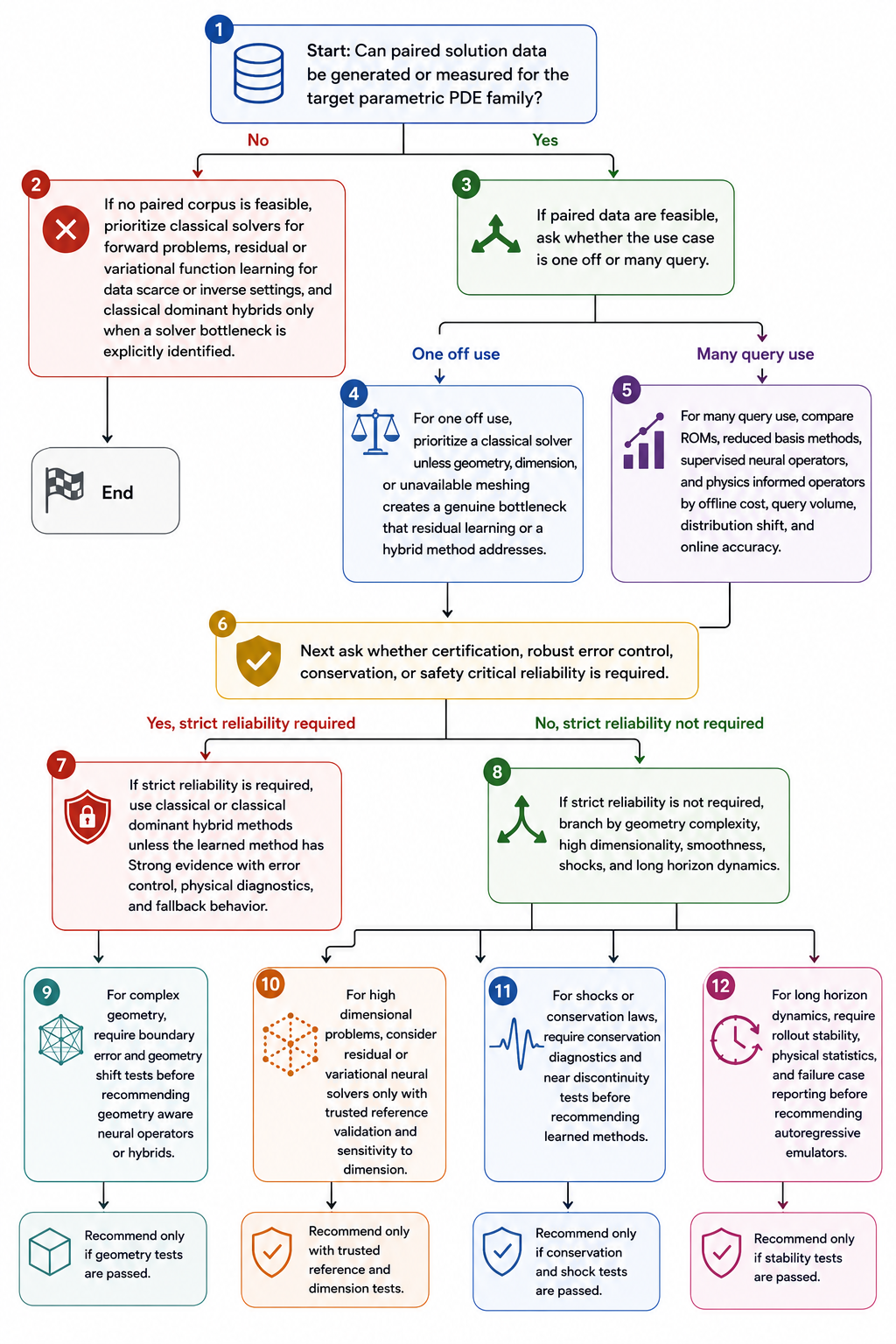}
\caption{Practical solver selection flowchart. The flowchart begins with paired solution data, then separates one off and many query use, then applies certification, geometry, dimensionality, smoothness, shock, and long horizon reliability filters.}
\label{fig:solver_selection_flowchart}
\end{figure}

The central practical synthesis is the decision matrix (Table~\ref{tab:decision_matrix}) and its flowchart (Figure~\ref{fig:solver_selection_flowchart}), which begin with data availability, separate one off from many-query use, and raise the evidence threshold when certification, conservation, or long horizon reliability is required. Data requirement is a central conclusion in its own right, because the families draw on fundamentally different information: classical solvers consume the model directly, function learners use residual or variational signals that are not automatically cheap, operator learners need a paired corpus whose generation cost must be counted before any speedup is accepted, and hybrids must be judged by where learning enters the solver.

Credible claims of ML superiority therefore require more than a favorable accuracy number: fair classical baselines, reduced basis baselines for many-query tasks whose parameter dependence admits them, transparent reference generation, separate reporting of data generation, training, and inference cost at stated query volumes, multiple seeds with confidence intervals, out of distribution tests matched to deployment, and PDE specific diagnostics such as conservation, near boundary and near discontinuity error, and rollout stability. The results of this review show why. The amortization law and the measured Darcy study show that break-even is set by offline cost: no FNO model below 1\% error broke even before about $3\times10^4$ queries, finer resolution raised break-even at least thirtyfold, and inference speed mattered mainly through the training cost it implies. The same data show that the reduced basis is not a universal many-query comparator: on the thresholded Darcy benchmark it needs 125--335 modes to match FNO and has no cheap certified online stage, so the choice between the two depends on parametric structure. The evidence ladder and the proxy gap show that each family's documented failures lie where its reported metric departs from its decision quantity, so evidence at one rung does not establish a claim at the next. Published benchmarks do not agree on the winner, so no single suite can support a cross family recommendation. The three case studies of Section~\ref{sec:case_studies} show what the resulting standard does in practice: one rubric supports in-distribution accuracy at Moderate quality for at least some cases of each method, while speedup, generalization, efficiency, and structural claims receive the weakest grades or negative verdicts. Establishing the rubric's inter-rater reliability on a pre-specified corpus is the natural next step.

The findings have different consequences for different readers. \emph{Method developers} should report data and training cost alongside inference speedup, compute $N^*$ for the use case they claim, and support structural claims with direct structural tests rather than accuracy. \emph{Benchmark designers} should include a tuned classical baseline and, for many-query tasks, a reduced basis baseline or a POD decay test that justifies its absence, fix deployment-matched shift splits, and report each family's decision quantity rather than only its training proxy. \emph{Practitioners} should start from the decision matrix, check whether their own query volume exceeds a method's break-even, and treat a Not supported or Not established verdict as a reason to stay with the classical workflow. \emph{Reviewers and editors} can apply the ten questions in Table~\ref{tab:reviewer_checklist}, which condense the evidence ladder, the amortization law, and the rubric into a form usable during peer review.

\begin{table}[H]
\centering
\caption{Ten questions for reviewing a claim that a learned or hybrid PDE solver outperforms a classical method.}
\label{tab:reviewer_checklist}
\scriptsize
\setlength{\tabcolsep}{4pt}
\renewcommand{\arraystretch}{1.15}
\begin{tabularx}{\linewidth}{p{0.03\linewidth}X}
\hline
\textbf{\#} & \textbf{Question} \\
\hline
1 & What type of claim is made, and which rung of the evidence ladder (P0--P4) or which structural test supports it? \\
2 & Is error measured against an independent reference with a stated tolerance and resolution, rather than by the training residual? \\
3 & Is the classical baseline the one appropriate to the regime (for example tuned multigrid or preconditioned Krylov, conservative finite volume or DG, adaptive FEM), and is it named? \\
4 & For a many-query claim, is a reduced basis or reduced order baseline included, or its absence justified? \\
5 & Is the comparison made at equal accuracy? \\
6 & Are data generation, training and tuning, and inference costs reported separately, with hardware? \\
7 & Is the break-even query volume $N^*$ given, and does the stated use case exceed it? \\
8 & Do the test splits include the distribution shifts relevant to deployment? \\
9 & Are at least three seeds with confidence intervals, failed runs, code, and data reported? \\
10 & Is every structural claim (conservation, symmetry, discretization invariance, interpretability) supported by a direct structural test? \\
\hline
\end{tabularx}
\end{table}

The next advance should therefore be better cross family evaluation infrastructure rather than more architectures: benchmark tasks with documented reference solvers, strong classical and ROM baselines, fixed in-distribution and out of distribution splits, standardized cost accounting, reproducibility checklists, physical diagnostics, and failure case reporting. With this infrastructure, classical, ML only, and hybrid solvers can be compared as decision alternatives rather than as isolated research communities, and solver choice can follow regime and evidence quality rather than method novelty.

\bibliographystyle{ACM-Reference-Format}
\bibliography{interactapasample}

\end{document}


\title{Supplementary Material for: Classical, Machine Learning, and Hybrid Solvers for Partial Differential Equations: An Evidence-Weighted Review and Decision Framework}

\author{Yingchao Huang}
\authornote{Corresponding author.}
\affiliation{%
  \institution{University of Alberta}
  \department{Faculty of Engineering, Electrical \& Computer Engineering Department}
  \city{Edmonton}
  \state{AB}
  \postcode{T6G 2H5}
  \country{Canada}
}
\email{yingcha1@ualberta.ca}

\author{Xin Wang}
\affiliation{%
  \institution{Saskatchewan Polytechnic}
  \department{Faculty of Digital Innovation, Arts \& Sciences}
  \city{Regina}
  \state{SK}
  \postcode{S4S 5X1}
  \country{Canada}
}

\author{Shanshan Yao}
\affiliation{%
  \institution{University of Alberta}
  \department{Civil and Environmental Engineering}
  \city{Edmonton}
  \state{AB}
  \postcode{T6G 2H5}
  \country{Canada}
}

\author{Fanhua Zeng}
\affiliation{%
  \institution{University of Regina}
  \department{Energy Systems Engineering, Faculty of Engineering and Applied Science}
  \city{Regina}
  \state{SK}
  \postcode{S4S 0A2}
  \country{Canada}
}

\author{Wei Peng}
\affiliation{%
  \institution{University of Regina}
  \department{Industrial Systems Engineering, Faculty of Engineering and Applied Science}
  \city{Regina}
  \state{SK}
  \postcode{S4S 0A2}
  \country{Canada}
}

\settopmatter{printacmref=false, printccs=false}

\maketitle

This electronic supplement accompanies the main article and contains the full protocols and case-study records summarized there. Section, table, and figure numbers with an S prefix refer to this supplement; numbers without the prefix refer to the main article.

\section{Reproducibility checklist}
\label{app:reproducibility_checklist}

The reproducibility checklist is included to make evidence extraction consistent across classical, ML only, and hybrid PDE solver studies, because solver recommendations depend on the reporting of PDE specifications, reference solutions, data generation, cost accounting, random seeds, and failure cases. The checklist is intended to be applied before a reported result is used in the evidence rubric, because missing information about solver tolerance, data splits, hardware, or baseline tuning can change the decision confidence assigned to a method.

\begin{table}[H]
\centering
\caption{Reproducibility checklist.}
\label{tab:reproducibility_checklist}
\scriptsize
\setlength{\tabcolsep}{4pt}
\renewcommand{\arraystretch}{1.18}
\begin{tabularx}{\linewidth}{p{0.20\linewidth}X}
\hline
Checklist field & Required reporting content \\
\hline
PDE statement & State the governing PDE, dependent variables, coefficients, source terms, parameter ranges, and whether the problem is deterministic, parametric, inverse, or time dependent. \\
\hline
Boundary and initial conditions & Report boundary condition type, boundary values, initial condition generation rule, compatibility assumptions, and whether boundary or initial conditions are shifted in test cases. \\
\hline
Geometry and mesh & Report domain geometry, mesh type, mesh resolution, grid spacing, element type, geometry encoding, and any interpolation between reference solver and learned model representations. \\
\hline
Reference solver & Report finite difference, finite volume, finite element, spectral, reduced order, or experimental reference source, including discretization order, time integrator, nonlinear solver, solver tolerance, and validation checks. \\
\hline
Training, validation, and test splits & Report split construction, number of PDE instances, parameter sampling rule, trajectory windows, whether time windows are correlated, and whether in distribution and out of distribution splits are separated. \\
\hline
Hardware and software & Report processor, accelerator, memory, software versions, solver library versions, neural framework versions, and operating system when timing or reproducibility claims are made. \\
\hline
Hyperparameters & Report network architecture, optimizer, learning rate schedule, batch or collocation sampling rule, loss weights, stopping criteria, regularization, and model selection rule. \\
\hline
Random seeds & Report at least three random seeds for learned methods and explain which sources of randomness are controlled, including initialization, sampling, batching, and data generation. \\
\hline
Code and data availability & Report public repository, commit hash, data location, license, scripts for data generation, scripts for training, scripts for evaluation, and any missing proprietary components. \\
\hline
Timing and memory & Report reference solution generation time, offline data generation time, training time, inference time, reduced basis construction time when relevant, memory use, and amortized cost at stated query volumes. \\
\hline
Physical and PDE metrics & Report problem appropriate metrics such as relative \(L^2\) error, \(H^1\) error, residual, near boundary error, conservation error, shock location error, rollout stability, and quantities of interest. \\
\hline
Failure cases & Report failed seeds, unstable rollouts, convergence failures, boundary failures, distribution shift failures, sensitivity to hyperparameters, and cases where the learned method underperforms a classical baseline. \\
\hline
\end{tabularx}
\end{table}

Table~\ref{tab:reproducibility_checklist} gives the minimum information needed to reproduce a solver claim and to decide whether the reported result can support a recommendation in Table~\mainref{tab:decision_matrix}. The checklist also makes data requirement explicit by requiring the source and cost of any solution corpus, trajectory archive, residual training signal, solver state, learned correction target, or reduced basis snapshot set.

\section{Evidence extraction schema}
\label{app:evidence_extraction_schema}

The evidence extraction schema is designed to convert heterogeneous papers into comparable evidence records, because classical solvers, residual trained neural solvers, supervised operator learners, autoregressive emulators, and hybrid methods report different information by default. The schema should be filled for each study before assigning claim strength, because missing extraction fields often reveal whether a result supports solver selection or only method exploration.

Table~\ref{tab:evidence_extraction_schema} supports conflict aware synthesis because it records differences in PDE regime, train test split, reference solver, baseline strength, cost accounting, and out of distribution testing that can explain contradictory results. The schema also prevents supervised operator learning results from being interpreted without solution corpus accounting, and it prevents residual trained results from being interpreted without optimization and validation evidence.

\begin{table}[H]
\centering
\caption{Evidence extraction schema.}
\label{tab:evidence_extraction_schema}
\scriptsize
\setlength{\tabcolsep}{4pt}
\renewcommand{\arraystretch}{1.18}
\begin{tabularx}{\linewidth}{p{0.20\linewidth}X}
\hline
Extraction field & Information to extract from each study \\
\hline
Bibliographic record & Extract authors, year, title, venue, DOI or URL, code repository, data repository, and date accessed when software or data are used. \\
\hline
PDE type & Extract whether the problem is elliptic, parabolic, hyperbolic, mixed type, incompressible flow, reaction diffusion, chaotic dynamics, high dimensional, or inverse. \\
\hline
Task and decision regime & Extract whether the task is one off solve, many query surrogate, high dimensional approximation, geometry transfer, long horizon rollout, or data informed inverse modeling. \\
\hline
Geometry and boundary conditions & Extract domain type, mesh or grid representation, boundary condition family, initial condition family, and whether geometry or boundary shifts are tested. \\
\hline
Solver family & Extract whether the method is classical, ROM, function learning, operator learning, autoregressive emulator, classical dominant hybrid, or ML dominant hybrid. \\
\hline
Learned component & For hybrid methods, extract whether learning enters as a residual term, discretization, correction, closure, preconditioner, prolongation operator, warm start, or iterative update. \\
\hline
Data requirement & Extract whether the method uses no training data, residual or variational supervision, paired solution corpus, trajectory data, solver states, correction targets, matrices, measurements, or mixed supervision. \\
\hline
Reference solver or measurement source & Extract the source of ground truth, including numerical method, resolution, tolerance, validation, experimental measurement details, and preprocessing. \\
\hline
Baseline set & Extract classical baselines, ROM baselines, neural baselines, hybrid baselines, tuning effort, and whether baseline compute budgets are matched. \\
\hline
Metrics & Extract accuracy metrics, residual metrics, boundary metrics, conservation metrics, physical quantities, rollout metrics, stability metrics, and uncertainty metrics. \\
\hline
Cost accounting & Extract reference generation cost, data generation cost, training cost, inference cost, reduced basis construction cost, memory, hardware, and amortized cost. \\
\hline
Out of distribution testing & Extract parameter shift, coefficient shift, boundary shift, geometry shift, resolution shift, forcing shift, noise shift, and rollout horizon shift. \\
\hline
Statistical reporting & Extract number of random seeds, confidence intervals, bootstrap intervals, failed runs, and hyperparameter sensitivity. \\
\hline
Reproducibility level & Extract code availability, data availability, dependency versions, commit hash, scripts, hyperparameters, solver tolerances, and licensing restrictions. \\
\hline
Evidence grade & Assign Strong, Moderate, Weak, or Red Flag evidence using Table~\ref{tab:claim_strength_operational}. \\
\hline
Decision implication & Record whether the result supports a recommendation, a conditional recommendation, exploratory promise, or insufficient evidence for decision guidance. \\
\hline
\end{tabularx}
\end{table}

\section{Operational claim strength rubric}
\label{app:claim_strength_rubric}

The operational claim strength rubric formalizes the Strong, Moderate, Weak, and Red Flag evidence quality levels used in Section~\mainref{sec:benchmark_landscape} and Section~\mainref{sec:decision_matrix}, because performance numbers must be translated into decision confidence through baseline fairness, statistical evidence, cost accounting, out of distribution testing, data generation transparency, and reproducibility \citep{takamoto2022pdebench,hao2024pinnacle}. The rubric is intentionally conservative because overconfident solver recommendations can arise when a method is evaluated only on narrow in distribution benchmarks or against under tuned baselines.

Table~\ref{tab:claim_strength_operational} is the operational version of the rubric used in Table~\mainref{tab:evidence_rubric}, and it should be applied to each claim rather than to an entire method family. A method may have Strong evidence for interpolation over one parametric family and Weak evidence for geometry transfer or long horizon deployment, so decision confidence should always be tied to the specific claim and regime \citep{kovachki2023neuraloperator,takamoto2022pdebench}. Quality is recorded separately from the verdict (Supported, Not supported, Mixed, or Not established) described in Section~\mainref{subsec:evidence_rubric}: a well-measured negative result is Strong quality with a Not supported verdict, and Red Flag is reserved for evidence too incomplete to judge.

\begin{table}[H]
\centering
\caption{Detailed operational claim strength rubric. This table elaborates Table~\mainref{tab:evidence_rubric} in \S\mainref{subsec:evidence_rubric} with expanded operational criteria for each evidence level.}
\label{tab:claim_strength_operational}
\scriptsize
\setlength{\tabcolsep}{3pt}
\renewcommand{\arraystretch}{1.18}
\begin{tabularx}{\linewidth}{p{0.12\linewidth}p{0.18\linewidth}p{0.18\linewidth}p{0.18\linewidth}X}
\hline
Evidence level & Operational meaning & Minimum evidence requirements & Typical decision language & Downgrade triggers \\
\hline
Strong & The claim is credible for decision guidance in the stated PDE regime and cost model. & Verified source, strong classical baseline, ROM baseline when relevant, tuned neural baselines, matched data and compute accounting, at least three seeds, confidence intervals, meaningful out of distribution tests, transparent data generation cost, and reproducible code or data. & Use direct recommendation language while stating failure conditions and scope. & Downgrade if a relevant baseline is missing, if amortized cost is unclear, if out of distribution tests do not match deployment, or if reproduction information is incomplete. \\
\hline
Moderate & The claim is plausible but conditional because one or more important evidence components are incomplete. & Verified source, reasonable baselines, partial statistical reporting, partial cost accounting, some out of distribution testing, and enough implementation detail for approximate reproduction. & Use conditional recommendation language and identify the missing evidence component. & Downgrade if the missing evidence affects the central deployment claim, such as no ROM baseline for a many query operator learning claim. \\
\hline
Weak & The claim is exploratory and should not be used as a default solver recommendation. & Narrow benchmark, limited baseline tuning, accuracy focused reporting, weak train test split, or incomplete data generation cost. & Use cautionary language and describe the result as motivation for further testing. & Downgrade to Red Flag if the study makes strong replacement or speedup claims without enough information for verification. \\
\hline
Red Flag & The claim is insufficient for decision guidance and should not be converted into a positive recommendation. & Speedup without amortization accounting, generalization from weak splits, weak or outdated baselines, missing reference solver information, missing solution corpus cost, physics consistency assessed only by training loss, or unreproducible evaluation. & Do not recommend and state what evidence would be needed. & Remains Red Flag until the evaluation is redesigned with appropriate baselines, cost accounting, and reproducibility information. \\
\hline
\end{tabularx}
\end{table}

\section{GON replication protocol}
\label{app:gon_replication}

This section reports a computational replication study for GON \citep{gu2025explainable}, structured as a targeted claim audit that separates faithful reproduction, classical baseline comparison, cost accounting, and evidence grading by claim type and PDE family. The executed portion covers faithful reproduction of all five original problem cases at the original architectures with three random seeds, an independent finite element reference and cost anchor, out-of-distribution evaluation, amortized cost analysis, a data efficiency sweep, and a Green's function structure audit. The one module that was not executed is the retuning of the neural baselines: the original FNO, PINN, and DeepONet implementations are not publicly released and could not be retrained within the available GPU budget, so all neural baseline values below are those reported by Gu et al.\ \citep{gu2025explainable}.

The replication has two purposes that should remain separate in reporting. The first purpose is faithful reproduction, where the original PDE families, train and test protocol, neural baselines, model settings, and reported metrics are reproduced as closely as the public information allows. The second purpose is decision stress testing, where the method is compared against strengthened neural baselines, classical reference solvers, data efficiency sweeps, out of distribution splits, and interpretability checks that make the solver recommendation more operational. 

Table~\ref{tab:gon_replication_modules} defines the modules required before the GON study can be used as high confidence decision evidence. The table also requires cost normalization with classical finite element or finite volume references, because a neural operator speedup claim should separate one-off cost, offline data generation, training, inference, and amortized cost at stated query volumes.

\paragraph{Predictive accuracy.} Table~\ref{tab:gon_reproduction_shell} reports the faithful reproduction of all five original cases. Every case was trained at its original layer widths, including Pipe and Stokes, without mixed precision, gradient accumulation, or any other memory-saving change; earlier reduced-width runs are superseded. The training settings follow the original paper (2000 Adam epochs with a constant learning rate for the Poisson cases; 7000 Adam epochs, a constant learning rate, and no weight decay for Pipe), and the only code additions are seed control and per-epoch logging. The evaluation metric is the unmodified GON \texttt{val\_mse}, so reproduced and original values are directly comparable. A preliminary single-seed 3D Poisson run had given $5.00\times10^{-5}$, but its logs were not retained and the value was not recovered by the three-seed protocol, so it is superseded by the values in the table.

The reported accuracy is not recovered on any non-trivial case. With the same metric, the three-seed means are about 40 times larger than the original value on 3D Poisson ($1.06\times10^{-3}$ versus $2.64\times10^{-5}$), about 1.8 times larger on Plate, about 2.8 times larger on Pipe, and about 2.3 times larger on Stokes. The 2D Poisson case moves in the opposite direction ($1.01\times10^{-6}$ versus $3.75\times10^{-4}$), but this is not evidence of a better method: the case is a thin 3D slab whose mesh is defined at unit scale while the data are at millimetre scale, so the source term vanishes on the mesh and the task reduces to interpolating boundary data. Against the neural baselines reported by Gu et al., GON does not beat FNO on 3D Poisson (the 95\% interval $[4.8\times10^{-4}, 1.6\times10^{-3}]$ contains the FNO value $6.32\times10^{-4}$), but it remains below the reported baseline on Plate, Pipe, and Stokes, where the upper end of each interval lies below the baseline value. Because those baselines were reported rather than retrained under matched splits, seeds, and hardware, the comparison mixes reproduced and reported numbers, and the accuracy grades below are capped at Moderate.

Table~\ref{tab:replication_ratios} collects the ratio of reproduced to originally reported error for every reproduction discussed in this review, our own and the independent FNO re-run of Lu et al.\ \citep{lu2022comprehensive}. The ratios are not one-sided: our re-implementation of the FNO Darcy study, with its own solver and coefficient fields, reached lower errors than Li et al.\ reported, while the faithful GON reproduction, with the original code and metric, reached higher errors on every non-degenerate case. A reproduction ratio is therefore evidence about a specific reported number, not about a method family.

\paragraph{Classical reference layer.} An independent P1 continuous Galerkin tetrahedral finite element solver (scikit-fem, sparse direct solve with SciPy) was built for the 3D Poisson heat-sink case, $-\Delta u=f$ with unit coefficient and Dirichlet boundary conditions. Over all 100 training and test instances it reproduces the stored reference solutions to a mean relative $L^2$ error of $2.44\times10^{-5}$ (maximum $1.28\times10^{-4}$, gate threshold $10^{-3}$), which confirms that the reference data follow the standard weak form with no hidden coefficient or sign change. A manufactured-solution test recovers the textbook P1 rate of about 2 on a clean cube; on the heat-sink geometry, uniform refinement to 9{,}108, 59{,}738, and 427{,}234 degrees of freedom gives relative $L^2$ errors of $1.55\times10^{-3}$, $7.65\times10^{-4}$, and $3.32\times10^{-4}$ (rates 1.13 and 1.27), a reduction attributed to element quality under uniform tetrahedral refinement rather than to a solver error. The production mesh has 1{,}578 degrees of freedom, of which 47 are free, and one solve takes 4.38~ms on average over 30 instances (4.08~ms assembly, 0.31~ms solve). The audit also exposed a formulation property that shapes the remaining modules: the heat sink is thin-walled and fully Dirichlet-bounded, and its deepest interior node is only 2.12~mm (4.2\% of the domain) from a Dirichlet face, so the stored solution is essentially the harmonic extension of the boundary data and the source term is nearly inert. This was established geometrically and confirmed on refined meshes; a numeric source-to-boundary norm ratio was not recorded. The finite element error is measured as relative $L^2$ over all nodes, and an MSE restricted to the GON interior evaluation nodes was not computed, so the two accuracy figures are not directly comparable.

\begin{table}[H]
\centering
\caption{GON replication modules and their evidence purpose.}
\label{tab:gon_replication_modules}
\scriptsize
\setlength{\tabcolsep}{3pt}
\renewcommand{\arraystretch}{1.18}
\begin{tabularx}{\linewidth}{p{0.16\linewidth}p{0.24\linewidth}p{0.22\linewidth}X}
\hline
Replication module & Protocol instruction & Evidence purpose & Reporting requirement \\
\hline
Faithful reproduction & Reproduce the original PDE families, data generation process, train and test protocol, neural baselines, model settings, and reported metrics as closely as possible. & Tests whether the original result can be recovered under documented conditions. & Report reproduced metrics, deviations, unavailable details, dependency versions, hardware, and failed reproduction cases. \\
\hline
Fair baseline audit & Retune PINN, DeepONet, physics informed DeepONet, FNO, and other reported neural baselines under matched data splits and matched compute budgets. [\textit{Not executed: baseline code not released; reported baseline values used.}] & Tests whether the reported advantage depends on weak or under tuned neural baselines. & Report baseline search space, training cost, validation rule, final test metrics, and confidence intervals. \\
\hline
Classical reference layer & Add at least one strong finite element, finite volume, finite difference, or spectral solver per PDE family as reference solution and cost anchor. & Distinguishes neural surrogate accuracy from practical solver replacement. & Report solver type, discretization, tolerance, resolution, hardware, wall clock time, memory, and validation diagnostics. \\
\hline
FEM or finite volume cost normalization & Report classical wall clock time, GON training time, GON inference time, offline data generation time, and reference solution generation time separately. & Prevents raw speedup claims from hiding offline costs or hardware differences. & Report same machine timing when possible, otherwise report both hardware configurations and avoid unnormalized speedup claims. \\
\hline
Amortized cost analysis & Compare one off solve cost separately from amortized cost per query at representative query volumes such as (N=100) and (N=1000). & Tests whether many query deployment is required before the learned operator is cost effective. & Report total offline cost, online cost per query, break even query volume, and comparison with direct classical solves and ROMs when applicable. \\
\hline
Data efficiency sweep & Train GON and baselines under reduced paired data budgets while holding evaluation splits fixed. [\textit{Executed for GON on two cases; baselines not swept.}] & Links the method directly to the data requirement axis in Section~\mainref{sec:decision_matrix}. & Report accuracy, variance, training stability, and cost versus number of training PDE instances. \\
\hline
Out of distribution tests & Evaluate parameter shift, boundary condition shift, resolution transfer, geometry shift when applicable, and noise sensitivity. [\textit{Executed on 3D Poisson; geometry shift not applicable.}] & Tests whether the method supports deployment claims beyond interpolation in the solution corpus. & Report each shift separately and identify which shifts change the evidence grade. \\
\hline
Interpretability audit & On a simple problem with known or well approximated Green's kernel, test symmetry, source behavior, boundary influence, and superposition where expected. [\textit{Executed on a controlled homogeneous Dirichlet case.}] & Tests whether the Green's function guidance is structurally meaningful rather than only predictive. & Report kernel visualizations, quantitative structure checks, and cases where predictive accuracy and interpretability disagree. \\
\hline
Statistical reporting & Use at least three random seeds and report confidence intervals or bootstrap intervals. & Separates stable effects from optimization or data sampling variation. & Report mean, interval, failed seeds, training instability, and hyperparameter sensitivity. \\
\hline
Repository and version control & Record repository URL, commit hash, date accessed, dependency versions, and deviations from original instructions. & Supports independent reproduction and explains implementation differences. & Report environment file, commit hash, data scripts, evaluation scripts, and unavailable components. \\
\hline
Evidence grading link & Grades for all executed modules are assigned in Table~\ref{tab:gon_grades} and summarized in \S\mainref{sec:case_studies}. The neural baseline retuning module was not executed and is marked as incomplete in the narrative. & Connects the replication directly to the review's claim strength rubric. & Report the final grade for each claim separately, including predictive accuracy, speedup, data efficiency, interpretability, and generalization. \\
\hline
\end{tabularx}
\end{table}

\begin{table}[H]
\centering
\caption{GON (GreensONet) faithful reproduction results at the original architectures. All metrics are test-set MSE computed with the unmodified GON evaluation code (\texttt{val\_mse}), so reproduced and original values use the same metric. Reproduction on an NVIDIA RTX~3090 (24~GB) with seeds 0, 1, and 2; intervals are two-sided Student's $t$ 95\% confidence intervals ($\mathrm{dof}=2$). Original GON and baseline values are those reported by Gu et al.\ \citep{gu2025explainable}; the baselines were not retrained. The Stokes model has three branch networks, one per velocity component, and the reported value is the $x$ component.}
\label{tab:gon_reproduction_shell}
\scriptsize
\setlength{\tabcolsep}{3pt}
\renewcommand{\arraystretch}{1.25}
\begin{tabularx}{\linewidth}{p{0.14\linewidth}p{0.11\linewidth}ccccp{0.17\linewidth}X}
\hline
\textbf{Case} & \textbf{Layers} & \textbf{Orig.\ GON} & \textbf{Seed 0} & \textbf{Seed 1} & \textbf{Seed 2} & \textbf{Reprod.\ mean [95\% CI]} & \textbf{Reported baseline} \\
\hline
\multicolumn{8}{l}{\textit{Steady heat conduction}} \\
2D Poisson (thin slab) & {[6,12,12,12,1]} & $3.75\times10^{-4}$ & $5.61\times10^{-7}$ & $1.22\times10^{-6}$ & $1.26\times10^{-6}$ & $1.01$ [$0.04$, $2.0$] $\times10^{-6}$ & $2.56\times10^{-2}$ \\
3D Poisson (heat sink) & {[6,12,12,12,1]} & $2.64\times10^{-5}$ & $1.33\times10^{-3}$ & $9.26\times10^{-4}$ & $9.16\times10^{-4}$ & $1.06$ [$0.48$, $1.6$] $\times10^{-3}$ & $6.32\times10^{-4}$ (FNO) \\
\hline
\multicolumn{8}{l}{\textit{Heterogeneous diffusion}} \\
Plate & {[6,12,12,12,1]} & $5.22\times10^{-4}$ & $1.03\times10^{-3}$ & $9.92\times10^{-4}$ & $8.65\times10^{-4}$ & $0.96$ [$0.75$, $1.2$] $\times10^{-3}$ & $2.89\times10^{-3}$ (FNO) \\
Pipe & {[6,24,24,24,1]} & $4.63\times10^{-4}$ & $1.46\times10^{-3}$ & $1.32\times10^{-3}$ & $1.09\times10^{-3}$ & $1.29$ [$0.83$, $1.7$] $\times10^{-3}$ & $2.14\times10^{-3}$ \\
\hline
\multicolumn{8}{l}{\textit{Stokes equations}} \\
Velocity ($x$) & {[6,12,24,12,1]} & $5.83\times10^{-4}$ & $1.33\times10^{-3}$ & $1.28\times10^{-3}$ & $1.49\times10^{-3}$ & $1.36$ [$1.1$, $1.6$] $\times10^{-3}$ & $2.17\times10^{-3}$ \\
\hline
\end{tabularx}
\end{table}

\paragraph{Out-of-distribution robustness.} Because the source term is nearly inert, the out-of-distribution tests on the 3D Poisson case perturb the boundary data as the physically active input (Table~\ref{tab:gon_ood}). Noise was added separately to the source $f$, the boundary data $g$, and both, at 1\% and 5\%. The extrapolation split retrains on the 80 instances with the lowest boundary mean (at most 0.614) and tests on the 20 with the highest (0.622 to 0.912). The resolution and boundary-shift tests use new finite element references on meshes regenerated with gmsh, which cannot reproduce the internal fin-wall Dirichlet constraints of the original COMSOL meshes. These tests therefore use an outer-Dirichlet-only definition with a different absolute baseline (91 interior nodes, $1.66\times10^{-3}$, versus 47 interior nodes, $1.06\times10^{-3}$), and the finding is the trend across shifts rather than the absolute level. GON is unaffected by input noise, mild extrapolation, a sixfold refinement of the interior mesh, and shifts of the boundary mean, but a fourfold increase in boundary amplitude raises the error about 16 times. Mesh coarsening could not be tested meaningfully because only two interior nodes survive on the thin geometry. The robustness that is observed follows from the problem being a linear boundary-harmonic map with an inert source, so it is structural rather than evidence of general out-of-distribution generalization.

\begin{table}[H]
\centering
\caption{Replication ratios: reproduced error divided by originally reported error, in the metric of the original paper. For GON the metric is test MSE, so the ratio of root-mean-square errors is the square root of the ratio shown (given in parentheses). For FNO the metric is relative $L^2$ error. Our runs use $N_{tr}=1000$, and each of our grids is paired with the nearest resolution of Li et al.'s Table~4. Lu et al.'s values are means over five seeds with and without output normalization (Tables~4, 5, and 10 of their paper), each paired with the nearest setting of Li et al.'s Tables~1, 3, and 4.}
\label{tab:replication_ratios}
\scriptsize
\setlength{\tabcolsep}{4pt}
\renewcommand{\arraystretch}{1.15}
\begin{tabularx}{\linewidth}{p{0.16\linewidth}p{0.17\linewidth}cccX}
\hline
\textbf{Study} & \textbf{Case} & \textbf{Reported} & \textbf{Reproduced} & \textbf{Ratio} & \textbf{Reproduction} \\
\hline
GON \citep{gu2025explainable} & 3D Poisson (heat sink) & $2.64\times10^{-5}$ & $1.06\times10^{-3}$ & 40 (6.3) & Original code, architecture, and metric; three seeds (Table~\ref{tab:gon_reproduction_shell}) \\
 & Plate & $5.22\times10^{-4}$ & $9.6\times10^{-4}$ & 1.8 (1.4) & \\
 & Pipe & $4.63\times10^{-4}$ & $1.29\times10^{-3}$ & 2.8 (1.7) & \\
 & Stokes ($x$ velocity) & $5.83\times10^{-4}$ & $1.36\times10^{-3}$ & 2.3 (1.5) & \\
 & 2D Poisson (thin slab) & $3.75\times10^{-4}$ & $1.01\times10^{-6}$ & 0.003 & Degenerate: the source term vanishes on the mesh, so the task reduces to interpolating boundary data \\
\hline
FNO \citep{li2021fno}, Darcy & $s=85$ / ours $64^2$ & 1.08\% & 0.78\% & 0.72 & Re-implementation with our own finite volume solver and coefficient fields, original architecture and optimizer; three seeds (Section~\ref{app:measured_nstar}) \\
 & $s=141$ / ours $128^2$ & 1.09\% & 0.67\% & 0.61 & \\
 & $s=211$ / ours $256^2$ & 1.09\% & 0.66\% & 0.61 & \\
 & $s=421$ / ours $512^2$ & 0.98\% & 0.58\% & 0.59 & \\
\hline
FNO \citep{li2021fno}, re-run by Lu et al.\ \citep{lu2022comprehensive} & Burgers, $s=256$ / Lu 128 & 1.49\% & 1.93\% / 2.23\% & 1.3 / 1.5 & Original dataset on a coarser grid; with / without output normalization; five seeds \\
 & Darcy (piecewise constant), $s=85$ / Lu $29^2$ & 1.08\% & 2.41\% / 4.83\% & 2.2 / 4.5 & Original dataset on a much coarser grid \\
 & Navier--Stokes, $\nu=10^{-3}$ & 0.86\% (3D) / 1.28\% (2D) & 1.81\% / 2.62\% & 1.4--2.1 / 2.0--3.0 & Same dataset and resolution ($64^2$), but Lu et al.\ predict the vorticity at $T=20$ from the first ten steps, whereas Li et al.\ predict the trajectory to $T=50$ \\
\hline
\end{tabularx}
\end{table}

\begin{table}[H]
\centering
\caption{GON out-of-distribution results on the 3D Poisson heat-sink case (test MSE, mean over seeds 0, 1, and 2, with the seed range in brackets). Ratios are relative to the in-distribution anchor of the same block.}
\label{tab:gon_ood}
\scriptsize
\setlength{\tabcolsep}{4pt}
\renewcommand{\arraystretch}{1.2}
\begin{tabularx}{\linewidth}{p{0.30\linewidth}p{0.32\linewidth}cX}
\hline
\textbf{Shift} & \textbf{Test MSE [seed range]} & \textbf{Ratio} & \textbf{Outcome} \\
\hline
\multicolumn{4}{l}{\textit{Original meshes (no new reference needed)}} \\
In-distribution anchor & $1.056\times10^{-3}$ [$9.16\times10^{-4}$, $1.33\times10^{-3}$] & 1.00 & -- \\
Noise on $f$, $g$, or both (1\%, 5\%) & $1.0561$--$1.0569\times10^{-3}$ & $\le 1.001$ & Robust \\
Boundary-tail extrapolation & $1.025\times10^{-3}$ [$5.55\times10^{-4}$, $1.27\times10^{-3}$] & 0.97 & Robust \\
\hline
\multicolumn{4}{l}{\textit{Regenerated meshes with new FEM references (outer Dirichlet only)}} \\
Training resolution (91 interior nodes) & $1.656\times10^{-3}$ [$1.51\times10^{-3}$, $1.94\times10^{-3}$] & 1.00 & -- \\
Finer resolution (557 interior nodes) & $1.717\times10^{-3}$ [$1.64\times10^{-3}$, $1.88\times10^{-3}$] & 1.04 & Robust \\
Coarser resolution (2 interior nodes) & $5.59\times10^{-4}$ & -- & Degenerate \\
Boundary-shift anchor (mean 0.47) & $1.333\times10^{-3}$ [$1.26\times10^{-3}$, $1.47\times10^{-3}$] & 1.00 & -- \\
Boundary mean shifted to 0.15 & $1.359\times10^{-3}$ [$1.27\times10^{-3}$, $1.52\times10^{-3}$] & 1.02 & Robust \\
Boundary mean shifted to 1.50 & $1.386\times10^{-3}$ [$1.34\times10^{-3}$, $1.48\times10^{-3}$] & 1.04 & Robust \\
Boundary amplitude $\times4$ & $2.127\times10^{-2}$ [$2.02\times10^{-2}$, $2.32\times10^{-2}$] & 15.95 & Fails \\
\hline
\end{tabularx}
\end{table}

\paragraph{Amortized cost.} Both solvers were timed on the same machine, an NVIDIA RTX~3090 GPU with an Intel i7-12700K CPU (Table~\ref{tab:gon_cost}). The finite element time is the median over 10 test cases of the per-case median of 20 repeats after one warm-up solve, which gives 4.348~ms; this differs slightly from the 4.38~ms mean reported above because the two runs aggregate differently. GON inference was timed the same way with CUDA synchronization around each timed region: the operator forward pass takes 2.816~ms, and the end-to-end query takes 28.06~ms because interpolating the inputs onto the model's quadrature points adds 25.24~ms. Training takes 24{,}698~s (6.86~h) for 2000 epochs on 80 instances, taken as the mean over the three seeds to represent the cost of one deployable model. Offline data generation is approximated by 80 finite element solves (0.35~s); the original COMSOL generation cost is not documented and was almost certainly higher, which would strengthen the conclusion. Even counting only the forward pass, GON recovers its training cost after about 16 million queries, and with the end-to-end query time it never does, because each query is slower than a finite element solve. This outcome is regime-dependent: with only 47 free degrees of freedom, a classical solve is too cheap for amortization to pay off, and the result does not rule out a crossover on problems with costly classical solves.

\begin{table}[H]
\centering
\caption{GON amortized cost against a direct finite element solve on the 3D Poisson heat-sink case (same machine). $N^\ast$ is the query volume at which total GON cost equals total FEM cost, $N^\ast=C_{\mathrm{offline}}/(t_{\mathrm{FEM}}-t_{\mathrm{GON}})$.}
\label{tab:gon_cost}
\scriptsize
\setlength{\tabcolsep}{4pt}
\renewcommand{\arraystretch}{1.2}
\begin{tabularx}{\linewidth}{Xccccc}
\hline
\textbf{Method} & \textbf{Offline cost} & \textbf{Online per query} & \textbf{Total, $N=100$} & \textbf{Total, $N=1000$} & \textbf{$N^\ast$} \\
\hline
Direct FEM (CPU) & 0 & 4.348~ms & 0.435~s & 4.348~s & -- \\
GON, operator forward only & 24{,}698~s & 2.816~ms & 24{,}698~s & 24{,}701~s & $1.61\times10^{7}$ \\
GON, end-to-end & 24{,}698~s & 28.06~ms & 24{,}701~s & 24{,}726~s & never \\
\hline
GON / FEM total & -- & -- & $\approx 56{,}800\times$ & $\approx 5{,}680\times$ & -- \\
\hline
\end{tabularx}
\end{table}

\begin{figure}[H]
\centering
\begin{minipage}{0.49\linewidth}
\centering
\includegraphics[width=\linewidth]{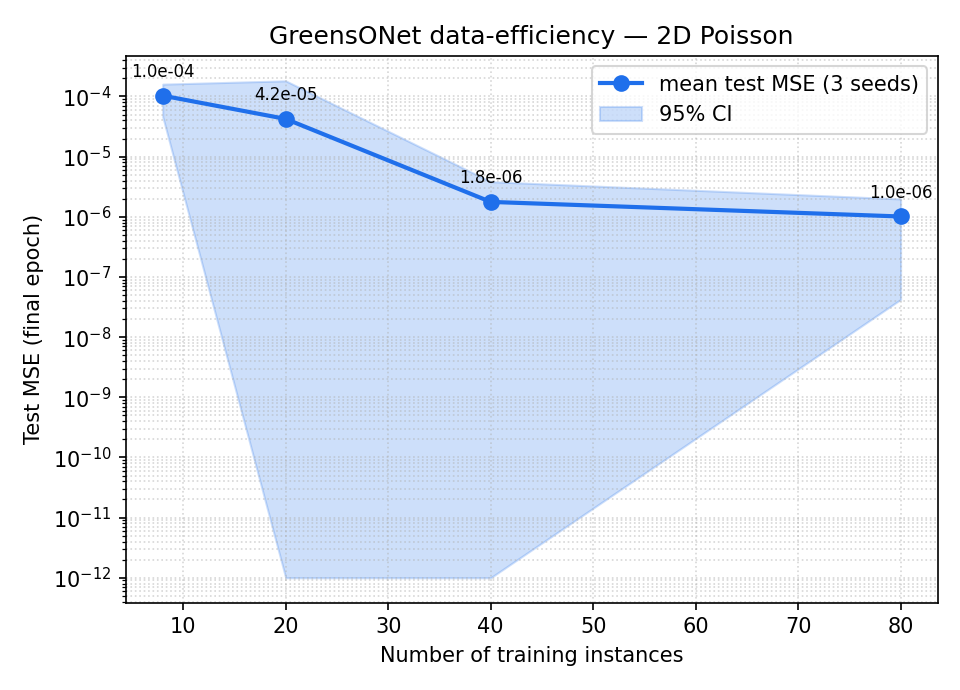}
\end{minipage}\hfill
\begin{minipage}{0.49\linewidth}
\centering
\includegraphics[width=\linewidth]{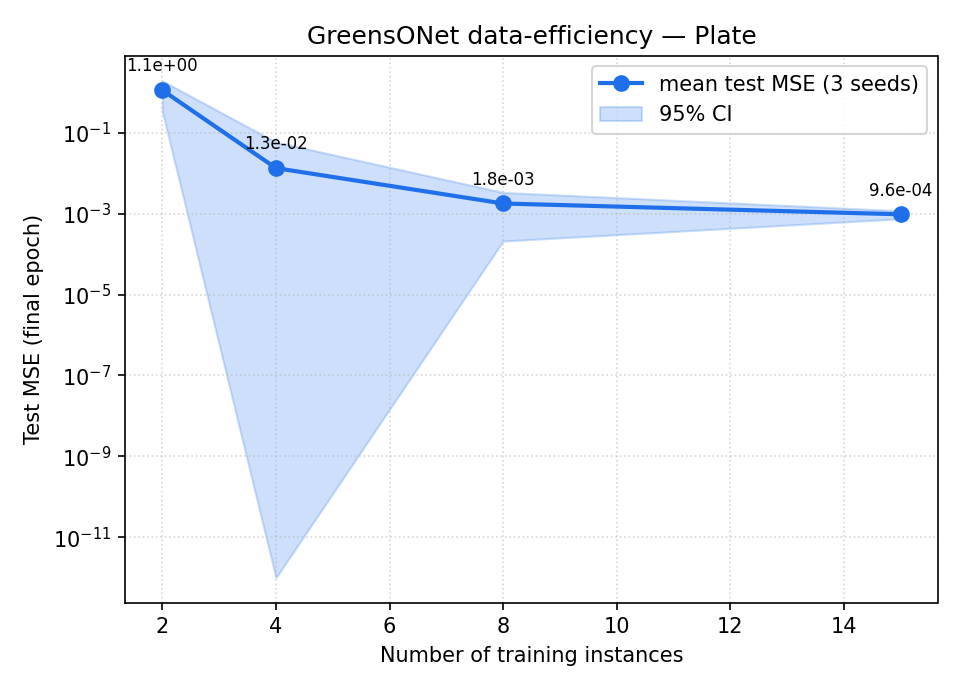}
\end{minipage}
\caption{GON learning curves for 2D Poisson (left) and Plate (right): final-epoch test MSE against the number of training instances, mean over seeds 0, 1, and 2 with the Student's $t$ 95\% interval shaded. Where the half-width exceeds the mean, the lower edge of the band is clipped at $10^{-12}$ for display on the logarithmic axis; the values in Table~\ref{tab:gon_dataeff} are unclipped.}
\label{fig:gon_dataeff}
\end{figure}

\paragraph{Data efficiency.} The training corpus was subsampled per seed while the test split was held fixed, for two representative cases: 2D Poisson (full corpus 80 instances) and Plate (full corpus 15 instances). Pipe and Stokes were not swept because they would roughly double the compute without changing the grade, and the neural baselines were not swept because their code is unavailable. Table~\ref{tab:gon_dataeff} and Figure~\ref{fig:gon_dataeff} show a strong, monotone dependence on corpus size: the MSE falls about 100 times from the smallest to the full corpus on 2D Poisson and about 1000 times on Plate, where the model fails outright with two training instances (MSE of order one). Seed variance is large at small corpus sizes and narrows as data grow. Because the 2D Poisson case has a vanishing source term, its learning curve measures learning from boundary data alone. Training time scales with corpus size; for example, a Plate seed took about 2.5~h with 2 instances and about 6.2~h with 8.

\begin{table}[H]
\centering
\caption{GON learning curves (final-epoch test MSE, mean over seeds 0, 1, and 2 with Student's $t$ 95\% half-width).}
\label{tab:gon_dataeff}
\scriptsize
\setlength{\tabcolsep}{5pt}
\renewcommand{\arraystretch}{1.2}
\begin{tabular}{lcccc}
\hline
\textbf{Case} & \textbf{Fraction} & \textbf{Training instances} & \textbf{Mean test MSE} & \textbf{95\% half-width} \\
\hline
2D Poisson & 10\% & 8 & $1.03\times10^{-4}$ & $5.7\times10^{-5}$ \\
2D Poisson & 25\% & 20 & $4.24\times10^{-5}$ & $1.4\times10^{-4}$ \\
2D Poisson & 50\% & 40 & $1.77\times10^{-6}$ & $2.1\times10^{-6}$ \\
2D Poisson & 100\% & 80 & $1.01\times10^{-6}$ & $9.7\times10^{-7}$ \\
\hline
Plate & 13\% & 2 & $1.12$ & $7.9\times10^{-1}$ \\
Plate & 27\% & 4 & $1.33\times10^{-2}$ & $4.5\times10^{-2}$ \\
Plate & 53\% & 8 & $1.78\times10^{-3}$ & $1.6\times10^{-3}$ \\
Plate & 100\% & 15 & $9.63\times10^{-4}$ & $2.2\times10^{-4}$ \\
\hline
\end{tabular}
\end{table}

\begin{figure}[H]
\centering
\includegraphics[width=\linewidth]{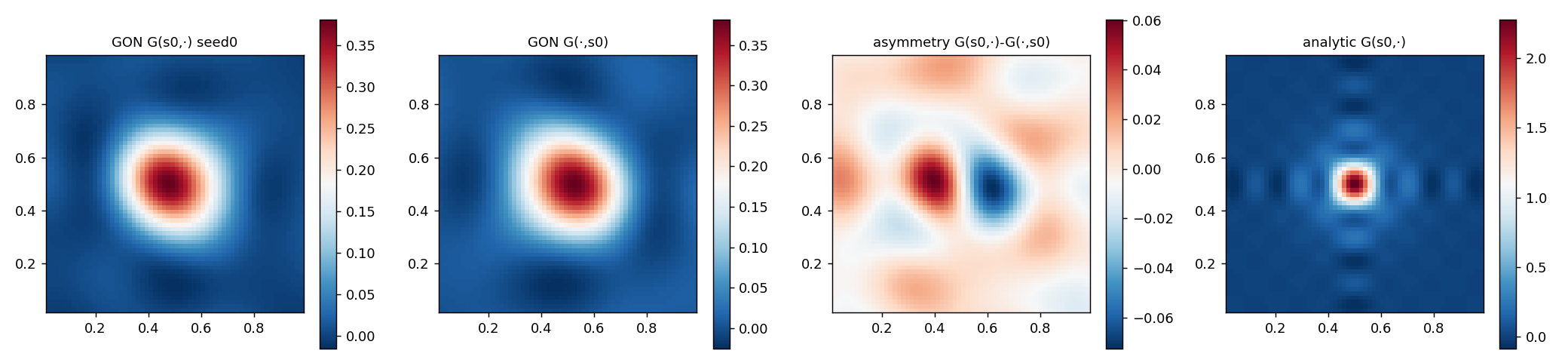}\\[2pt]
\includegraphics[width=\linewidth]{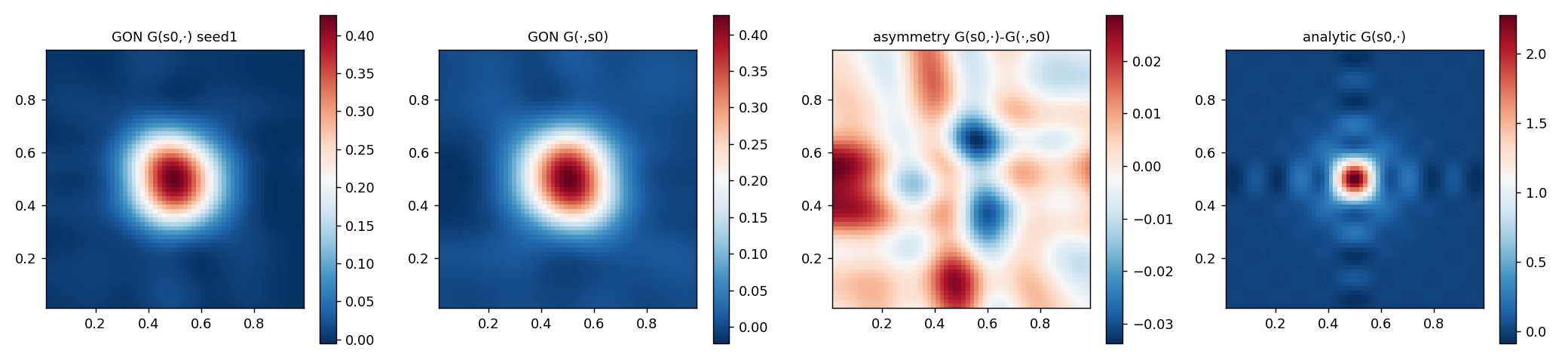}\\[2pt]
\includegraphics[width=\linewidth]{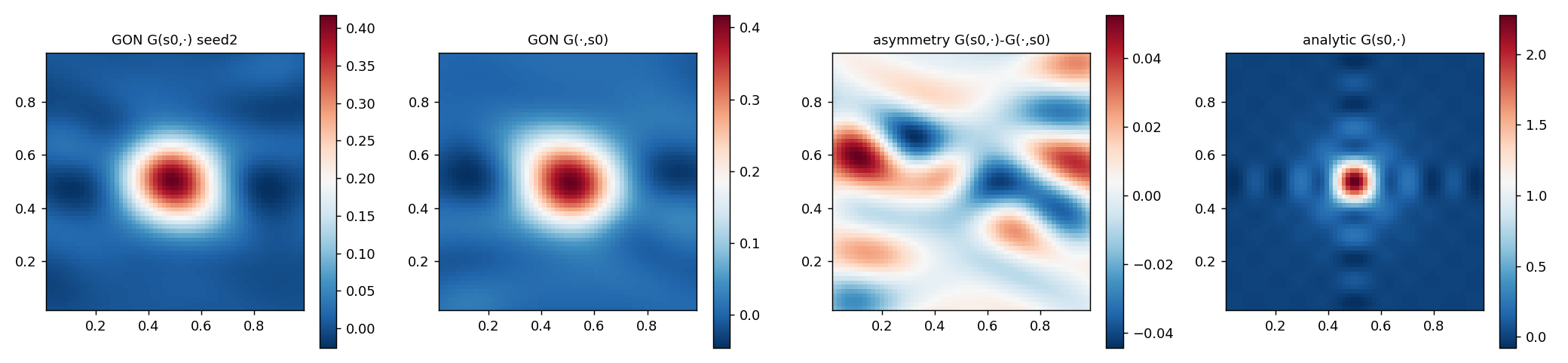}
\caption{Learned GON kernel on the controlled homogeneous Dirichlet case for seeds 0, 1, and 2 (top to bottom), shown on the mid-plane slice $z=0.15$ over $x,y\in[0.02,0.98]$ for the central source $s_0=(0.5,0.5,0.15)$. From left to right: the learned kernel $G(s_0,\cdot)$, its transpose $G(\cdot,s_0)$, the asymmetry $G(s_0,\cdot)-G(\cdot,s_0)$, and the analytic box Green's function $G(s_0,\cdot)$. The learned kernels are shown in GON's native units without rescaling, so colour scales differ between panels and are not comparable with the analytic panel.}
\label{fig:gon_kernel}
\end{figure}

\paragraph{Green's function structure audit.} Because the shipped 2D Poisson case has a vanishing source, a controlled case was built for the interpretability audit: $-\Delta u=f$ on the box $[0,1]\times[0,1]\times[0,0.3]$ with homogeneous Dirichlet conditions ($g\equiv0$), with data generated by P1 finite elements. In this setting the whole prediction comes from the interior integral, so an accurate model must encode the Green's function in its interior kernel. The extracted kernel was compared with the continuous box Green's series after an optimal signed scaling (for which about 0.20 is the floor set by the coarse vertical discretization) and with the exact discrete Green's matrix $K^{-1}$ of the finite element system; the truncation order of the series was not recorded. Table~\ref{tab:gon_kernel} and Figure~\ref{fig:gon_kernel} show that the model is highly accurate (mean MSE $1.12\times10^{-7}$) and reconstructs solutions by superposition to within about 5\%, yet its kernel is not a valid Green's function: the symmetry error is about 0.18 where it should be zero, and the kernel differs from the analytic and discrete Green's functions by about 47\% and 30\%. The kernel also does not enforce the homogeneous Dirichlet condition: taken over all interior sources and all boundary vertices on the six faces of the box, its largest magnitude with either argument on the boundary is comparable to its largest interior magnitude (ratios 0.999 and 1.016, where a Green's function gives 0). This is a worst-case measure, dominated by sources close to a face and by the top and bottom faces of the thin slab, which lie close to every interior point; for the deep-interior source plotted in Figure~\ref{fig:gon_kernel} the kernel is small near the side walls of the mid-plane slice. The slices show the same pattern for every seed: the learned kernel is a smooth, broad bump with a peak near 0.4, whereas the analytic Green's function is sharply concentrated at the source with a peak above 2, and the difference between $G(s_0,\cdot)$ and $G(\cdot,s_0)$ reaches about 10--15\% of the kernel peak, with a pattern that changes from seed to seed. Accurate prediction therefore coexists with a structurally invalid embedded kernel, even on a problem constructed so that a correct operator would have to encode the true Green's function.

\begin{table}[H]
\centering
\caption{Green's function structure audit on the controlled homogeneous Dirichlet case. The two boundary metrics are the largest kernel magnitude with the evaluation point (eval.) or the source point (src.) on the boundary, over all interior partners and all boundary vertices on the six faces, divided by the largest interior-to-interior kernel magnitude; a Dirichlet Green's function gives 0. The analytic and discrete errors are computed after an optimal signed scaling of the kernel; all errors other than the boundary metrics are relative $L^2$.}
\label{tab:gon_kernel}
\scriptsize
\setlength{\tabcolsep}{4pt}
\renewcommand{\arraystretch}{1.2}
\begin{tabular}{lccccccc}
\hline
\textbf{Seed} & \textbf{Test MSE} & \textbf{Symm.} & \textbf{Bndry (eval.)} & \textbf{Bndry (src.)} & \textbf{vs analytic} & \textbf{vs discrete} & \textbf{Recon.} \\
\hline
0 & $1.53\times10^{-7}$ & 0.183 & 0.999 & 1.024 & 0.468 & 0.299 & 0.057 \\
1 & $6.25\times10^{-8}$ & 0.160 & 0.997 & 1.018 & 0.473 & 0.296 & 0.041 \\
2 & $1.22\times10^{-7}$ & 0.205 & 1.000 & 1.006 & 0.481 & 0.312 & 0.047 \\
\hline
Mean & $1.12\times10^{-7}$ & 0.183 & 0.999 & 1.016 & 0.474 & 0.302 & 0.048 \\
Ideal & -- & 0 & 0 & 0 & $\approx0.20$ (floor) & 0 & 0 \\
\hline
\end{tabular}
\end{table}

\paragraph{Reproducibility record.} The replication used the authors' public GON repository \citep{gu2025explainable} through a fork whose first commit is dated 23 February 2026; the upstream commit hash was not recorded, so a mechanical comparison against upstream is not possible, and the deviations are instead documented in the code (training settings listed above, seed control, and logging). The software stack was PaddlePaddle 3.0.0 with CUDA runtime 12.0 (driver 12.2) and cuDNN 8.9 on a GPU of compute capability 8.6. The versions of Python, SciPy, scikit-fem, and gmsh on the training machine, and the peak GPU and finite element memory, were not recorded. These gaps are reported rather than reconstructed, and they are the main reason the reproducibility evidence stops short of a complete record under the checklist in Section~\ref{app:reproducibility_checklist}.

\paragraph{Evidence grades.} Table~\ref{tab:gon_grades} assigns each claim an evidence quality grade and, separately, a verdict. The executed protocol supports the accuracy claims for heterogeneous diffusion and Stokes at Moderate quality and does not support the heat conduction accuracy claim, while the speedup and Green's function interpretability claims are not supported on Strong evidence. The original paper's accuracy figures are not recovered, and its neural baselines could not be retrained, so even the Moderate grades rest partly on reported rather than reproduced comparisons. Overall, GON is a reproducible operator learner whose accuracy advantage over the reported baselines is confined to diffusion and Stokes, which offers no speedup in the regime tested, and whose distinctive interpretability claim fails a direct structural test; both negative verdicts rest on Strong evidence.

\begin{table}[H]
\centering
\caption{Evidence quality and verdict for GON \citep{gu2025explainable} by claim, and evidence quality of the supporting dimensions.}
\label{tab:gon_grades}
\scriptsize
\setlength{\tabcolsep}{4pt}
\renewcommand{\arraystretch}{1.2}
\begin{tabularx}{\linewidth}{p{0.25\linewidth}p{0.10\linewidth}p{0.12\linewidth}X}
\hline
\textbf{Claim or dimension} & \textbf{Quality} & \textbf{Verdict} & \textbf{Basis} \\
\hline
\multicolumn{4}{l}{\textit{Claims}} \\
Accuracy: steady heat conduction & Moderate & Not supported & 3D Poisson does not beat the reported FNO baseline; the 2D Poisson case has a vanishing source term. \\
Accuracy: heterogeneous diffusion & Moderate & Supported & Plate and Pipe lie below the reported baselines, but the baselines were not retrained. \\
Accuracy: Stokes (velocity $x$) & Moderate & Supported & Below the reported baseline at the original width; velocity components only, no pressure output. \\
Recovery of the reported accuracy & Strong & Not supported & Same metric, original architectures, three seeds; reproduced MSE is 2--40 times higher on every non-trivial case. \\
Speedup and amortized cost & Strong & Not supported & Same-machine timing with training counted; no crossover end-to-end, $N^\ast\approx1.6\times10^{7}$ for the forward pass alone. \\
Data efficiency & Moderate & Mixed & Strong monotone dependence on corpus size on two cases; failure at two Plate training instances. \\
Out-of-distribution robustness & Moderate & Mixed & Robust to noise, resolution, extrapolation, and mean shifts on one case; fails about 16 times on boundary amplitude. \\
Green's function interpretability & Strong & Not supported & Controlled case, three seeds: accurate model, but its kernel is asymmetric, fails the boundary-vanishing check, and differs from the true Green's function by 30--47\%. \\
\hline
\multicolumn{4}{l}{\textit{Evidence dimensions}} \\
Code reproducibility & Strong & -- & All five cases run to completion at original widths with three seeds. \\
Statistical reporting & Strong & -- & Three seeds with 95\% intervals for every accuracy, out-of-distribution, and learning-curve value. \\
Classical baseline fairness & Weak & -- & The original study has no classical baseline; the added FEM reference reproduces the data to $2.4\times10^{-5}$ relative $L^2$. \\
Neural baseline fairness & Not assessed & -- & Baselines reported by the original authors; code not released. \\
Data generation transparency & Moderate & -- & Formulation verified independently; original COMSOL generation cost undocumented. \\
\hline
\end{tabularx}
\end{table}

\section{Evidence extraction audit: Fourier Neural Operator (Li et al.\ 2021)}
\label{app:fno_audit}

This section is the second of three case studies illustrating the evidence rubric. In contrast to the GON study in Section~\ref{app:gon_replication}, which is a full computational replication, this case study is an \emph{evidence extraction audit}: the schema of Section~\ref{app:evidence_extraction_schema} is applied to a published paper by reading its reported evidence against each schema field, without re-executing the original experiments. The audit target is the Fourier Neural Operator paper of Li et al.\ \citep{li2021fno}, selected because it is the most reproduced operator learning paper in the PDE surrogate literature and is the origin of the resolution-transfer and speedup claims examined in Sections~\mainref{subsec:operator_learning} and~\mainref{subsec:cost_comparison}. Independent re-evaluations of the same architecture in PDEBench \citep{takamoto2022pdebench} and CFDBench \citep{luo2023cfdbench} provide external evidence used to cross-check the grades assigned here.

The purpose of an extraction audit is to demonstrate that the rubric can be applied efficiently to the existing literature, since a practitioner cannot re-run every published method but must still judge which reported claims support a solver decision. The audit separates the paper's distinct claims by type, because a single paper can present Strong evidence for one claim and Red Flag evidence for another.

\begin{table}[H]
\centering
\caption{Evidence extraction record for Li et al.\ \citep{li2021fno}, following the schema of Section~\ref{app:evidence_extraction_schema}.}
\label{tab:fno_extraction}
\scriptsize
\setlength{\tabcolsep}{4pt}
\renewcommand{\arraystretch}{1.18}
\begin{tabularx}{\linewidth}{p{0.22\linewidth}X}
\hline
Extraction field & Finding for Li et al.\ (2021) \\
\hline
Bibliographic record & Li, Kovachki, Azizzadenesheli, Liu, Bhattacharya, Stuart, Anandkumar. ``Fourier Neural Operator for Parametric Partial Differential Equations.'' ICLR 2021. The paper states that code and datasets are available but gives no repository URL in the text; full architecture specifications and hyperparameters are listed. Repository completeness for every reported experiment was not checked in this audit. \\
\hline
PDE type & 1D viscous Burgers (nonlinear convection--diffusion), $\partial_t u + \partial_x (u^2/2) = \nu \partial_{xx} u$ on $(0,1)$ with $\nu = 0.1$; 2D Darcy flow (steady linear elliptic, parametric coefficient), $-\nabla \cdot (a(x)\nabla u(x)) = f(x)$ on $(0,1)^2$ with fixed forcing $f \equiv 1$; and 2D incompressible Navier--Stokes in vorticity form on $(0,1)^2$ at viscosities $\nu = 10^{-3}, 10^{-4}, 10^{-5}$. \\
\hline
Task and decision regime & Many-query parametric surrogate over a fixed input distribution (Gaussian random field coefficients / initial conditions). \\
\hline
Geometry and boundary conditions & Regular periodic or unit-square grids; no irregular geometry. Burgers: periodic on $(0,1)$. Darcy: homogeneous Dirichlet, $u = 0$ on $\partial(0,1)^2$. Navier--Stokes: periodic on the unit torus. The non-periodic Darcy boundary is handled by the local linear term $W$ acting in physical space rather than by the Fourier layer. \\
\hline
Solver family & Supervised operator learning (spectral convolution in the Fourier domain). \\
\hline
Data requirement & Paired input--output solution corpus generated by a numerical solver: a split-step Fourier method for Burgers (heat part solved exactly in Fourier space, nonlinear part advanced by forward Euler), a second-order finite difference scheme on a $421 \times 421$ grid for Darcy, and a pseudo-spectral stream-function solver with Crank--Nicolson time stepping ($\Delta t = 10^{-4}$) for Navier--Stokes. Standard corpus of $N = 1{,}000$ training and $200$ test pairs; Navier--Stokes at $\nu = 10^{-4}$ is also run with $N = 10{,}000$. \\
\hline
Reference solver / ground truth & The data-generating solvers above serve as reference: Burgers on $2^{13} = 8{,}192$ points (downsampled to $256$--$8{,}192$); Darcy on $421 \times 421$ (downsampled to $85$--$421$); Navier--Stokes on $256 \times 256$ (downsampled to $64 \times 64$). Solver tolerances are not reported. \\
\hline
Baseline set & Compared primarily against other neural surrogates. Burgers (Table~3): NN, GCN, FCN, PCANN, GNO, LNO, MGNO. Darcy (Table~4): NN, FCN, PCANN, RBM, GNO, LNO, MGNO. Navier--Stokes (Table~1): ResNet, U-Net, TF-Net, FNO-2D versus FNO-3D. The classical POD-based reduced basis method (RBM) appears only as one entry in the Darcy table, with reported relative errors of $2.44$--$2.59\%$ across resolutions; no tuned adaptive-FEM or multigrid baseline is included for the elliptic case, and no classical solver is compared at matched accuracy. \\
\hline
Metrics & Relative $L^2$ error only. No $H^1$/gradient, flux, boundary residual, or physical conservation metrics are reported. \\
\hline
Cost accounting & Inference wall-clock compared against a single traditional solver for Navier--Stokes ($0.005$\,s versus $2.2$\,s per instance, about $440\times$); the abstract and introduction state qualitatively that FNO is ``up to three orders of magnitude faster'' than traditional PDE solvers. Offline corpus generation is not counted in this headline figure; it is disclosed separately in Section~5.5, where data generation plus training ($\approx 12$\,h) is amortized over a function-space MCMC Bayesian inverse problem requiring $30{,}000$ forward evaluations ($25{,}000$ samples plus $5{,}000$ burn-in). Training cost is reported per epoch in Table~1 (e.g.\ $38.99$\,s for FNO-3D and $127.80$\,s for FNO-2D, over $500$ epochs). \\
\hline
Out-of-distribution testing & Grid-resolution transfer is tested on the same input distribution, including zero-shot super-resolution from $64 \times 64 \times 20$ training to $256 \times 256 \times 80$ evaluation on Navier--Stokes. No coefficient, forcing, or geometry shift is tested; test instances are drawn from the same input measure $\mu$ as the training data. \\
\hline
Statistical reporting & All entries in Tables~1, 3, and~4 are single point values; no multi-seed averages, standard deviations, or confidence intervals are reported. \\
\hline
Reproducibility level & Code and datasets stated as available. Hardware: single NVIDIA V100 (16\,GB). Hyperparameters: Adam for $500$ epochs, initial learning rate $10^{-3}$ halved every $100$ epochs; four Fourier layers with ReLU and batch normalization; width $d_v = 64$ (1D) / $32$ (2D); modes $k_{\max,j} = 16$ (1D) / $12$ (2D). Random seeds not specified. \\
\hline
\end{tabularx}
\end{table}

Table~\ref{tab:fno_extraction} shows the characteristic pattern of a strong architecture paper read as decision evidence: the accuracy reporting is careful and the code is available, but the comparison layer is oriented toward other neural models, with a single reduced basis entry on Darcy and no matched-accuracy classical solver, and the headline speedup separates online inference from the offline cost that the paper discloses only later. Table~\ref{tab:fno_claim_grades} converts this record into per-claim grades using the rubric of Table~\ref{tab:claim_strength_operational}.

\begin{table}[H]
\centering
\caption{Per-claim evidence quality and verdicts for Li et al.\ \citep{li2021fno} under the rubric of Section~\ref{app:claim_strength_rubric}. Grades are derived from the extraction record in Table~\ref{tab:fno_extraction}.}
\label{tab:fno_claim_grades}
\scriptsize
\setlength{\tabcolsep}{4pt}
\renewcommand{\arraystretch}{1.18}
\begin{tabularx}{\linewidth}{p{0.22\linewidth}p{0.13\linewidth}p{0.12\linewidth}X}
\hline
Claim & Quality & Verdict & Reasoning \\
\hline
In-distribution accuracy on Darcy, Burgers, Navier--Stokes & Moderate & Supported & Low relative $L^2$ error is credibly demonstrated against several neural baselines, but the only reduced-order comparison is a single RBM entry on Darcy. That entry is the strongest baseline in the Darcy table, at 2.44--2.59\% against FNO's 0.98--1.09\%, so FNO's error is 2.3--2.6 times lower rather than ``nearly one order of magnitude lower'' as the text states. No tuned classical baseline is included, and no multi-seed confidence intervals are reported, so the claim cannot reach Strong for a solver-replacement decision. \\
\hline
Speedup versus traditional solvers & Weak $\to$ Red Flag & Mixed & Inference time is compared against a single numerical solver; the only measured benchmark gives about $440\times$, below the ``up to three orders of magnitude'' stated in the abstract, and the headline speedup excludes the offline cost of data generation and training ($\approx 12$\,h). The paper does amortize this cost over the $30{,}000$ forward evaluations of an MCMC inverse problem in Section~5.5, but the headline figure is the one that is usually cited, which conflates offline and online phases as described in Section~\mainref{subsec:cost_comparison}. \\
\hline
Zero-shot super-resolution / resolution transfer & Moderate (math.) / Weak (deployment) & Supported (math.) / Not established (deployment) & Discretization-transfer across grids on a fixed input distribution supports the mathematical sense of Section~\mainref{subsec:operator_learning}, but no input-distribution shift is tested, so the empirical deployment claim is in-distribution interpolation across grids. \\
\hline
Generalization across the parametric family & Weak & Not established & Training and test coefficients are drawn from the same Gaussian random field; no coefficient, forcing, or geometry shift is evaluated, so generalization evidence is confined to interpolation within the training distribution. \\
\hline
\end{tabularx}
\end{table}

The audit verdict is that Li et al.\ \citep{li2021fno} provides Moderate evidence for in-distribution accuracy of Fourier neural operators on smooth parametric PDEs, but Weak-to-Red-Flag evidence for the operational speedup and out-of-distribution generalization claims that are most often cited when the method is proposed as a replacement for a classical solver. This split verdict is consistent with the independent benchmark re-evaluations in PDEBench and CFDBench, which report that FNO accuracy is strongly regime dependent and that zero-shot transfer degrades once the evaluation departs from the training regime \citep{takamoto2022pdebench,luo2023cfdbench}. The decision consequence, carried into Table~\mainref{tab:decision_matrix}, is that FNO on many-query Darcy flow warrants a Moderate-confidence recommendation contingent on the practitioner counting corpus generation cost and either comparing against a reduced basis alternative or showing by a decay test, as in Section~\ref{app:measured_nstar}, that none applies; the single RBM entry reported in the paper is not sufficient. A single relative-error table might suggest a High-confidence reading, but the extraction record does not support it.

\section{Evidence extraction audit: Physics-Informed Neural Networks (Raissi et al.\ 2019)}
\label{app:pinn_audit}

This section is the third case study illustrating the evidence rubric, and like the Fourier neural operator audit of Section~\ref{app:fno_audit} it is an evidence extraction audit rather than a computational replication. The target is the physics-informed neural network paper of Raissi et al.\ \citep{raissi2019pinn}, selected because it is the foundational reference for residual-trained function learning and the origin of the forward and inverse PINN demonstrations that Sections~\mainref{subsec:function_learning} and~\mainref{sec:decision_matrix} evaluate. Auditing the two foundational papers of the operator learning and function learning families in parallel gives the case-study series a clean structure: one full replication (GON, an interpretability-guided operator) and two audits of the papers that anchor the two dominant ML-only families. The independent failure-mode analyses of Krishnapriyan et al.\ \citep{krishnapriyan2021characterizing} and De Ryck et al.\ \citep{deryck2024numerical}, and the controlled PINN-versus-FEM comparison of Grossmann et al.\ \citep{grossmann2023canpinn} provide the external evidence used to cross-check the grades.

Two points of interpretive care are needed for this audit. First, Raissi et al.\ is explicitly a proof-of-concept paper introducing a framework, and it makes comparatively measured claims; the audit therefore grades the evidence as reported, and separately notes where downstream citation practice has treated the paper as stronger deployment evidence than it claims to be. Second, the paper's inverse (data-driven discovery) results are a distinct claim type from its forward (data-driven solution) results and are graded separately.

\begin{table}[H]
\centering
\caption{Evidence extraction record for Raissi et al.\ \citep{raissi2019pinn}, following the schema of Section~\ref{app:evidence_extraction_schema}.}
\label{tab:pinn_extraction}
\scriptsize
\setlength{\tabcolsep}{4pt}
\renewcommand{\arraystretch}{1.18}
\begin{tabularx}{\linewidth}{p{0.22\linewidth}X}
\hline
Extraction field & Finding for Raissi et al.\ (2019) \\
\hline
Bibliographic record & Raissi, Perdikaris, Karniadakis. ``Physics-informed neural networks: A deep learning framework for solving forward and inverse problems involving nonlinear partial differential equations.'' Journal of Computational Physics, 2019. Public GitHub repository (\url{https://github.com/maziarraissi/PINNs}) with TensorFlow code and datasets for all continuous and discrete-time benchmarks. \\
\hline
PDE type & Continuous-time models: 1D Burgers (forward and inverse), 1D Schr\"odinger (complex-valued, forward), 2D Navier--Stokes (inverse). Discrete-time models (implicit Runge--Kutta): 1D Burgers (forward and inverse), 1D Allen--Cahn (forward), 1D Korteweg--de Vries (dispersive, inverse). \\
\hline
Task and decision regime & Two regimes: (i) forward per-instance solve (data-driven solution of a specified PDE); (ii) data-informed inverse identification of PDE parameters from observations. \\
\hline
Geometry and boundary conditions & Simple 1D spatial domains and 2D cylinder wake; boundary and initial conditions enforced exclusively via soft MSE/SSE loss penalty terms (no hard constraints). Dirichlet or periodic BCs in 1D; interior velocity observations without explicit BC loss for 2D Navier--Stokes. \\
\hline
Solver family & Function learning (strong-form residual training with automatic differentiation). \\
\hline
Data requirement & Forward: PDE residual plus initial/boundary points ($N_0=50$, $N_b=50$, $N_f=20{,}000$ for Schr\"odinger; $N_n=200$--$250$ for discrete Allen--Cahn/Burgers). Inverse: sparse observations ($N=5{,}000$ points, 1\% of the total available data, for 2D Navier--Stokes; $N=2{,}000$ for 1D Burgers; 2 snapshots of $\approx 200$ points for discrete KdV). \\
\hline
Reference solver / ground truth & High-fidelity reference solutions: analytical exact solution for 1D Burgers; Chebfun (Fourier spectral, 256--512 modes, explicit RK4) for Schr\"odinger, Allen--Cahn, and KdV; spectral/hp-element solver NekTar (412 triangular elements, 10th-order Jacobi polynomials) for 2D Navier--Stokes. Solver tolerances not reported. \\
\hline
Baseline set & No benchmark comparison against tuned classical FEM/FDM/spectral solvers on matched accuracy or wall-clock cost; no comparison against adjoint-based PDE-constrained optimization or variational data assimilation (4D-Var) baselines for inverse cases. \\
\hline
Metrics & Relative $L^2$ error against reference solution for forward solves; percentage error in estimated parameters $\lambda$ for inverse discovery. No boundary error, gradient/$H^1$ error, or physical conservation law diagnostics separately reported. \\
\hline
Cost accounting & Training time reported only in an isolated instance ($\approx 60$~seconds on an NVIDIA Titan X GPU for 1D Burgers); not reported for the other benchmarks. No wall-clock comparison against classical numerical solvers at matched accuracy. \\
\hline
Out-of-distribution testing & Not applicable in the operator sense; the method solves a specified instance. No systematic study of failure across increasing PDE difficulty (this gap is precisely what Krishnapriyan et al.\ \citep{krishnapriyan2021characterizing} later fills). \\
\hline
Statistical reporting & Single-run point estimates reported across all benchmarks; no multi-seed averages, standard deviations, confidence intervals, or optimization-variability studies. \\
\hline
Reproducibility level & Public GitHub code and released datasets; TensorFlow implementations provided. Network depth, layer width, activation functions, optimizers (L-BFGS / Adam), and RK stage counts ($q$) reported per problem; random seeds omitted; hardware specs reported for 1D Burgers only. \\
\hline
\end{tabularx}
\end{table}

Table~\ref{tab:pinn_extraction} records a foundational framework paper whose evidence establishes feasibility on selected nonlinear PDEs but does not include the classical baselines, cost accounting, or optimization-variability reporting needed to support a solver-replacement decision. Table~\ref{tab:pinn_claim_grades} assigns per-claim grades.

\begin{table}[H]
\centering
\caption{Per-claim evidence quality and verdicts for Raissi et al.\ \citep{raissi2019pinn} under the rubric of Section~\ref{app:claim_strength_rubric}. Grades are derived from the extraction record in Table~\ref{tab:pinn_extraction}.}
\label{tab:pinn_claim_grades}
\scriptsize
\setlength{\tabcolsep}{4pt}
\renewcommand{\arraystretch}{1.18}
\begin{tabularx}{\linewidth}{p{0.22\linewidth}p{0.13\linewidth}p{0.12\linewidth}X}
\hline
Claim & Quality & Verdict & Reasoning \\
\hline
Forward solution feasibility on selected nonlinear PDEs & Moderate & Supported & Low relative $L^2$ error against reference solutions is credibly demonstrated for the shown instances, establishing feasibility; but no tuned classical baseline, no cost comparison, and no multi-seed reporting prevent a Strong grade for solver replacement. \\
\hline
Inverse parameter identification from sparse data & Moderate & Supported & The Navier--Stokes and KdV discovery examples are striking demonstrations of combining residual constraints with sparse observations, but no comparison against adjoint-based PDE-constrained optimization or variational assimilation is provided, as flagged for the inverse regime in Section~\mainref{sec:taxonomy}. \\
\hline
Efficiency / practical solver replacement & Weak & Not established & No wall-clock or accuracy-matched comparison against a classical solver is reported; the controlled study of Grossmann et al.\ \citep{grossmann2023canpinn} later finds PINNs do not beat FEM on combined time and accuracy. \\
\hline
Robustness across PDE difficulty & Weak & Not established & Results are shown on configurations that train successfully; the systematic failure modes on harder related PDEs are documented later by Krishnapriyan et al.\ \citep{krishnapriyan2021characterizing} and explained by De Ryck et al.\ \citep{deryck2024numerical}. \\
\hline
\end{tabularx}
\end{table}

The audit verdict is that Raissi et al.\ \citep{raissi2019pinn} provides Moderate evidence for the feasibility of residual-trained forward and inverse PDE solving, which is what a foundational framework paper should establish, but Weak evidence for the efficiency and robustness claims that later citation practice often attributes to it. This split is consistent with the review's treatment of function learning in Section~\mainref{subsec:function_learning}: PINNs are most defensible in data-scarce and inverse regimes where a residual or variational signal is available, and least defensible as a drop-in replacement for a tuned classical solver without optimization-reliability and cost evidence. Carried into Table~\mainref{tab:decision_matrix}, the audit supports a Moderate-to-Low confidence recommendation for PINN-style function learning contingent on reporting reference error, boundary error, multiple seeds, and a classical baseline.

Read together, the three case studies show that the rubric produces differentiated, claim-specific grades across solver families rather than a uniform verdict: the GON replication supports two of three accuracy claims at Moderate quality and finds the speedup and interpretability claims of an interpretability-guided operator not supported on Strong evidence, the Fourier neural operator audit yields Moderate accuracy evidence but Weak-to-Red-Flag cost and generalization evidence, and the physics-informed network audit yields Moderate feasibility evidence but Weak efficiency and robustness evidence. The differentiated verdicts, produced by one instrument applied to three foundational methods spanning operator learning, function learning, and interpretability-guided operators, illustrate how the evidence methodology proposed in this review separates claims.

\section{Justification of the landscape table marks}
\label{app:landscape_justification}

Table~\ref{tab:landscape_justification} records the basis for each mark in the review landscape table of the main article. Column definitions are those of its caption: ``cross family'' means classical, ML only, and hybrid solvers compared within one framework (\cmark), only two families or one slice (\textsl{partial}), or one family only (\xmark); ``evidence graded'' means claims weighed against fair baselines and cost accounting (\cmark), for a subset only (\textsl{partial}), or not at all (\xmark).

\begin{table}[H]
\centering
\caption{Basis for the cross family and evidence graded marks of the landscape table.}
\label{tab:landscape_justification}
\scriptsize
\setlength{\tabcolsep}{4pt}
\renewcommand{\arraystretch}{1.15}
\begin{tabularx}{\linewidth}{p{0.22\linewidth}p{0.10\linewidth}X}
\hline
\textbf{Review} & \textbf{Marks} & \textbf{Basis} \\
\hline
Thom{\'e}e (2001) \citep{thomee2001fdtofe} & \xmark{} / \xmark & Scope stated in title and abstract: historical synthesis of finite difference and finite element methods; no learned solvers, no evidence grading. \\
Xu \& Zikatanov (2017) \citep{xu2017amg} & \xmark{} / \xmark & Algebraic multigrid theory only; classical scope, no grading of empirical claims. \\
Cuomo et al.\ (2022) \citep{cuomo2022scientific} & \xmark{} / \xmark & Review of physics-informed neural networks only; no classical or hybrid comparison under shared criteria, no claim grading. \\
Hafiz et al.\ (2024) \citep{hafiz2024largedata} & \xmark{} / \xmark & Data-driven PDE solvers only; no grading against classical baselines. \\
Farea et al.\ (2024) \citep{farea2024understanding} & \xmark{} / \xmark & PINN techniques, applications, and challenges only. \\
Luo et al.\ (2025) \citep{luo2025pinnreview} & \xmark{} / \xmark & PINN review only. \\
Liu et al.\ (2025) \citep{liu2025neuraloperators} & \xmark{} / \xmark & Neural operator architectures only. \\
Huang et al.\ (2025) \citep{huang2025pdemeetdnn} & \textsl{partial} / \xmark & Sections II-B, II-C, and III-C and Table III cover operator learning, including PDE-constrained operator learning that combines mechanism-based models with networks; the review is qualitative and does not weigh claims against fair baselines or account for cost. \\
Heinlein et al.\ (2021) \citep{heinlein2021combining} & \textsl{partial} / \xmark & Abstract: ``our focus is on a very specific area within SciML given by the combination of domain decomposition methods (DDMs) with machine learning techniques''; Sections 2.7 and 4 survey algorithms without grading empirical evidence. \\
Klawonn et al.\ (2024) \citep{klawonn2024mldd} & \textsl{partial} / \xmark & Abstract: the survey examines ``the combination of machine learning (ML) and domain decomposition methods (DDMs)''; Sections 1 and 6 give a methodological taxonomy without claim-by-claim grading or cost accounting. \\
McGreivy \& Hakim (2024) \citep{mcgreivy2024weakbaselines} & \textsl{partial} / \cmark & Compares ML surrogates with classical numerical baselines but not hybrid families as a class; abstract: ``79\% (60/76) make a comparison with a weak baseline''; recommendation 3 and Eq.~(1) account for data generation and training cost. \\
Grossmann et al.\ (2024) \citep{grossmann2023canpinn} & \textsl{partial} / \cmark & Head-to-head comparison of PINNs and the finite element method only (Sections 1 and 3); Section 7: ``Considering the solution time and accuracy, PINNs are not able to beat FEM in our study.'' \\
Nooraiepour et al.\ (2026) \citep{nooraiepour2026pdeage} & \cmark{} / \textsl{partial} & Abstract: classical and machine learning approaches examined ``through a unified evaluative framework''; hybrid models in Section 5.4. Section 5.6.2 discusses weak baselines and Section 5.6.3 (Eq.~50) gives a break-even formula, but individual claims are not graded and published results are not reproduced. Preprint. \\
\hline
\end{tabularx}
\end{table}

\section{Rank concordance in published benchmark suites}
\label{app:leaderboard}

Method rankings were computed from the main result tables of four suites, transcribed into machine-readable form: PDEBench \citep{takamoto2022pdebench} (RMSE), PINNacle \citep{hao2024pinnacle} (L2RE, Table~3 of the NeurIPS 2024 version, checked cell by cell; LBFGS, which did not converge on 7 of 22 tasks, is excluded, and including it ranked last where it failed gives $W=0.11$), APEBench \citep{koehler2024apebench} (nRMSE, Table~13), and CFDBench \citep{luo2023cfdbench} (NMSE, Tables~12--15). Within each task, methods were ranked by error (ties averaged). PINNacle marks some cells ``NaN'' (did not converge), which were ranked last, and others ``--'' (method not suitable); tasks with ``--'' cells or with unreported values were removed so that every remaining task ranks every method. Kendall's $W$ measures agreement of the rankings across tasks ($W=1$: identical rankings; $W=0$: no agreement), and the Friedman test asks whether any method ranks consistently better than others. The analysis script and transcribed tables are released with the article.

\begin{table}[H]
\centering
\caption{Rank concordance of methods across tasks in four benchmark suites.}
\label{tab:leaderboard_concordance}
\tiny
\setlength{\tabcolsep}{3.5pt}
\renewcommand{\arraystretch}{1.15}
\begin{tabular}{lcccccl}
\hline
\textbf{Suite} & \textbf{Tasks used / total} & \textbf{Methods} & \textbf{Kendall's $W$} & \textbf{Friedman $p$} & \textbf{Distinct winners} & \textbf{Best mean rank (tasks won)} \\
\hline
PDEBench & 8 / 11 & 3 & 1.00 & $3.4\times10^{-4}$ & 1 & FNO (100\%) \\
PINNacle & 18 / 22 & 11 & 0.10 & 0.059 & 8 & LRA (22\%) \\
APEBench & 15 / 15 & 5 & 0.15 & 0.063 & 5 & ResNet (40\%) \\
CFDBench & 28 / 28 & 7 & 0.32 & $5.9\times10^{-10}$ & 4 & FNO (43\%) \\
CFDBench, without identity baseline & 28 / 28 & 6 & 0.29 & $1.0\times10^{-7}$ & 4 & FNO (43\%) \\
\hline
\multicolumn{7}{l}{FNO beats U-Net on 11/11 PDEBench tasks, 12/28 CFDBench tasks, and 3/15 APEBench tasks.} \\
\end{tabular}
\end{table}

PDEBench's high concordance reflects a main table with only three methods of very different kinds (FNO, U-Net, and PINN). Across the eleven PINN variants of PINNacle, rankings agree only weakly from task to task ($W=0.10$, Friedman $p=0.059$). Ranks discard effect sizes, so we also decomposed $\log_{10}$ L2RE over the 18 complete tasks into task, method, and task-by-method components. With the one non-converged cell (LRA on NS 2d-C) set to L2RE $=1$, the shares of variance are 0.51, 0.06, and 0.43, and the method effect is significant ($F=2.41$, $p=0.011$); dropping that task gives 0.52, 0.08, and 0.40 ($F=3.13$, $p=0.001$). The variants therefore differ on average, but the task-by-method interaction is five to seven times the method effect. Relative to vanilla PINN, the geometric-mean error ratios range from 0.54 (NTK) to 1.54 (vPINN), and only LAAF's interval excludes 1 (0.78, 95\% interval 0.64--0.96). No variant is reliably best, because the best variant depends on the task. Because each pair of suites shares at most two methods (FNO and U-Net), a cross-suite rank correlation is not meaningful; the pairwise comparison in the last row shows instead that the same two architectures reverse their order between suites.

\section{Evidence trace for the decision matrix}
\label{app:decision_trace}

Tables~\ref{tab:decision_trace_a} and~\ref{tab:decision_trace_b} trace each confidence level in the decision matrix of the main article to the evidence behind it, so that a reader can check, and dispute, every cell. Evidence is tagged by type: [T] established theory, [X] an external study cited in the main article, [G] a claim graded in this review's case studies (evidence quality $\cdot$ verdict), and [M] a quantity measured or derived in this review. The last column names the missing evidence that keeps the confidence from being higher. Confidence follows the rules of the main article: High requires Strong evidence across at least two PDE families or mature theory plus benchmark evidence, and Moderate or Low reflects the gaps listed.

\begin{table}[H]
\centering
\caption{Evidence trace for the decision matrix, part 1.}
\label{tab:decision_trace_a}
\scriptsize
\setlength{\tabcolsep}{3.5pt}
\renewcommand{\arraystretch}{1.15}
\begin{tabularx}{\linewidth}{p{0.13\linewidth}p{0.15\linewidth}Xp{0.20\linewidth}}
\hline
\textbf{Regime} & \textbf{Confidence in Table~\mainref{tab:decision_matrix}} & \textbf{Supporting evidence} & \textbf{What caps the confidence} \\
\hline
One off high fidelity solve & High for classical; Low for most ML replacement claims & [T] a posteriori estimation and adaptive FEM convergence and optimality \citep{ainsworth2000aposteriori,bonito2024adaptive}; near-linear multigrid complexity \citep{trottenberg2001multigrid}. [X] Controlled study: FEM faster at the same or higher accuracy than PINNs \citep{grossmann2023canpinn}; 79\% of audited superiority claims used weak baselines \citep{mcgreivy2024weakbaselines}. [G] PINN efficiency: Weak $\cdot$ Not established; GON speedup: Strong $\cdot$ Not supported (\S\mainref{subsec:case_audits}, \S\mainref{subsec:case_gon}). & No equal-accuracy, cost-matched study in which a learned solver beats a tuned classical workflow on a one off solve. \\
\hline
Many query parametric family & High for ROMs in favorable structure; Moderate for neural operators with full evidence & [T] Certified a posteriori bounds and analyzable offline cost for reduced basis methods \citep{hesthaven2016certified,rozza2008reduced}. [M] Measured FNO break-even against multigrid on Darcy: at least $3\times10^4$ queries for errors below 1\%, rising thirtyfold from $64^2$ to $512^2$, and never at $512^2$ with 500 or fewer samples (Table~\mainref{tab:measured_nstar}); on the same problem a POD basis needs 125--335 modes to match FNO and the coefficient does not compress, so no reduced basis with a cheap certified online stage exists (Table~\ref{tab:pod_check}). [G] FNO in-distribution accuracy: Moderate $\cdot$ Supported; FNO speedup: Weak$\to$Red Flag $\cdot$ Mixed, about $1.5\times$ end to end in its own application. & FNO generalization: Weak $\cdot$ Not established; the FNO paper reports one reduced basis entry and no tuned classical comparison. \\
\hline
High dimensional or mesh free solve & Moderate to Low & [X] Deep BSDE and Deep Galerkin solves in hundreds of dimensions \citep{han2018solving,sirignano2018dgm}; PINN failure on harder PDEs \citep{krishnapriyan2021characterizing}. [G] PINN forward feasibility: Moderate $\cdot$ Supported; robustness: Weak $\cdot$ Not established. & Few trusted high-dimensional references or dimension-scaling studies (\S\mainref{sec:open_problems}). \\
\hline
Complex or irregular geometry & High for mature FEM; Moderate to Low for geometry learning & [T] Variational error analysis on unstructured meshes \citep{ern2004theory,brennerscott2008mathematical}. [X] Geometry-aware operators \citep{li2023geofno,li2023gino}; large errors or error accumulation under geometry shift on some CFDBench tasks \citep{luo2023cfdbench}. [M] CFDBench rankings: Kendall's $W=0.32$, best model wins $43\%$ of tasks (Table~\ref{tab:leaderboard_concordance}). & Boundary error and geometry-shift splits rarely reported; mesh and data-generation cost seldom counted. \\
\hline
\end{tabularx}
\end{table}

\begin{table}[H]
\centering
\caption{Evidence trace for the decision matrix, part 2.}
\label{tab:decision_trace_b}
\scriptsize
\setlength{\tabcolsep}{3.5pt}
\renewcommand{\arraystretch}{1.15}
\begin{tabularx}{\linewidth}{p{0.13\linewidth}p{0.15\linewidth}Xp{0.20\linewidth}}
\hline
\textbf{Regime} & \textbf{Confidence in Table~\mainref{tab:decision_matrix}} & \textbf{Supporting evidence} & \textbf{What caps the confidence} \\
\hline
Long time nonlinear or chaotic rollout & High for classical reference; Moderate to Low for emulators and hybrids & [X] ML accelerated CFD matched higher-resolution baselines with substantial speedups on two-dimensional turbulence \citep{kochkov2021mlaccelerated}; rollout instability traced to neglected frequencies \citep{lippe2023pderefiner}. [M] APEBench rankings: Kendall's $W=0.15$; FNO beats U-Net on $20\%$ of tasks (Table~\ref{tab:leaderboard_concordance}). & One step error dominates reporting; long-horizon statistics and attractor tests are rare. \\
\hline
Data informed inverse regime & High for mature adjoint workflows; Moderate for PINN style inverse use with validation & [T] Adjoint-based PDE-constrained optimization and variational assimilation with established convergence and regularization theory (\S\mainref{sec:taxonomy}). [G] PINN inverse identification: Moderate $\cdot$ Supported (Navier--Stokes and KdV discovery). & No comparison against an adjoint or variational assimilation baseline in the audited PINN work. \\
\hline
Shock dominated or conservation critical regime & High for classical FVM; Low to Moderate for learned methods & [T] Conservation and shock capture built into finite volume discretizations \citep{leveque2002finitevolume,toro2009riemann}; linear-reconstruction operators bounded below at order $1/p$ on advection and Burgers \citep{lanthaler2023nonlinear}. [X] DeepONet $28.5\%$ versus FNO $1.57\%$ error on inviscid Burgers \citep{lanthaler2023nonlinear}; conservation-aware neural methods in constrained settings \citep{discacciati2020controlling}. & No standardized cross-family demonstration of conservation-preserving learned shock capture at finite volume accuracy. \\
\hline
Limited paired data but known PDE & High for classical; Moderate for residual methods with evidence & [G] PINN forward feasibility: Moderate $\cdot$ Supported; efficiency: Weak $\cdot$ Not established. [X] PINN failure modes and their operator-dependent explanation \citep{krishnapriyan2021characterizing,deryck2024numerical}. [M] PINNacle: eleven PINN variants have eight different winners on 18 tasks, Kendall's $W=0.10$, and the task-by-method interaction in log error is five to seven times the method effect, so no variant is reliably best (Table~\ref{tab:leaderboard_concordance}). & Optimization reliability and classical cost comparison rarely reported; variant choice does not transfer across tasks. \\
\hline
\end{tabularx}
\end{table}

The trace makes two features of the matrix explicit. Every High confidence entry rests on theory plus independent empirical evidence, and no learned-method entry is above Moderate, because each lacks at least one equal-accuracy, cost-matched comparison or deployment-shift test. Every entry will change if the evidence changes, and the trace identifies which study would move it.

\section{Measured break-even study: protocol and full results}
\label{app:measured_nstar}

\paragraph{Problem.} $-\nabla\cdot(a\nabla u)=1$ on $(0,1)^2$ with $u=0$ on the boundary; $a=12$ where a Gaussian random field $\psi$ is non-negative and $a=3$ elsewhere, with $\psi$ expanded in a cosine basis (64 modes per direction) with eigenvalues $\pi^2(k^2+l^2)+\tau^2$, $\tau=3$, and decay exponent $\alpha=2$, following the Darcy setup of Li et al.\ \citep{li2021fno}. Each sample index fixes one field, which is evaluated exactly on any grid, so every resolution solves the same instances.

\paragraph{Classical solver.} Cell-centred finite volumes on an $n\times n$ grid with harmonic face averaging and Dirichlet data imposed at half-cell distance; conjugate gradients preconditioned with smoothed-aggregation algebraic multigrid (PyAMG 5.3), relative residual tolerance $10^{-6}$. A query comprises assembly, multigrid setup, and solve. Times are single-threaded on an Intel i7-12700K, the median over ten instances of the per-instance median of five repeats after one warm-up.

\paragraph{FNO.} Four Fourier layers, 12 modes, width 32, Adam with learning rate $10^{-3}$ halved every 100 epochs, relative $L^2$ loss, 500 epochs, pointwise Gaussian normalization, inputs $(a,x,y)$, and domain padding; batch size 20 up to $256^2$ and 8 at $512^2$. No hyperparameter search was performed. Training and inference ran on one NVIDIA RTX 3090 (PyTorch 2.3.1, CUDA 12.1); one query is a single forward pass at batch size one, timed with device synchronization.

\paragraph{Errors and matching.} Test errors (200 instances) are relative $L^2$ errors against solutions on a $2n\times2n$ grid averaged onto the $n\times n$ grid. For each FNO model, the classical comparator is the coarsest grid $m\in\{16,32,\ldots,n\}$ whose error, after bilinear interpolation to $n$, is no larger than the FNO error. Because the grids double, this comparator is at least as accurate as FNO and therefore more expensive than necessary, which favours FNO. The interpolated column instead interpolates $\log$ error and $\log$ cost linearly in $\log m$ between adjacent grids to obtain the classical cost at exactly the FNO error. $N^*=(N_{tr}c_{\mathrm{gen}}+C_{\mathrm{train}})/(c_{\mathrm{cl}}-c_{\mathrm{nn}})$ with $c_{\mathrm{gen}}$ the classical solve time at $n$. The training cost in network evaluations per training sample, $E_{\mathrm{eff}}=C_{\mathrm{train}}/(N_{tr}c_{\mathrm{nn}})$, has mean (coefficient of variation over the 15 runs at each resolution) 359 (0.96\%) at $64^2$, 983 (0.34\%) at $128^2$, 1185 (0.13\%) at $256^2$, and 1310 (0.90\%) at $512^2$.

\paragraph{Sanity checks.} With $10^3$ training samples, FNO errors (0.58--0.78\%) are of the same order as the 0.98--1.09\% reported by Li et al.; observed classical convergence orders between successive grids are 1.4--2.1; the classical error at $n$ is below every FNO error, so the reference resolves it; and all trainings ran on the GPU. No check failed, and all 60 trainings completed. Table~\ref{tab:measured_nstar_full} gives the full results, from which Table~\mainref{tab:measured_nstar} of the main article is drawn; the full logs, environment record, and analysis scripts are released with the article.

\begin{table}[H]
\centering
\caption{Full results of the measured break-even study (median over three seeds). $m^*$: matched classical grid; $m_{\mathrm{int}}$: grid size at which the interpolated classical error equals the FNO error.}
\label{tab:measured_nstar_full}
\scriptsize
\setlength{\tabcolsep}{3pt}
\renewcommand{\arraystretch}{1.1}
\begin{tabular}{rrcrrrrrrr}
\hline
$n$ & $N_{tr}$ & FNO error & $m^*$ & $c_{\mathrm{cl}}$ (ms) & $c_{\mathrm{nn}}$ (ms) & Data (s) & Training (s) & $N^*$ & $N^*$ (interp., $m_{\mathrm{int}}$) \\
\hline
64 & 100 & 4.81\% & 32 & 4.52 & 0.69 & 0.9 & 25 & $5.6\times10^{3}$ & $1.0\times10^{4}$ (17) \\
64 & 250 & 2.06\% & 32 & 5.25 & 0.69 & 2.2 & 62 & $1.4\times10^{4}$ & $1.5\times10^{4}$ (29) \\
64 & 500 & 1.18\% & 64 & 8.91 & 0.69 & 4.5 & 123 & $1.6\times10^{4}$ & $2.4\times10^{4}$ (39) \\
64 & 1000 & 0.78\% & 64 & 8.91 & 0.68 & 8.9 & 245 & $3.1\times10^{4}$ & $4.0\times10^{4}$ (47) \\
64 & 2000 & 0.60\% & 64 & 8.91 & 0.68 & 17.8 & 491 & $6.2\times10^{4}$ & $7.1\times10^{4}$ (54) \\
128 & 100 & 5.52\% & 16 & 3.82 & 0.78 & 2.4 & 77 & $3.5\times10^{4}$ & $3.5\times10^{4}$ (16) \\
128 & 250 & 2.33\% & 32 & 5.35 & 0.78 & 6.0 & 191 & $4.3\times10^{4}$ & $4.9\times10^{4}$ (28) \\
128 & 500 & 1.17\% & 64 & 9.10 & 0.78 & 12.1 & 381 & $4.7\times10^{4}$ & $6.4\times10^{4}$ (45) \\
128 & 1000 & 0.67\% & 128 & 24.16 & 0.78 & 24.2 & 763 & $3.4\times10^{4}$ & $9.3\times10^{4}$ (65) \\
128 & 2000 & 0.46\% & 128 & 24.16 & 0.78 & 48.3 & 1523 & $6.7\times10^{4}$ & $1.4\times10^{5}$ (78) \\
256 & 100 & 5.97\% & 16 & 3.79 & 2.60 & 10.1 & 308 & $7.0\times10^{5}$ & $7.0\times10^{5}$ (16) \\
256 & 250 & 2.46\% & 32 & 5.26 & 2.60 & 25.2 & 772 & $3.0\times10^{5}$ & $3.9\times10^{5}$ (28) \\
256 & 500 & 1.18\% & 64 & 9.14 & 2.60 & 50.3 & 1541 & $2.4\times10^{5}$ & $3.5\times10^{5}$ (47) \\
256 & 1000 & 0.66\% & 128 & 24.44 & 2.60 & 100.6 & 3081 & $1.5\times10^{5}$ & $4.2\times10^{5}$ (69) \\
256 & 2000 & 0.45\% & 128 & 24.44 & 2.60 & 201.3 & 6167 & $2.9\times10^{5}$ & $5.3\times10^{5}$ (90) \\
512 & 100 & 4.66\% & 32 & 5.31 & 9.52 & 43.7 & 1256 & never & never (18) \\
512 & 250 & 1.59\% & 64 & 8.92 & 9.52 & 109.2 & 3143 & never & never (38) \\
512 & 500 & 0.94\% & 64 & 8.92 & 9.54 & 218.3 & 6293 & never & never (55) \\
512 & 1000 & 0.58\% & 128 & 23.86 & 9.70 & 436.7 & 12579 & $9.2\times10^{5}$ & $6.1\times10^{6}$ (78) \\
512 & 2000 & 0.44\% & 128 & 23.86 & 9.72 & 873.4 & 25154 & $1.8\times10^{6}$ & $4.2\times10^{6}$ (96) \\
\hline
\end{tabular}
\end{table}

\begin{figure}[H]
\centering
\includegraphics[width=0.72\linewidth]{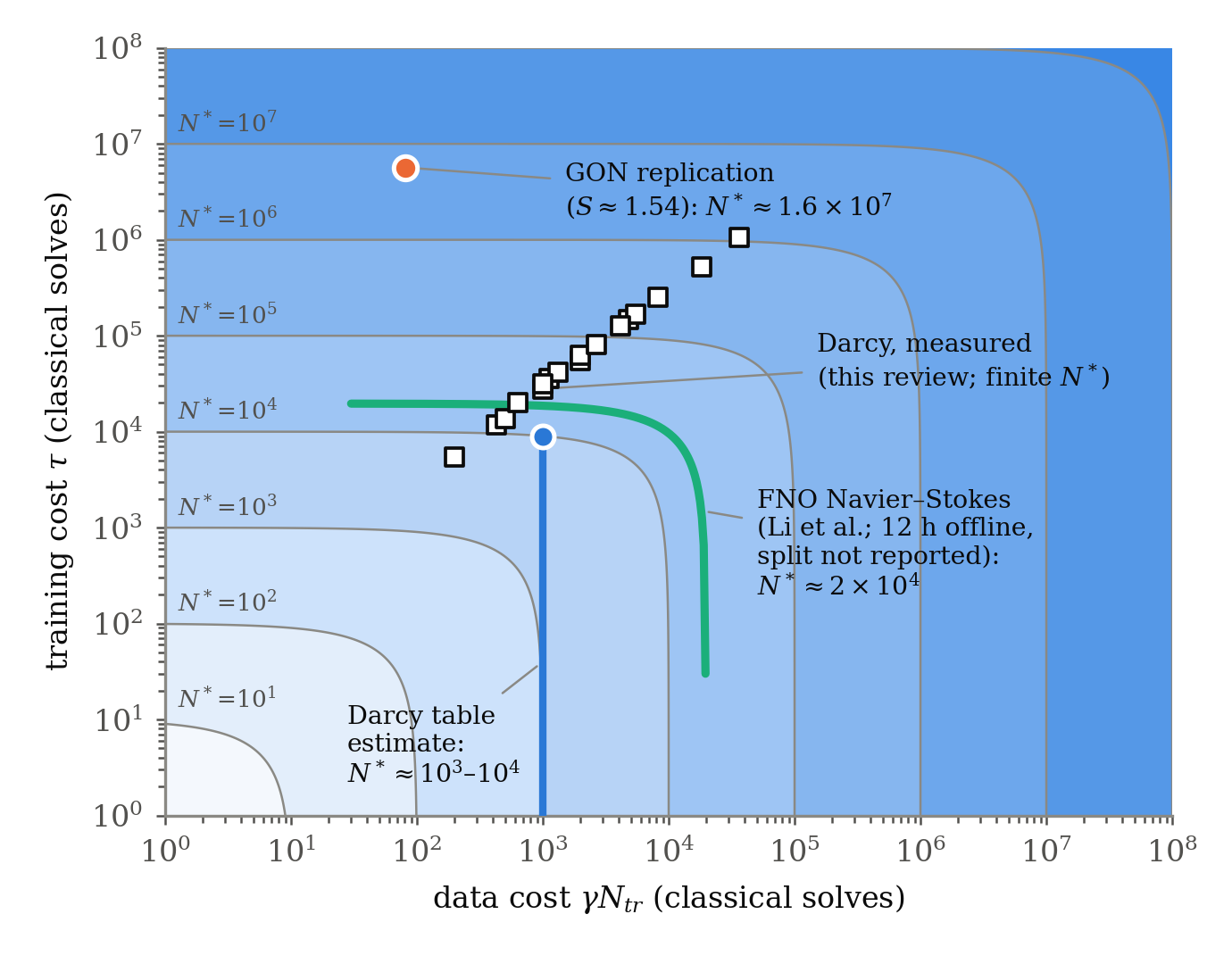}
\caption{Amortization phase diagram for Equation~(\mainref{eq:amortization_law}). Axes give the data cost $\gamma N_{tr}$ and the training cost $\tau$, both in classical solves at matched accuracy; contours show $N^*$ in the limit $S\to\infty$, and at fixed $\tau$ any $S \ge 10$ raises $N^*$ by at most 11\%; a faster network moves a point down, because $\tau = E_{\mathrm{eff}} N_{tr}/S$. The vertical bar is the Darcy operator estimate of Section~\mainref{subsec:cost_comparison}, whose training cost is not reported separately (the bar spans training costs up to $9\times10^3$ solves). The green segment is the original FNO Navier--Stokes study \citep{li2021fno}: its 12\,h of data generation and training equal about $2\times10^4$ classical solves, but their split is not reported, so every split on the segment is possible; with $S=440$, its break-even $N^*\approx2\times10^4$ lies below the $3\times10^4$ queries of the paper's MCMC application. The GON replication lies at $\tau \approx 5.7\times10^6$; its low speedup ($S\approx1.54$) raises its break-even to $N^*\approx1.6\times10^7$. Open squares are the measured Darcy models of Section~\mainref{subsec:measured_nstar} with finite break-even; they lie close to a line on which training costs about thirty times as much as data generation.}
\label{fig:amortization_phase_diagram}
\end{figure}

\paragraph{Phase diagram.} Figure~\ref{fig:amortization_phase_diagram} places the terms of the amortization law on one diagram, with the measured models of this study, the original FNO Navier--Stokes application, and the GON replication.

\paragraph{Accuracy scaling.} Conjecture~1 of the main text uses three exponents fitted to these data. The FNO data-scaling exponent $\alpha$, from a log-log fit of test error against $N_{tr}$ over all seeds, is 0.70, 0.84, 0.88, and 0.78 at $n=64$, 128, 256, and 512 over $N_{tr}=100$--$2000$, but 0.39, 0.53, 0.57, and 0.42 over $N_{tr}=1000$--$2000$: returns diminish at the accurate end, possibly because the architecture is fixed. The classical order $p$ between successive grids is 1.38--1.55 across the four series; the last step before $n$, 1.8--2.1, is inflated by proximity to the reference grid and is excluded. The classical cost exponent $q$, the slope of log cost against log inverse error, is 0.4--1.0 on the grids $m=16$--$128$ that match FNO, because multigrid setup dominates on coarse grids, and rises to 1.1--1.4 between $m=128$ and 512, approaching $d/p\approx1.4$. With $q$ and $\alpha$ taken over the matched range, $N^*\propto\varepsilon^{\,q-1/\alpha}$ predicts exponents of about $-1.0$ at $64^2$ ($q\approx0.43$, $1/\alpha=1.43$) and $-0.6$ at $128^2$ ($q\approx0.6$, $1/\alpha=1.19$); log-log fits of the measured $N^*$ against FNO error give $-0.88$ and $-0.54$ with the interpolated comparator ($-1.03$ and $-0.16$ with the coarsest-grid comparator, whose grid steps make $c_{\mathrm{cl}}$ piecewise constant). At $256^2$, $S$ is between 1.5 and 9.4, and the factor $1/(1-1/S)$ adds about $+0.4$ to the exponent, predicting about 0, against a measured $+0.11$; at $512^2$ only two training-set sizes break even, so no fit is made. Because $\alpha$ and $q$ are fitted to the same runs as $N^*$, this agreement checks the decomposition of $N^*$ rather than providing an independent prediction; the independent test is in three dimensions.

\paragraph{Reduced basis check.} To test the reduced basis estimate of Table~\ref{tab:cost_comparison_darcy} on the same problem, we computed the POD of the 2000 training solutions at each resolution by the method of snapshots (Euclidean inner product, float64), reading the snapshot files generated for the FNO study (script \texttt{pod\_decay.py}). Nothing was regenerated: the stored coefficient fields reproduce the field generator exactly, and a fresh multigrid solve reproduces the stored solutions to a relative error of $3\times10^{-8}$. Test errors are relative $L^2$ errors against the 200 reference solutions used for FNO. Table~\ref{tab:pod_check} reports the best approximation by $k$ POD modes, a POD--Galerkin reduced basis assembled from the full finite volume operator, and the POD tails of the coefficient $a$ and of the face transmissibilities, which bound how many affine terms empirical interpolation would need. The projection error is the same at every resolution, as expected, because every grid solves the same instances. It is also consistent with the only published reduced basis result for this benchmark: the POD-based reduced basis entry in Li et al.'s Darcy table has an error of 2.4--2.6\% \citep{li2021fno}, between our best approximation (2.1\%) and our POD--Galerkin error (3.8--4.4\%) at $k=30$. Matching values of $k$ are interpolated log-linearly between the computed $k\in\{100,200,500\}$. The numerical rank of every snapshot set is the full 2000. The online Galerkin times recorded by the script are not reported, because they reused the $k=200$ projection for every $k$ and ran with multithreaded linear algebra while the classical solve was timed on one core; the assembly time alone, reported in the last column, is a lower bound on any reduced basis query without an affine decomposition.

\begin{table}[H]
\centering
\caption{POD reduced basis check on the measured Darcy problem (2000 snapshots). Errors are mean relative $L^2$ errors on the 200 test instances. $k_{1000}$ and $k_{2000}$: POD dimension at which the best approximation matches the FNO error with $N_{tr}=1000$ and $2000$; in parentheses, the same with only 500 snapshots. Tails: number of POD modes needed for a 1\% energy tail of the coefficient $a$ and of the face transmissibilities. Assembly: one finite volume assembly in classical solves at $n$.}
\label{tab:pod_check}
\tiny
\setlength{\tabcolsep}{4pt}
\renewcommand{\arraystretch}{1.1}
\begin{tabular}{rcccccccc}
\hline
$n$ & Projection, $k=30$ & Galerkin, $k=30$ & Galerkin, $k=200$ & FNO error ($N_{tr}=10^3$/$2\times10^3$) & $k_{1000}$ & $k_{2000}$ & Tails $a$ / faces & Assembly \\
\hline
64 & 2.10\% & 3.75\% & 1.18\% & 0.78\% / 0.60\% & 125 (158) & 183 & 1851 / 1844 & 0.05 \\
128 & 2.13\% & 4.11\% & 1.56\% & 0.67\% / 0.46\% & 166 (233) & 298 & 1924 / 1903 & 0.06 \\
256 & 2.13\% & 4.30\% & 1.79\% & 0.66\% / 0.45\% & 173 (251) & 320 & 1931 / 1919 & 0.06 \\
512 & 2.13\% & 4.40\% & 1.91\% & 0.58\% / 0.44\% & 213 (348) & 335 & 1932 / 1925 & 0.06 \\
\hline
\end{tabular}
\end{table}

POD itself is cheap, 6--35 classical solves for 2000 snapshots, so the offline cost of a reduced basis is essentially its snapshot cost, the same $\gamma N_{tr}$ term that FNO pays for its training data. The two methods differ in what follows: FNO adds a training cost about thirty times its data cost, while the reduced basis adds almost nothing offline but cannot reach a cheap online stage, because the coefficient does not compress. Reaching FNO's accuracy with a POD--Galerkin basis would need several hundred to several thousand modes on extrapolation from $k\le200$, which we did not compute.

\section{Tables and figures moved from the main article}
\label{app:moved}

\begin{figure}[H]
\centering
\includegraphics[width=0.85\linewidth]{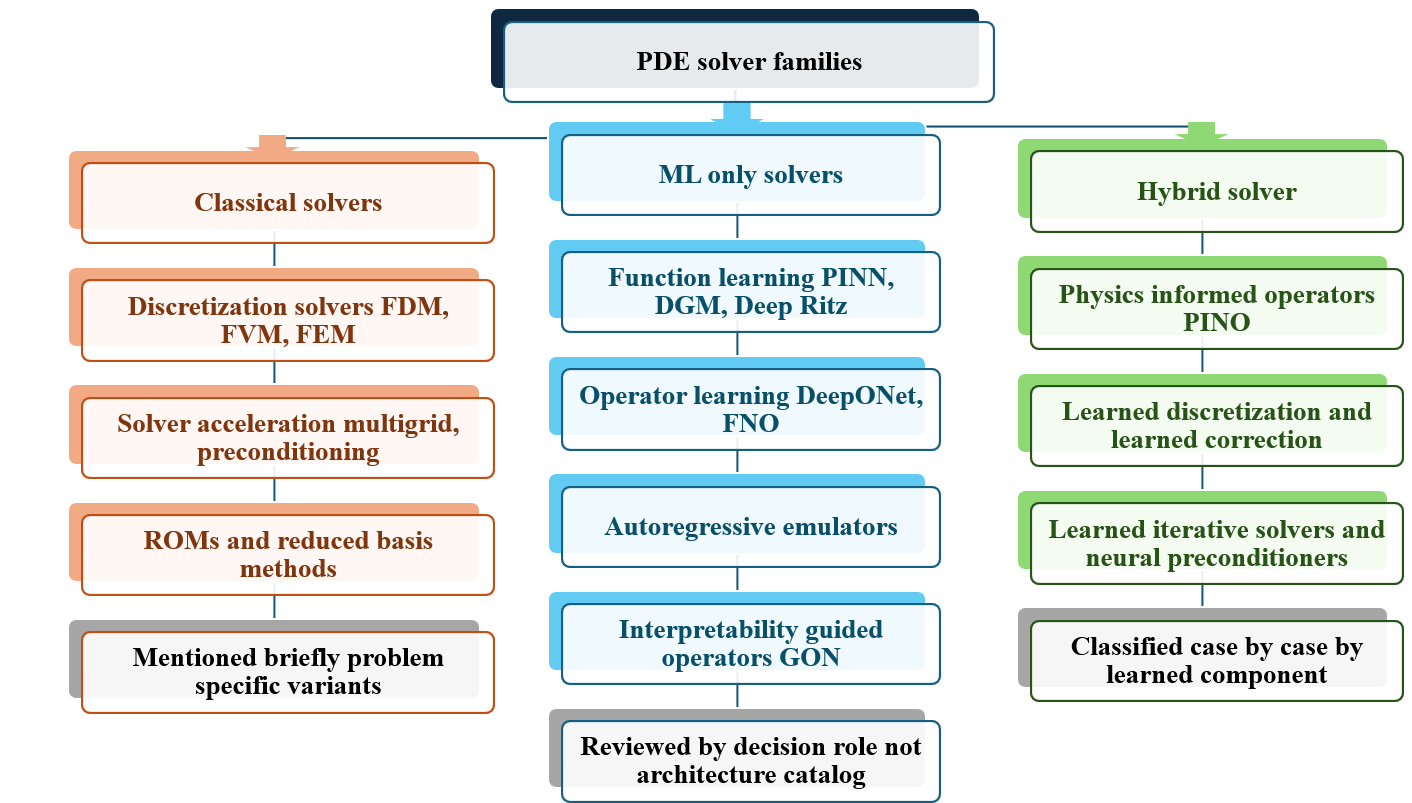}
\caption{Taxonomy of classical, ML only, and hybrid PDE solvers. Solid boxes indicate families reviewed in depth because they recur in solver selection decisions, while lighter boxes indicate methods or variants mentioned only when they clarify a decision regime.}
\label{fig:taxonomy}
\end{figure}

\begin{figure}[H]
\centering
\includegraphics[width=\linewidth]{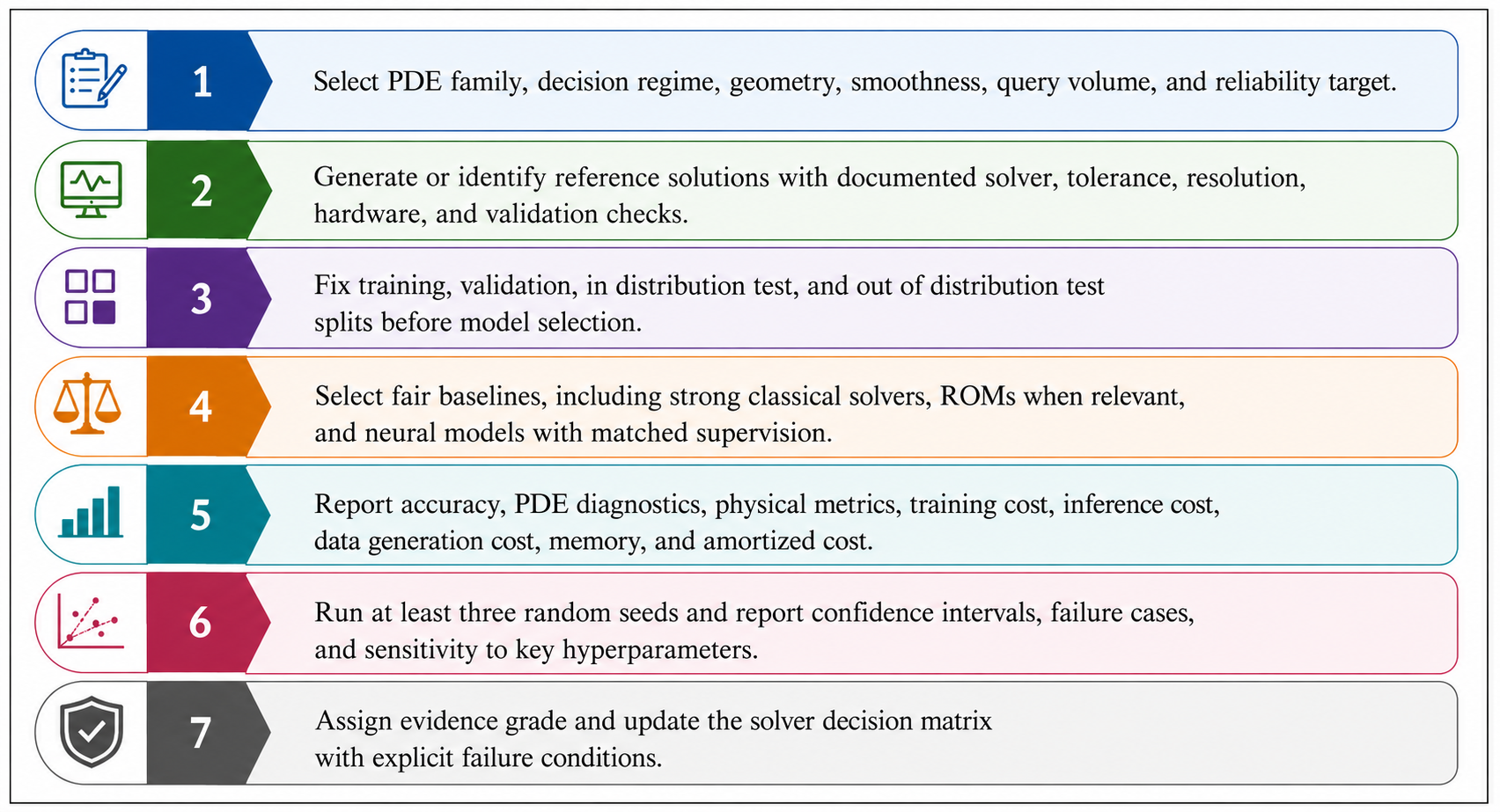}
\caption{Evidence weighted benchmarking protocol. The protocol links PDE selection, reference solution construction, baseline fairness, cost accounting, out of distribution tests, evidence grading, and decision matrix updates.}
\label{fig:evidence_protocol}
\end{figure}

The tables and figures below were moved from the main article for length; the main text summarizes each and points here. They appear in the order of the main-text sections they support: the solver taxonomy (Figure~\ref{fig:taxonomy}, Section~\mainref{sec:taxonomy}); the benchmark suites, the recommended core PDE problems, and the evidence weighted benchmarking protocol (Tables~\ref{tab:benchmark_suites} and~\ref{tab:core_pde_benchmarks} and Figure~\ref{fig:evidence_protocol}, Section~\mainref{sec:benchmark_landscape}); the order-of-magnitude Darcy cost estimate (Table~\ref{tab:cost_comparison_darcy}, Section~\mainref{subsec:cost_comparison}); the GRADE mapping (Table~\ref{tab:grade_mapping}, Section~\mainref{subsec:evidence_rubric}); and the research priorities (Table~\ref{tab:open_problem_priorities}, Section~\mainref{sec:open_problems}).

\begin{table}[H]
\centering
\caption{Benchmark suites and their evidence coverage. ROM baselines are not a central focus in any listed suite.}
\label{tab:benchmark_suites}
\tiny
\setlength{\tabcolsep}{2pt}
\renewcommand{\arraystretch}{1.0}
\begin{tabularx}{\linewidth}{R{0.85}R{1.1}R{1.0}R{0.9}R{1.0}R{0.95}R{0.75}R{1.45}}
\hline
\textbf{Benchmark} & \textbf{PDE families} & \textbf{Solver families} & \textbf{OOD tests} & \textbf{Cost metrics} & \textbf{Classical baselines} & \textbf{Code/data} & \textbf{Main limitation} \\
\hline
PDEBench & Advection, Burgers, diffusion-reaction, diffusion-sorption, Darcy, Navier--Stokes, shallow water & Neural operators, CNN baselines, other scientific ML & Parameter and resolution shifts & Timing present; full amortized cost needs careful interpretation & Numerical simulations as reference & Provided & Strong for broad ML comparison; ROM and full offline cost accounting still needed \\
\hline
PINNacle & Diverse PDE tasks for PINNs & PINN variants and training strategies & Problem-dependent stress tests & Training performance central; classical replacement cost not the focus & Reference solvers used for evaluation & Available & Strong for PINN comparison; cross-family recommendation needs added classical baselines \\
\hline
CFDBench & Lid-driven cavity, tube flow, dam flow, Karman vortex & DeepONet, FNO, U-Net, feed-forward surrogates & BC, property, and geometry generalization & Inference metrics reported; full generation cost a decision gap & Numerical CFD as reference & Available & Strong for CFD generalization; limited to selected regimes and surrogate families \\
\hline
FlowBench & Complex flow and heat transfer, 2D and 3D geometries & FNO, CNO, WNO, DeepONet, foundation models & Geometry, Re, Gr, and flow condition variation & Field and engineering quantity metrics; full deployment cost separate & DNS as reference & Public dataset & Strong for complex geometry surrogates; classical and ROM comparisons require extra experiments \\
\hline
APEBench & Autoregressive emulator tasks in 1D, 2D, 3D & Autoregressive emulators and neural operator baselines & Multiple PDE and rollout configurations & Runtime and rollout evaluation emphasized & Pseudo-spectral simulator & JAX code & Strong for rollout stress; periodic pseudo-spectral settings limit geometry coverage \\
\hline
The Well & Diverse spatiotemporal physical systems & Surrogate and foundation-style ML models & Dataset-dependent shift evaluation & Dataset scale central; solver cost varies by source & Numerical simulations as reference & Public & Strong for scale and diversity; solver recommendation needs task-specific metadata and baselines \\
\hline
AL4PDE & Parametric time-dependent PDEs for active learning & Neural PDE solvers with active learning & Active learning query policies & Data acquisition cost part of motivation & Numerical solvers for data generation & Available & Strong for data efficiency; not a general cross-family solver benchmark \\
\hline
RealPDEBench & Real measured and paired simulated physical systems & Scientific ML baselines for sim-to-real tasks & Sim-to-real transfer and dataset shift & Metrics reported; solver replacement evidence emerging & Numerical simulations paired with measurements & Preprint & Useful measurement focus; evidence preliminary until broader replication \\
\hline
\end{tabularx}
\end{table}

\begin{table}[H]
\centering
\caption{Recommended core PDE problems for cross family evaluation.}
\label{tab:core_pde_benchmarks}
\tiny
\setlength{\tabcolsep}{3pt}
\renewcommand{\arraystretch}{1.18}
\begin{tabularx}{\linewidth}{R{0.7}R{0.7}R{1.05}R{1.0}R{1.05}R{1.05}R{1.2}R{1.0}R{1.25}}
\hline
PDE problem & PDE type & Regime stressed & Suitable classical baseline & Suitable ML baseline & Suitable hybrid baseline & Data requirement & Metrics & Reason for inclusion \\
\hline
Poisson & Elliptic & Boundary handling, geometry, and error control & FEM, adaptive FEM, multigrid & PINN, Deep Ritz, DeepONet, FNO & Learned multigrid or neural preconditioner & No paired data for classical and residual solvers, paired corpus for operator learning & Relative $L^2$, $H^1$, boundary error, residual, wall clock & Tests whether learning improves a regime where classical solvers are mature \\
\hline
Darcy flow & Elliptic parametric & Heterogeneous coefficients and many query inference & FEM, multigrid, reduced basis where applicable & DeepONet, FNO, U-NO, GON style variants & PINO or learned correction & Paired coefficient solution corpus for supervised operator learning & Relative $L^2$, flux error, coefficient OOD, data generation cost & Standard neural operator test where ROM comparison and corpus cost matter \\
\hline
Heat or diffusion & Parabolic & Smooth time dependent dynamics and stability & FDM, FEM, implicit time stepping, multigrid & PINN, DeepONet, FNO, autoregressive emulator & PINO or learned time stepping correction & Residual signal or paired trajectories depending on method & Relative $L^2$, gradient error, rollout error, training and inference cost & Provides a controlled parabolic regime before harder nonlinear dynamics \\
\hline
Burgers & Nonlinear convection diffusion & Shocks, steep gradients, and viscosity dependence & FVM, spectral methods with dealiasing where appropriate & PINN, FNO, autoregressive emulator & Learned discretization or PINO & Paired trajectories for operator or emulator methods & Relative $L^2$, shock location, conservation, rollout stability, viscosity OOD & Tests whether models handle sharp structures rather than only smooth interpolation \\
\hline
Allen Cahn or Gray Scott & Nonlinear reaction diffusion & Interfaces, metastability, and pattern formation & FEM, FDM, spectral methods & PINN, neural operator, autoregressive emulator & Learned correction or physics informed operator & Paired trajectories or residual signal depending on method & Interface error, pattern statistics, rollout error, residual, cost & Tests nonlinear dynamics with spatial structure and sensitivity to time integration \\
\hline
Incompressible Navier Stokes & Fluid dynamics & Pressure velocity coupling, conservation, and long rollout & FVM, FEM, spectral solver, CFD reference solver & FNO, DeepONet, graph emulator, autoregressive emulator & ML accelerated CFD or learned closure & Paired flow trajectories are usually required for supervised surrogates & Field error, divergence, kinetic energy, drag or lift, rollout stability, cost & Tests physically meaningful fluid quantities and dynamic stability \\
\hline
Kuramoto Sivashinsky & Chaotic nonlinear dynamics & Long horizon rollout and attractor fidelity & Spectral time stepping reference & Autoregressive emulator, neural operator & Solver in the loop correction & Paired trajectories for learned rollouts & Short horizon error, long horizon statistics, spectrum, stability failures & Tests whether one step accuracy translates into useful long time behavior \\
\hline
High dimensional parabolic PDE & High dimensional PDE & Curse of dimensionality and mesh free approximation & Problem specific sparse grid or stochastic numerical baseline when available & DGM, PINN, Deep Ritz & Residual based hybrid or variance reduced solver in the loop method & Residual supervision or sampled terminal and boundary conditions & Relative error against trusted reference, variance, training cost, sensitivity to dimension & Tests the regime where neural function learning may address a genuine classical bottleneck \\
\hline
Helmholtz or Maxwell equations & Wave and electromagnetic & High frequency oscillations, indefinite operators, and conservation of divergence or curl & DG, mixed FEM with N\'ed\'elec or Raviart-Thomas elements, Helmholtz preconditioned Krylov solver & PINN, DeepONet, FNO & Learned preconditioner, PINO & Paired solutions at multiple wavenumbers for operator learners & Relative $L^2$, wavenumber sweep, divergence or curl error, conservation, cost versus wavenumber & Tests whether learned solvers handle high-frequency oscillations, indefiniteness, and structure-preservation at a level comparable to mixed FEM \\
\hline
Linear or nonlinear elasticity & Structural mechanics & Complex geometry, stress concentration, and engineering quantities of interest & FEM with stable elements, adaptive FEM for singularities, mixed FEM for near-incompressibility & DeepONet, FNO, graph operator & Learned correction or physics informed operator & Paired displacement and stress fields for operator learners & Relative $L^2$, $H^1$, stress error, locking diagnostics, cost, geometry OOD & Tests learned operators on a practical engineering domain with certification needs and near-incompressibility sensitivity \\
\hline
\end{tabularx}
\end{table}

\begin{table}[H]
\centering
\scriptsize
\setlength{\tabcolsep}{4pt}
\renewcommand{\arraystretch}{1.1}
\caption{Worked cost comparison for the 2D Darcy flow parametric benchmark ($64\times64$ grid, $10^3$ training instances). Costs are normalized relative to one direct FEM solve (CPU, AMG preconditioner, $64^2$ grid). All figures are representative order-of-magnitude estimates from published benchmarks \citep{li2021fno,lu2021deeponet,hesthaven2016certified,takamoto2022pdebench}; actual values depend on hardware, implementation, and problem dimension. GPU and CPU costs are \emph{not} normalized; the table reports cost ratios rather than absolute wall-clock times. Crossover query $N^*$ is measured against direct FEM: it is the number of online queries at which offline~$+$~$N^* \times$ online equals $N^*$ direct FEM solves. \textdagger~The certified reduced basis provides an a posteriori error bound at each query; FNO and DeepONet do not. This certification gap is not captured in the $N^*$ column. The certified RB row assumes affine, low dimensional parameter dependence; the thresholded coefficient of the standard benchmark has neither, and the measured POD row (Table~\ref{tab:pod_check}) shows what a reduced basis achieves on it.}
\label{tab:cost_comparison_darcy}
\begin{tabularx}{\linewidth}{R{1.05}R{1.0}R{1.0}R{1.0}R{1.0}R{1.0}R{0.95}}
\hline
\textbf{Method} & \textbf{Offline cost ($\times$\,FEM)} & \textbf{Online cost per query ($\times$\,FEM)} & \textbf{Rel.\ $L^2$ error} & \textbf{Corpus cost counted?} & \textbf{Crossover query $N^*$} & \textbf{Evidence grade} \\
\hline
Direct FEM ($64^2$, AMG) & 1 (reference) & 1 & $< 10^{-4}$ (solver tol.) & N/A & N/A & Strong \citep{ern2004theory} \\
Certified RB ($r\!=\!30$, greedy + SCM), affine parameter dependence assumed\textdagger & ${\sim}10^3$ & ${\sim}10^{-3}$ & ${\sim}10^{-3}$ (cert.\ bound) & N/A & ${\sim}10^3$ & Strong (certified) in favorable structure \citep{hesthaven2016certified} \\
POD--Galerkin RB on the thresholded coefficient (measured, Table~\ref{tab:pod_check}) & ${\sim}10^3$ (snapshots; POD ${\le}35$) & ${\ge}0.05$ (reassembly; no affine decomposition) & $3.8\%$ ($r\!=\!30$), $1.2\%$ ($r\!=\!200$) & Yes & Not established at FNO accuracy & Measured, no certificate \\
FNO ($k_{\max}\!=\!12$, 4 layers, 1000 training solves) & ${\sim}10^3$--$10^4$ (corpus + train) & ${\sim}10^{-2}$ & ${\sim}10^{-2}$ (relative $L^2$) & Rarely & ${\sim}10^3$--$10^4$ & Moderate (no cert.) \citep{li2021fno,takamoto2022pdebench} \\
DeepONet (unstacked, 1000 training solves) & ${\sim}10^3$--$10^4$ (corpus + train) & ${\sim}10^{-2}$ & ${\sim}10^{-2}$--$10^{-1}$ & Rarely & ${\sim}10^3$--$10^4$ & Moderate (no cert.) \citep{lu2021deeponet,takamoto2022pdebench} \\
\hline
\end{tabularx}
\end{table}

\begin{table}[H]
\centering
\caption{GRADE domains \citep{guyatt2008grade,guyatt2011grade3} and their counterparts in the claim strength rubric for PDE solvers.}
\label{tab:grade_mapping}
\scriptsize
\setlength{\tabcolsep}{4pt}
\renewcommand{\arraystretch}{1.15}
\begin{tabularx}{\linewidth}{p{0.26\linewidth}X}
\hline
\textbf{GRADE domain} & \textbf{Counterpart for PDE solver claims} \\
\hline
Study limitations (risk of bias) & Weak, untuned, or unpreconditioned classical baseline; tuning effort spent only on the learned method; reference solutions at a coarse tolerance \\
Inconsistency & Method ranking that changes across PDE families or benchmark suites (Section~\mainref{subsec:benchmark_suites}) \\
Indirectness & Benchmark PDE, distribution, or geometry different from the deployment; in-distribution splits only \\
Imprecision & Single seed, no confidence intervals, or differences within seed variance \\
Reporting (publication) bias & Only configurations that train successfully are reported; failure cases omitted \\
Upgrade: large effect & A large advantage that persists under an equal-accuracy, cost-matched protocol \\
Upgrade: dose--response & An advantage that grows with resolution, dimension, or query volume as theory predicts \\
Upgrade: confounding & An advantage that persists against a comparator tuned more strongly than the learned method \\
Addition: replication & Independent reproduction of the reported result within its stated uncertainty \\
\hline
\end{tabularx}
\end{table}

\begin{table}[H]
\centering
\caption{Research priorities linked to decision barriers.}
\label{tab:open_problem_priorities}
\tiny
\setlength{\tabcolsep}{3pt}
\renewcommand{\arraystretch}{1.0}
\begin{tabularx}{\linewidth}{p{0.13\linewidth}p{0.14\linewidth}p{0.18\linewidth}p{0.13\linewidth}p{0.18\linewidth}X}
\hline
Open problem & Affected solver families & Decision barrier created by the problem & Current evidence status & Promising research direction & Required benchmark or evidence \\
\hline
Out of distribution trustworthiness & Operator learning, autoregressive emulators, hybrid surrogates & A model can appear accurate on an in distribution test while failing under parameter, geometry, boundary, resolution, or forcing shifts & Moderate for benchmarked shifts, weaker for deployment level shifts & Fixed shift taxonomies, deployment matched splits, and physical diagnostics tied to each PDE family & Parameter shift, coefficient shift, boundary shift, geometry shift, resolution shift, noise shift, and query volume reporting \\
\hline
Uncertainty quantification & Neural operators, PINNs, hybrid solvers, Bayesian surrogates, ROM comparisons & Recommendations lack calibrated error awareness for design, inverse inference, and risk sensitive use; neural estimates are rarely calibrated against multilevel Monte Carlo or polynomial chaos baselines & Mature for reduced basis, stochastic Galerkin, and MLMC; incomplete for most neural solvers & Calibrate neural uncertainty against MLMC and stochastic Galerkin at matched cost, tied to residuals, data coverage, and reference error & Coverage probability, calibration curves, OOD detection, reference tolerance, and MLMC or stochastic Galerkin cost at matched accuracy \\
\hline
Geometry generalization & FEM competitors, geometry aware operators, graph solvers, hybrid correction methods & Learned solvers may not transfer from training geometries to deployment geometries or boundary conditions & Moderate for emerging geometry aware operators, strong for mature FEM on meshed domains & Mesh aware operators, boundary aware losses, geometry shift benchmarks, and cost accounting for meshing and data generation & Boundary error, geometry split definitions, mesh quality variation, finite element baselines, and data generation cost \\
\hline
Shock and discontinuity preservation & FVM baselines, neural operators, PINNs, learned discretizations, emulators & Low average error can hide shock displacement, conservation loss, or near discontinuity failure & Strong classical evidence for conservative schemes, weaker neural evidence across broad shock regimes & Conservation aware learning, shock diagnostics, limiter comparisons, and learned components inside conservative workflows & Conservation error, shock location error, near shock error, entropy diagnostics where appropriate, and finite volume baselines \\
\hline
Long horizon reliability & Autoregressive emulators, neural operators for dynamics, ML accelerated CFD, hybrid correction methods & One step accuracy may not imply stable or physically meaningful rollout & Moderate for selected emulator benchmarks, weaker for chaotic and deployment shifted regimes & Training and evaluation based on rollout stability, attractor fidelity, invariants, and physical statistics & Long horizon rollout, stability failures, invariant drift, spectra, attractor statistics, and shifted forcing tests \\
\hline
Hybrid convergence guarantees & Learned iterative solvers, neural preconditioners, learned multigrid, solver in the loop methods & Learned updates can cause false fixed points, stagnation, or convergence failure outside training regimes & Emerging and problem dependent & Residual monitored learned updates, fallback classical iterations, convergence aware training, and robustness tests across operator families & Residual histories, stopping criteria, iteration counts, failure cases, shifted coefficients, and strong preconditioned baselines \\
\hline
Data efficiency & Function learning, operator learning, hybrid solvers, ROMs & Offline solution generation or residual optimization can dominate the total cost & Moderate for selected benchmarks, incomplete for full amortized accounting & Learning curves tied to solution corpus size, residual samples, active sampling, and amortized query volume & Accuracy versus data budget, generation time, training time, inference time, query volume, and comparison with reduced basis methods \\
\hline
Solver in the loop data acquisition & Operator learning, hybrid correction, PINO, active learning based solvers & Uniform sampling can waste reference solves and miss failure regions & Emerging through active learning and differentiable solver studies & Use residuals, uncertainty, validation error, and physical diagnostics to guide new classical solves & Active data acquisition benchmarks, reference solve counts, failure region discovery, and total cost comparison \\
\hline
Reproducibility infrastructure & All solver families & Claims cannot be converted into decision confidence without code, data, tolerances, hardware, and baseline details & Improving through benchmark suites but incomplete for cross family decisions & Shared extraction schema, standardized cost reporting, fixed splits, and replication protocols & Code, data, random seeds, hardware, solver tolerances, hyperparameters, baseline tuning, and failure case logs \\
\hline
High dimensional PDE validation & Function learning, weak formulations, stochastic numerical baselines, selected hybrids & Classical references can be difficult while neural validation remains uncertain & Promising but narrow in many studies & Dimension scaling protocols, trusted low dimensional reductions, stochastic reference checks, and uncertainty aware validation & Error versus dimension, sampling sensitivity, independent references, random seeds, and comparison with sparse grid or stochastic baselines \\
\hline
Approximation theory for PDE solution spaces & Function learning, operator learning, hybrid surrogates, comparison with high-order FEM & The convergence rate of a learned solver relative to $hp$-FEM or spectral methods is not stated or tested in most studies, making it impossible to judge when learning is theoretically competitive & Emerging: exponential ReLU expressivity results for elliptic PDEs now available, but operator learning and function learning rates are often not matched to benchmark regimes & Connect neural approximation rates to solution regularity, match network design to singularity class, compare learning curves against $hp$ adaptive baselines & Accuracy versus depth and width, comparison with adaptive high-order classical scheme, singularity-resolved error, and training cost tied to approximation class \\
\hline
\end{tabularx}
\end{table}

\bibliographystyle{ACM-Reference-Format}
\bibliography{supplement}